\documentclass[fleqn,usenatbib]{mnras}

\usepackage{newtxtext,newtxmath}

\usepackage[T1]{fontenc}

\DeclareRobustCommand{\VAN}[3]{#2}
\let\VANthebibliography\thebibliography
\def\thebibliography{\DeclareRobustCommand{\VAN}[3]{##3}\VANthebibliography}

\usepackage{graphicx}	
\usepackage{amsmath}	

\title[AU Mic flare spectroscopy]{Spectroscopic observations of flares and superflares on AU~Mic\thanks{Based on observations collected at the ESO1.52m telescope at European Southern Observatory in La Silla.}}

\author[P. Odert et al.]{
P. Odert,$^{1}$\thanks{E-mail:petra.odert@uni-graz.at}
M. Leitzinger,$^{1}$
R. Greimel,$^{2}$
P. Kab\'ath,$^{3}$
J. Lipt\'ak,$^{3,4}$
P. Heinzel,$^{3,5}$
R. Karjalainen,$^{3}$
\newauthor
J. Wollmann,$^{3}$
E.W. Guenther,$^{6}$
M. Skarka,$^{3}$
J. Srba,$^{3}$
P. \v{S}koda,$^{3}$
J. Fr\'yda,$^{4}$
R. Brahm,$^{7,8,9}$
L. Vanzi,$^{10,11}$
\newauthor
J. Jan\'ik$^{12}$
\\
$^{1}$Institute of Physics/AGP, University of Graz, Universit\"atsplatz 5, A-8010 Graz, Austria\\
$^{2}$RG Science, Schanzelgasse 17, A-8010 Graz, Austria\\
$^{3}$Astronomical Institute, Czech Academy of Sciences, 25165 Ond\v{r}ejov, Czech Republic\\
$^{4}$Astronomical Institute of Charles University, V Hole\v{s}ovi\v{c}k\'{a}ch 2, 180 00 Prague, Czech Republic\\
$^{5}$University of Wroc{\l}aw, Center of Scientific Excellence – Solar and Stellar Activity, Kopernika 11, 51-622 Wroc{\l}aw, Poland\\
$^{6}$Th\"uringer Landessternwarte Tautenburg, Sternwarte 5, 07778 Tautenburg, Germany\\
$^{7}$Facultad de Ingeniera y Ciencias, Universidad Adolfo Ib\'a\~nez, Av. Diagonal las Torres 2640, Pe\~nalol\'en, Santiago, Chile\\
$^{8}$Millennium Institute for Astrophysics, Chile\\
$^{9}$Data Observatory Foundation, Chile\\
$^{10}$Centre of Astro-Engineering, Pontificia Universidad Catolica de Chile, Av. Vicu\~na Mackenna 4860, Santiago, Chile\\
$^{11}$Department of Electrical Engineering, Pontificia Universidad Catolica de Chile, Av. Vicu\~na Mackenna 4860, Santiago, Chile\\
$^{12}$Department of Theoretical Physics and Astrophysics, Faculty of Science, Masaryk University, Kotl\'{a}\v{r}sk\'{a} 2, Brno, CZ-611 37, Czech Republic
}

\date{Accepted XXX. Received YYY; in original form ZZZ}

\pubyear{2015}

\begin{document}
\label{firstpage}
\pagerange{\pageref{firstpage}--\pageref{lastpage}}
\maketitle

\begin{abstract}
The young active flare star AU~Mic is the planet host star with the highest flare rate from \textit{TESS} data. Therefore, it represents an ideal target for dedicated ground-based monitoring campaigns with the aim to characterize its numerous flares spectroscopically. We performed such spectroscopic monitoring with the ESO1.52m telescope of the PLATOSpec consortium. In more than 190~hours of observations, we find 24 flares suitable for detailed analysis. We compute their parameters (duration, peak flux, energy) in eight chromospheric lines (H$\alpha$, H$\beta$, H$\gamma$, H$\delta$, \ion{Na}{i}\,D1\&D2, \ion{He}{i}\,D3, \ion{He}{i}\,6678) and investigate their relationships. Furthermore, we obtained simultaneous photometric observations and low-resolution spectroscopy for part of the spectroscopic runs. We detect one flare in the \textit{g'}-band photometry which is associated with a spectroscopic flare. Additionally, an extreme flare event occurred on 2023-09-16 of which only a time around its possible peak was observed, during which chromospheric line fluxes were raised by up to a factor of three compared to the following night. The estimated energy of this event is around $10^{33}$\,erg in H$\alpha$ alone, i.e. a rare chromospheric line superflare.
\end{abstract}




\section{Introduction}
Stellar activity on late-type main-sequence stars seems to be very similar to what we understand on the Sun as solar activity. The most energetic activity phenomena on the Sun and on late-type main-sequence stars are outbreaks of radiation (flares) and mass expulsions into the helio-/astrosphere termed coronal mass ejections (CMEs). Both phenomena have been extensively observed and studied on the Sun. On the stellar side flares are much better investigated than CMEs. Especially since the satellite missions \textit{CoRoT} \citep{Baglin2006}, \textit{Kepler} \citep{Borucki10}, and now \textit{TESS} \citep{Ricker2015}, statistical investigations have become reality which gave the stellar flare research a significant boost. As CMEs can not be identified the same way as flares, their investigation still requires a significant amount of effort, although a lot has been published on this topic in the last decade \citep[][and references therein]{Moschou2019, Leitzinger2022, Osten2023, Tian2023}.

Flares on stars are known since the first half of the last century \citep{Hertzsprung1924, vanMaanen1940, Luyten1949, Joy1949}. From that time also the term ``UV Ceti'' type stars originates. Nowadays we know that stars throughout the main sequence show flares and even on giant stars flares have been reported. Spectroscopic investigations of active stars have been performed during the last decades \citep[e.g.][]{Pettersen1989}. Starting in the late seventies of the last century satellite missions operating in X-rays (\textit{Einstein Observatory}, \textit{EXOSAT}, \textit{ROSAT}, \textit{Beppo-SAX}, \textit{ASCA}, \textit{XMM-Newton}, \textit{Chandra}, \textit{Swift}, \textit{Astrosat}) provided insights into the high-energy emission of stellar flares. In addition, missions operating in the ultraviolet, such as \textit{IUE}, \textit{EUVE}, \textit{FUSE} and \textit{HST} enabled and enable coordinated multi-wavelength campaigns of stellar flares. Prominent flare stars such as Proxima~Cen, AD~Leo, YZ~CMi, EV~Lac, AT~Mic and later also fainter stars were targets of multi-wavelength campaigns aiming at the characterization of the multi-wavelength nature of stellar flares. Relations of flare fluxes in different wavelength regimes were established \citep[e.g.][]{Haisch1981, Kahler1982, Baliunas1984, Nelson1986, Doyle1988, Butler1988, Ambruster1989}, and the relation of spectral line to continuum fluxes were investigated \citep[e.g.][]{Houdebine1992, Hawley2003, Houdebine2003}. Apart from photometry, continuum enhancements in flares were detected spectroscopically \citep[e.g.][]{Eason1992, vandenOord1996, Kowalski2010, Kowalski2012, Kowalski2013, Muheki2020b}. Some stellar flares were found to exhibit the Neupert effect \citep[e.g.][]{Hawley1995, Guedel1996, Guedel2002, Mitra-Kraev2005, Fuhrmeister2011}, similar as in solar flares. Spectral line broadening and asymmetries during flares were investigated \citep[e.g.][]{Worden1984, Phillips1988, Doyle1988, Fuhrmeister2005, Fuhrmeister2011, Fuhrmeister2018, Vida2019, Muheki2020a, Muheki2020b, Koller2021, Wu2022, Namekata2022, Wollmann2023, Notsu2024, Namekata2024}, which also enabled the search for filament/prominence eruptions possibly accompanying the flares. Optical coronal lines could be identified in some flares as well \citep[e.g.][]{Fuhrmeister2007, Muheki2020b}. For a deeper read on stellar flares we refer the reader to reviews presented by \citet{Mullan1977, Pettersen1989, Bastian1994, Kowalski2024}.

One of the major topics in stellar flare research in the last decade was to better understand superflares, which are very energetic flares often simply defined by an energy threshold of $>$10$^{33}$erg \citep[e.g.][]{Maehara2012}. This threshold was chosen to define superflares as events larger than the most energetic flares of the present-day Sun, such as the Carrington event which had an estimated energy of a few 10$^{32}$\,erg \citep{Hayakawa23}. Apart from this simple definition, \citet{Mullan2018} suggested a possible physical difference between solar-like and larger flares, as indicated by a break in power law slopes of flare energy distributions occurring between 10$^{32}$ and 10$^{33}$\,erg for solar-like stars. That superflares may be a special kind of phenomenon has also been discussed in \citet{Cliver2022}. Superflares have been reported in the literature as single detections in the past \citep[see][]{Schaefer1997, Schaefer1998, Schaefer2000}. The \textit{Kepler} satellite enabled first statistical investigations of this phenomenon. \citet{Maehara2012} presented the first paper of a number of publications dedicated to the statistical determination of superflare frequency and energy distribution on solar-like stars using \textit{Kepler} data \citep[e.g.][]{Shibayama2013, Candelaresi2014, Maehara2015, Davenport2016, Balona2015, Akopian2017, Okamoto2021, Althukair2023b, Althukair2023a}. Consequently, also spectroscopic investigations of superflare stars were performed to see if those stars were by any means special \citep{Notsu2015a, Notsu2015b, Honda2015}. It was found that a fraction of the stars show values for effective temperature and surface gravity being roughly in the range of solar values, leading to the conclusion that superflares may also have occurred on our Sun.

The \textit{Kepler} mission observed a fixed field in Cygnus, whereas \textit{TESS} is an all-sky survey. Studies with similar goals were undertaken also with \textit{TESS} data covering a different number of \textit{TESS} sectors \citep[e.g.][]{Tu2020, Doyle2020, Gunther2020, Pietras2022}. For the identification of the most violent flaring/superflaring F, G, K, and M stars considering latest \textit{TESS} data, \citet{Greimel2024} conducted an automated search in \textit{TESS} sectors 1--72. It was found that the exoplanet host star with the largest number of flares/superflares is AU~Mic.

Spectroscopic investigations of superflares on main-sequence stars are rare. This is related to the fact that obtaining spectroscopic time series requires observational effort and is time consuming. \citet{Hawley1991} presented ultraviolet (UV) and optical spectroscopy (3560--4440\AA{}) of a superflare on the dMe star AD~Leo. In the recent past, superflares were captured spectroscopically on the young, nearby solar analogue EK~Dra \citep{Namekata2022a, Namekata2022b, Leitzinger2024}. Only the H$\alpha$ line could be studied though, due to the limited wavelength coverage of the involved instruments. \\
There are still some open questions regarding superflares. The origin of superflare emission is one of them and is discussed by \citet{Heinzel2018}. These authors posed the question if flaring loops can contribute to superflare emission, aside from footpoint emission which is the usually assumed source of white-light emission. Especially as on young and active stars the size of flare loops is larger than flare loops known from the present-day Sun, this is a reasonable hypothesis. The authors found that flaring loops on young and active stars may contribute significantly to superflare emission and may even dominate footpoint emission. With spectroscopic observations it could be possible to distinguish between footpoint and flare loop emission.

Another open question regarding superflares is which spectral lines are affected by superflares in contrast to the less energetic normal flares. To answer this open question, spectroscopic monitoring covering a broad spectral range is needed. Therefore, we utilized the ESO1.52m telescope situated on La Silla, Chile, operated by the PLATOSpec consortium and currently equipped with an Echelle spectrograph (see Section~\ref{Observations}), to monitor stars which are known to have high flare/superflare rates. Similar, but shorter campaigns have been previously carried out by our group utilizing the 2m Perek Telescope equipped with the Ond\v{r}ejov Echelle Spectrograph \citep[OES;][]{Kabath2020} monitoring the M dwarf AD~Leo \citep{Wollmann2023}, as well as the G/K dwarfs BY~Dra, EK~Dra, and V833~Tau \citep{Leitzinger2024b}, the latter in collaboration with observatories from Slovakia, Hungary and Austria. In this study we present spectroscopic and coordinated photometric monitoring of AU~Mic. We focus in this study on the flare detection, both in spectroscopic and photometric data, the flare parameter determination and analysis of eight prominent spectral lines, and the discussion of the contribution of flare loop emission to the overall flare emission. The detection and analysis of other spectral lines and spectral line asymmetries, being possibly related to flare plasma motions, will be presented in a separate study.

\subsection{The target star}
AU~Mic is a young active early-M dwarf which hosts two transiting planets~b and c \citep{Plavchan20, Martioli21, Zicher22}. A third planet~d was detected by transit-timing variations \citep{Wittrock22, Wittrock23}, as well as one further candidate planet~e from radial velocity observations \citep{Donati23}. On planet~b, planetary atmospheric escape of neutral hydrogen was detected \citep{Rockcliffe2023} using \textit{HST} observations. Star-planet interaction, especially for the hot Neptune in the system (planet~b), has been investigated with \textit{TESS} data \citep{Ilin2022} and was detected as a modulation in the \ion{He}{i}\,D3 spectral line \citep{Klein2022}. AU~Mic hosts an edge-on debris disc \citep{Kalas04}. Its magnetic field was studied utilizing Zeeman-Doppler imaging \citep{Kochukhov20,Klein21, Donati23}. Spot properties and coverage were investigated using light curve analyses, Doppler imaging and radial velocity data \citep{Martioli2020, Klein2022, Ikuta23, Waalkes24}. AU~Mic is a member of the $\sim$23\,Myr $\beta$~Pic moving group \citep{Mamajek14} and a wide companion to the AT~Mic system \citep{Caballero09}. According to \citet{Ibanez2019} it shows a chromospheric activity cycle of five years. AU~Mic is also a well-known flare star with flare observations in different wavelength ranges, like in the optical \citep[e.g.][]{Kunkel1970, Gershberg99, Gilbert22, Ikuta23}, UV \citep[e.g.][]{Robinson2001, Feinstein22}, X-rays \citep[e.g.][]{Magee2003, Pye15}, and radio \citep{MacGregor20, Bloot2024}. It was also a target of several multi-wavelength campaigns \citep[e.g.][]{Tsikoudi2000, Smith05, Tristan23}. Flares could affect the atmospheres of AU~Mic's planets by changing their atmospheric chemistry \citep{Segura2010, Tilley2019}, as well as affecting atmospheric ionization and escape \citep{Chadney2017, Odert2020}. There are also indications of possible CME activity \citep{Cully94, Veronig21}. Numerical modelling applied to the AU~Mic system predicts extreme space weather for the planets, with high magnetic pressures in quiescence and even more extreme during CME activity \citep{AlvaradoGomez2022}. The stellar parameters, as adopted for this paper, are summarized in Table\,\ref{tab:aumic}.

\begin{table}
	\centering
	\caption{Parameters of the target star AU~Mic.}
	\label{tab:aumic}
	\begin{tabular}{lcc}
		\hline
		spectral type & M0Ve &  \citet{Pecaut13}\\
		distance (pc) & 9.725$\pm$0.005 & \citet{Cifuentes20}\\
        luminosity (L$_{\sun}$) & 0.100$\pm$0.003 & \citet{Cifuentes20}\\
        effective temperature (K) & 3600 & \citet{Cifuentes20}\\
        mass (M$_{\sun}$) & 0.8340$\pm$0.0314 & \citet{Cifuentes20}\\
        radius (R$_{\sun}$) & 0.8132$\pm$0.0258 & \citet{Cifuentes20}\\
		rotation period (d) & 4.863 & \citet{Plavchan20}\\
        age (Myr) & 23$\pm$3 & \citet{Mamajek14}\\
		\hline
	\end{tabular}
\end{table}


\section{Observations}
\label{Observations}

\subsection{\textit{TESS}}
AU~Mic was observed with the Transiting Exoplanet Survey Satellite (\textit{TESS}) in sectors~1 (2018-07-25 to 2018-08-22) and 27 (2020-07-05 to 2020-07-30) with cadences of 120\,s (sectors 1\&27) and 20\,s (sector~27). We used data from both sectors to determine the occurrence rate of strong flares on AU~Mic.

\subsection{Spectroscopy and simultaneous photometry within the PLATOSpec project}
We performed the spectroscopic monitoring with the ESO1.52m (E152) telescope. The E152 telescope is located at La Silla, Chile and it is operated by the PLATOSpec consortium\footnote{The PLATOSpec consortium is led by the Astronomical Institute of the Czech Academy of Sciences and consisting of Th\"uringer Landessternwarte Tautenburg and Universidad Catolica de Chile as major partners and of Universidad Adolfo Ibanez and Masaryk University as minor partners and University of Graz as collaborating partner.}. It was refurbished in 2022 and now it hosts an interim spectrograph, PUCHEROS+\footnote{\url{https://stelweb.asu.cas.cz/plato/}}, dedicated to ground-based support observations of \textit{TESS} and later \textit{PLATO} targets with the upcoming high-resolution Echelle spectrograph PLATOSpec (foreseen 2024). 

PUCHEROS+ is an improved version of the PUCHEROS spectrograph \citep{Vanzi12}, with a resolving power of R$\approx$18,000. The detector is an Andor iKon M CCD which provides a wavelength coverage of about 400 to 700\,nm. Typically, the spectrograph's radial velocity stability is about 100\,m\,s$^{-1}$ over a month for cool main-sequence stars with \textit{V}=8\,mag. PUCHEROS+ spectra are reduced using the CERES+ pipeline, an updated version of CERES \citep{Brahm17}, in which the functionality was adapted to include the PUCHEROS+ instrument-specific data reduction. CERES+ produces wavelength calibrated and optimally extracted order-by-order spectra, and radial velocities are obtained using the cross-correlation method. While the data will be available in the ESO archive later, they are currently accessible upon request from the PLATOSpec team.

The E152 is equipped with two different finder telescopes, each with a lens aperture size of 15\,cm. Both finder telescopes are equipped with photometric cameras which are used typically for the simultaneous monitoring of spectroscopic targets, such as in this study. The photometric camera OndCam (Ond\v{r}ejov camera) is a C4 camera with a CMOS sensor (chip Gpixel GSENSE4040) with 4096$\times$4096\,pixels each of 9\,$\mu$m size. It has six filters (Sloan \textit{u'g'r'i'z-s'}, H$\alpha$) and one clear slot allowing for color photometry in seven different bands. The field-of-view is 1.37$^\circ$ and the pixel scale 1.21\,arcsec/pixel.

GrazCam (Graz camera) consists of four components, an ASI 2600MM PRO CMOS camera, an ESATTO 2 focuser, an ARCO 2 rotator, and an ASI EFW filter wheel. The filter wheel has eight slots and is equipped with seven filters (Sloan \textit{u'g'r'i'}, Johnson \textit{BV}, SA200). With the Star Analyser 200 (SA200) by Paton Hawksley, low-resolution slitless spectroscopy can be performed. The field-of-view of GrazCam is 52.2$\times$34.9\,arcmin and the pixel scale is 0.501\,arcsec/pixel.

\subsection{Data sets}
In this section, we will describe our acquired data from the spectroscopic and simultaneous photometric observations.

\subsubsection{Spectroscopy}
AU~Mic was observed from 2022-10-31 until 2023-09-21 for a total of 56 (partial) nights, resulting in a total observing time of about 190\,h with a net on-source time of about 160\,h. Typical exposure times vary between 300, 600, and 900\,s. The observations are summarized in Table\,\ref{tab:obslog}.

For further analysis, we use the wavelength-calibrated deblazed spectra generated by the CERES+ pipeline. First, we perform some additional processing of the spectra. The wavelengths are corrected for barycentric motion. Low signal-to-noise ratio (SNR) spectra are flagged, for which we choose a threshold of a mean SNR per pixel of 10 in the order containing the H$\alpha$ line. Cosmic rays are removed using the \texttt{scipy} function \texttt{find\_peaks} with appropriately adjusted threshold and width, selected to not affect the intrinsic stellar emission lines.

\subsubsection{Photometry}
Simultaneous photometric observations were carried out with \textit{OndCam} installed on one of the two 15\,cm finder telescopes. For the observations described here, we mainly used the Sloan \textit{g'}-filter, with a few nights using the \textit{r'}-filter. The data were taken in a time series simultaneously with the spectroscopic data. Typical exposure times are in the order of a few seconds. Parameters of the simultaneous photometry (exposure time and filter) are given in Table\,\ref{tab:obslog} as well. We note that photometry alone was taken during additional three nights in 2023.

The reduction is done using calibration images from a regularly updated library of darkframes and flatfields for the camera and instrument. A standard reduction process is performed using a custom-written pipeline \citep{fryda2023} performing standard flatfielding, bias and dark subtraction. We use Sextractor \citep{Bertin1996} to obtain aperture photometry for each reduced image. The light curves shown throughout the paper are differential photometry light curves. As comparison star we use HD~197673 which is an A9 dwarf. 

\subsubsection{Low-resolution spectroscopy}
\label{sec:lrs}
For two nights, we also obtained simultaneous low-resolution spectroscopy with the SA200 transmission grating of GrazCam to investigate continuum enhancements during flares/superflares \citep[see e.g.][]{Zhilyaev12}. The exposure time was 200\,s ensuring a reasonable signal. The first order spectrum is well aligned along pixel rows. A one-dimensional spectrum is extracted by selecting a proper aperture on the chip. Background apertures of the same size are selected above and below the target aperture. Background apertures are averaged and subtracted from the target aperture. The background-subtracted target aperture is then collapsed in spatial direction to obtain a one dimensional spectrum. To determine the wavelength resolution we apply the formula given in the SA200 manual\footnote{\url{https://www.shelyak.com/wp-content/uploads/Star-Analyser-200-Instructions-v1.2.pdf}}, where the dispersion is a function of aperture size, detector pixel size, distance of grating to detector, and lines per mm of the grating. This yields in our case a dispersion of 7\,\AA{}/pixel. To find the starting wavelength $\lambda_{0}$, we apply the detector response function to the flux-calibrated \textit{Gaia} DR3 spectrum of AU~Mic\footnote{taken from VizieR (\url{https://vizier.cds.unistra.fr}) catalogue I/355/spectra} \citep{GaiaCollaboration23}. This enables the identification of the same molecular bands in both spectra and therefore also wavelength values in the SA200 spectrum. After that, we are then also able to correct the SA200 spectrum for the detector response function. To flux-calibrate the spectrum, we fit both the SA200 and \textit{Gaia} DR3 spectra with Planck functions. With the ratio of both Planck fits we then flux-calibrate the SA200 spectrum (see upper panel of Fig.\,\ref{fig:sa200_spectrum}).  

\section{Results}

\subsection{Flare rates from \textit{TESS}}
\begin{figure}
	\includegraphics[width=\columnwidth]{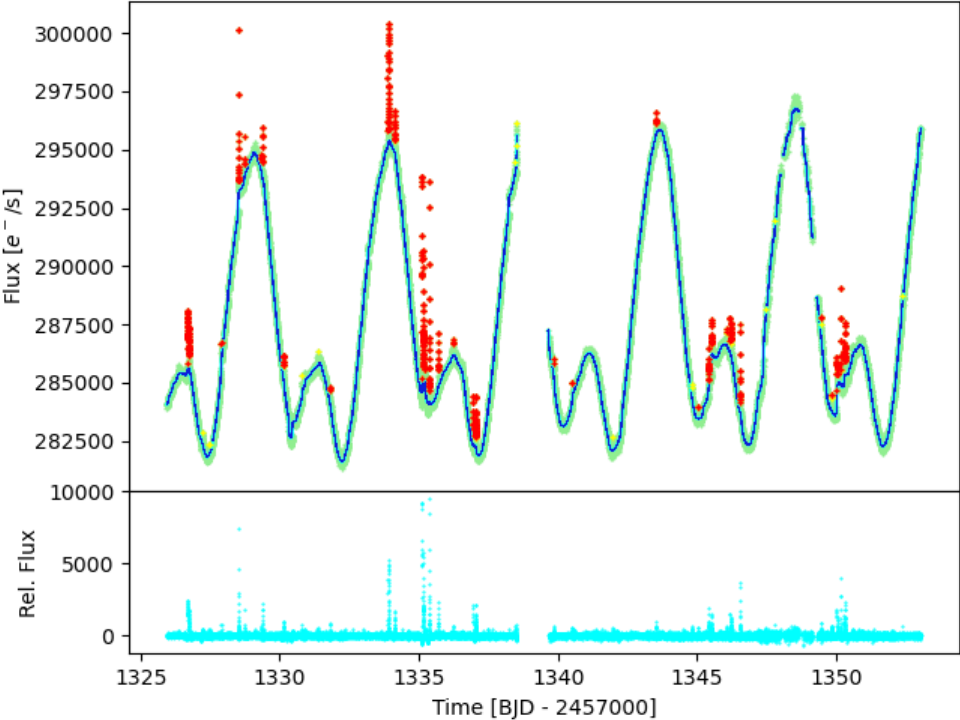}
    \includegraphics[width=\columnwidth]{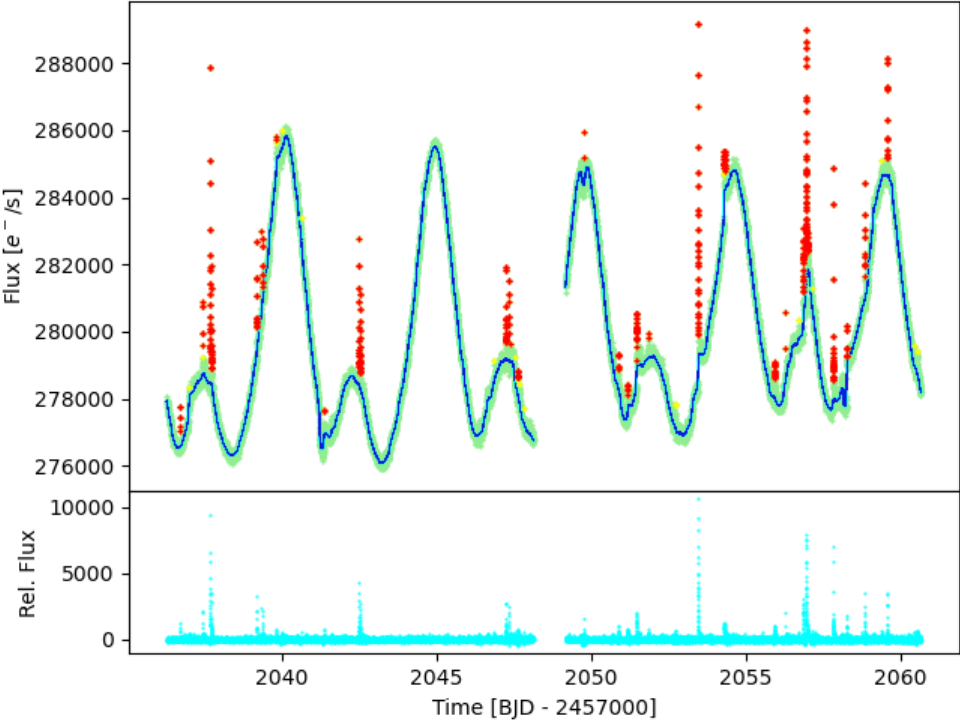}
    \caption{\textit{TESS} 2\,min cadence light curves of AU~Mic from sectors 1 (upper panel) and 27 (lower panel). Upper subpanels show the original \textit{TESS} PDCSAP flux (light green) and fitted curve (blue line), with identified flares marked in red. Yellow points denote enhanced values, but below the adopted threshold for flare detection. Lower subpanels show the detrended flux (cyan).}
    \label{fig:tess}
\end{figure}

We show the two available 120\,s light curves of AU~Mic (sectors~1 and 27) in Fig.\,\ref{fig:tess}. Detected flares, as identified with our flare detection algorithm \citep{Greimel2024}, are indicated in red. This algorithm first performs a flattening of the light curves by iterative application of a running median filter, then identifies data points above a certain threshold in the flattened light curves as flares.

We identify 25 flares in sector~1, 19 in sector~27 for 2\,min cadence, and 28 in sector~27 for 20\,s cadence. This results in a flare rate of 0.92 strong flares per day. The flare rate of AU~Mic in the two \textit{TESS} sectors was already determined in several other studies, including \citet[][only sector 1]{Gunther2020}, \citet{Martioli21}, \citet{Gilbert22}, \citet{Ilin2022}, \citet{Su2022} and \citet{Ikuta23}. These studies all applied different detection algorithms and thresholds. Therefore, the flare counts range from 31 to 162 in sector~1, and from 25 to 157 in sector~27 in the previous studies. As our flare search algorithm was designed to automatically identify strong flares on a large number of different stars, the total number of flares we found is slightly lower compared to some of these previous studies. We do however clearly see that AU~Mic produced stronger flares in sector~1 than in sector~27, and that more flares are detected in shorter cadence data.

\subsection{Flare detection from spectroscopy}
To identify flares, we compute the equivalent widths (EWs) of several prominent spectral lines known to exhibit chromospheric activity. This includes the first four Balmer lines, the cores of the sodium D lines, as well as two helium lines. The chosen line and continuum windows are summarized in Table\,\ref{tab:lines}. Line windows are chosen to fully cover the stellar lines even during the strongest flares with broadened wings, whereas continuum windows are placed in regions blue- and redward of the lines. We note that the selected continuum windows are in most cases placed close to the chosen line windows, because some of the lines are located rather close to the edges of the Echelle orders. The quiescent continuum fluxes around the spectral lines used for the computation of flare energies in Section~\ref{sec:flareparams} are also summarized in Table\,\ref{tab:lines}. These are computed from a flux-calibrated high-SNR low-activity spectrum (see Section~\ref{sec:flareparams}) as mean values of the fluxes in the blue and red continuum windows around each line, which should give a representative value for the continuum in the line window. We note that the selection of continuum windows does not have a relevant effect on our results, as we are interested in the relative enhancement of the stellar line fluxes by flares, not in measuring precise absolute values. This holds especially for the \ion{Na}{i}\,D1\&D2, as well as the \ion{He}{i}\,6678 lines, for which the chosen ``continuum'' values are only reference levels and do not represent the true continua in these wavelength regions.

\begin{table}
	\centering
	\caption{Wavelength windows for lines and continua in \AA. Average quiescent continuum fluxes around the selected lines (in units of $10^{-12}$\,erg\,cm$^{-2}$\,s$^{-1}$\,\AA$^{-1}$) are given in the last column.}
	\label{tab:lines}
	\begin{tabular}{lcccc}
		\hline
		Line & Window & Cont. blue & Cont. red & Cont. flux \\
		\hline
        H$\alpha$ & 6564.5$\pm$8 & 6540.5,6556.5 & 6572.5,6588.5 & 2.09\\
        H$\beta$  & 4862.6$\pm$5 & 4847.6,4857.6 & 4867.6,4877.6 & 0.78\\
        H$\gamma$ & 4341.6$\pm$4 & 4329.6,4337.6 & 4345.6,4353.6 & 0.48\\
        H$\delta$ & 4102.8$\pm$3 & 4093.8,4099.8 & 4105.8,4111.8 & 0.35\\
        \ion{Na}{i}\,D1   & 5897.5$\pm$1 & 5879.0,5885.0 & 5904.0,5910.0 & 1.28\\
        \ion{Na}{i}\,D2   & 5891.5$\pm$1 & 5879.0,5885.0 & 5904.0,5910.0 & 1.27\\
        \ion{He}{i}\,D3   & 5877.2$\pm$1 & 5870.2,5876.2 & 5878.2,5884.2 & 1.38\\
        \ion{He}{i}\,6678 & 6679.9$\pm$1 & 6662.9,6678.9 & 6680.9,6696.9 & 1.86\\
		\hline
	\end{tabular}
\end{table}

To compute the EWs, we first perform a linear fit between the median flux values in the respective blue and red continuum windows of the spectra. This assumes that locally the continuum can be represented by a straight line. We then calculate the EW as $EW=\int(C-F)/C d\lambda$, where $C$ is the linear continuum fit and $F$ is the flux in the line window. This applies the convention that the EW of emission lines is negative. The uncertainties of the EWs are calculated using Eq.\,7 of \citet{Vollmann06}.

\begin{figure*}
	\includegraphics[width=\textwidth]{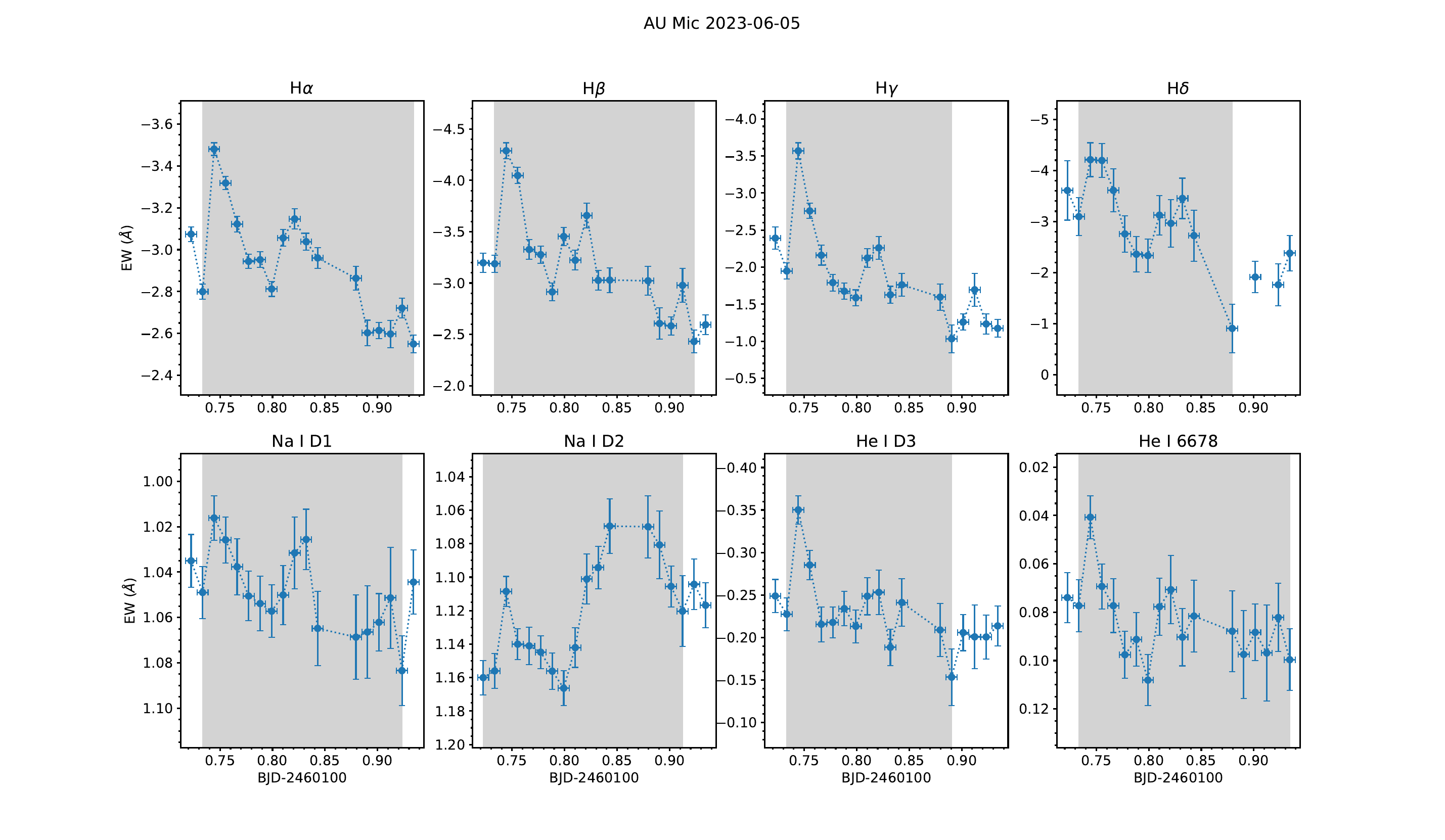}
    \caption{Flare \#13 on 2023-06-05 in all studied spectral lines. The flare was significantly detected in all spectral lines.}
    \label{fig:2023-06-05}
\end{figure*}

To identify flares, we select the maximum and minimum data points in the EW light curves of each line. We then compute the normalized amplitude, i.e. the normalized net flux at the flare peak (Eq.\,\ref{eq:fnetnorm}), and its error (see Section~\ref{sec:flareparams}) and require that the amplitude in H$\alpha$ is at least five times its error and the amplitudes in one or more other lines are at least three times their errors. We optimized and verified these criteria by visual inspection of the light curves. Applying these criteria to all available light curves, we select 24 flares (see Table\,\ref{tab:flares}, as well as Fig.\,\ref{fig:2023-06-05} and Appendix~\ref{app:lc}) for further analysis. For all these flares, we plot the EW light curves of all studied spectral lines and shade the time range that was identified as flaring. From these data, we can see that AU~Mic is frequently in a flaring state for the whole observation and there may be no true quiet level in the data. In one case (flare \#7) the detection significance in H$\beta$ is below three due to a noisier-than-usual non-flare spectrum, but we manually add its parameters to the list, since the flare can be clearly seen by visual inspection and it is significantly detected in the higher Balmer lines. We note that in several additional nights AU~Mic was likely flaring, but we omit these data if 1) we see only an elevated, but rather constant flux level, 2) the time series is too short, or 3) there are too many or too large data gaps. An exception is made for flare \#23 for which we observed an elevated flux level only, but of such large magnitude that it could have only been caused by a very energetic flare event.

\begin{figure*}
	\includegraphics[width=0.8\columnwidth]{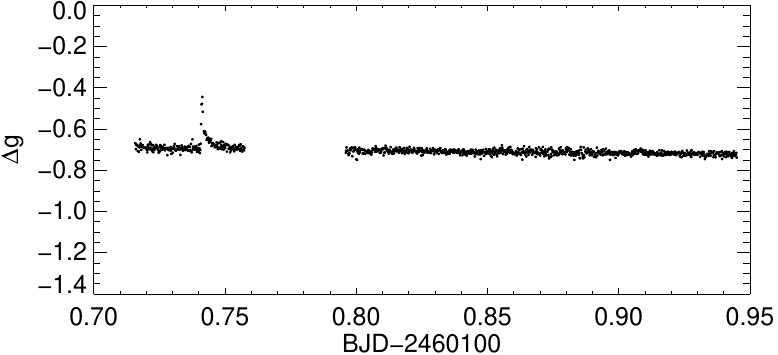}
    \includegraphics[width=0.6\columnwidth]{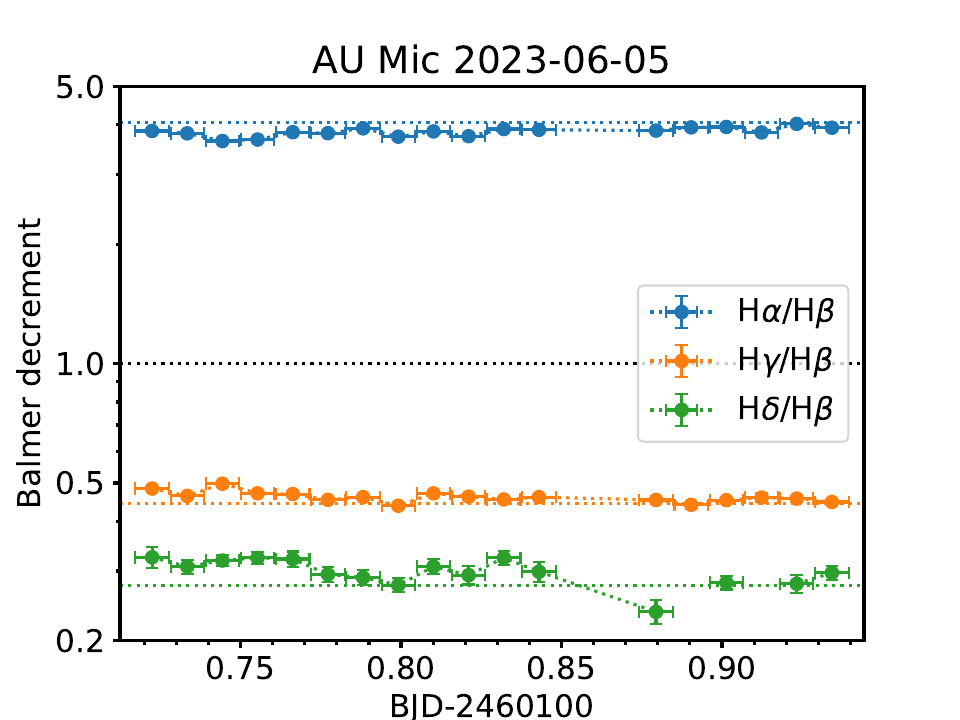}
    \caption{Left panel: Light curve (\textit{g'}-band magnitude difference relative to comparison star) of the flare on 2023-06-05 corresponding to the EW time series shown in Fig.\,\ref{fig:2023-06-05}. In the \textit{g'}-band the flare peaks at BJD-2460100=0.741. Left panel: Balmer decrements relative to H$\beta$ (filled circles) compared to their quiescent values (dotted lines).}
    \label{fig:phbd2023-06-05}
\end{figure*}

\subsection{Flares from photometry}
\label{sec:phot}
We detect one flare (\#13; see Fig.\,\ref{fig:phbd2023-06-05}) in the accompanying photometry. There are no additional flares detected in the photometry-only nights, and the nights with no clear spectroscopic flares. In the \textit{g'}-band, flare \#13 has a peak amplitude of 27\,per cent, a duration of about 11\,min, and an equivalent duration of 40.9\,s. This results in a \textit{g'}-band energy of ${\sim}10^{32}$\,erg, estimated using a quiescent \textit{g'}-band magnitude of 9.579\,mag \citep{Cifuentes20}. Following \citet{Shibayama2013} and assuming that the flare emission corresponds to a blackbody flux with a temperature of 9000\,K, we estimate the total white-light energy of this flare to be about $10^{33}$\,erg, i.e. a superflare. The flare area at the peak is estimated to be 0.18\,per cent of the stellar disc with the same assumptions \citep{Shibayama2013}.

\subsection{Slitless spectroscopy}
\begin{figure}
	\includegraphics[width=\columnwidth]{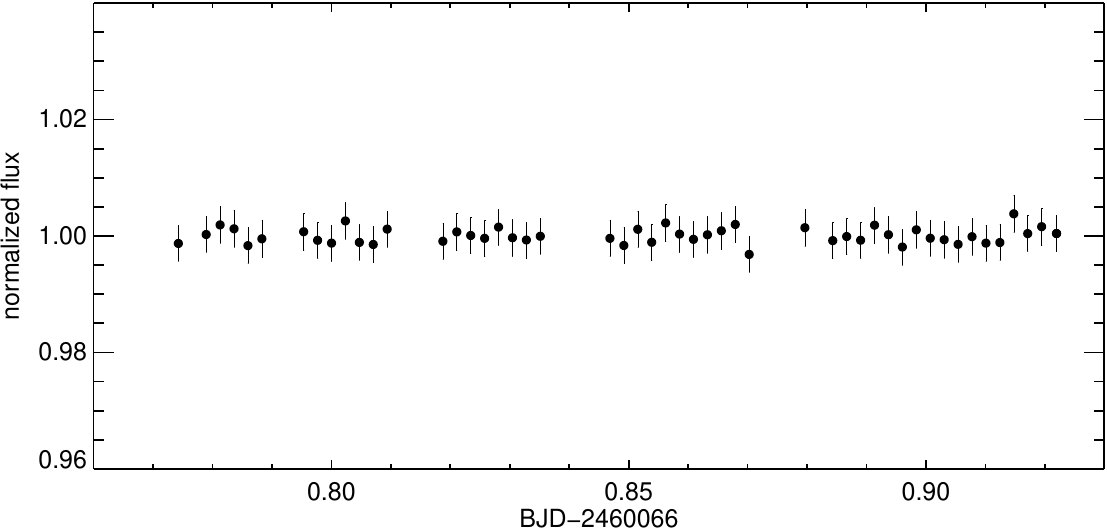}
    \caption{Light curve obtained from SA200 spectroscopy on 2023-05-02.}
    \label{fig:sa200_lightcurve}
\end{figure}

For a few nights, GrazCam was used to obtain slitless spectroscopy coordinated with PUCHEROS+ and OndCam observations. Those were the nights of 2023-05-01 and 2023-05-02. At the end of the night of 2023-05-02, the rising phase of a flare (\#8) is clearly seen in H$\alpha$, H$\beta$, and H$\gamma$ (see Fig.\,\ref{fig:2023-05-02}). The coordinated \textit{g'}-band photometry does not show any flare-like variations, however, the noise in this night is rather large. To check for flares, we also deduce a light curve from the SA200 data. To deduce the light curve, we sum all pixels from the background-subtracted spectrum, corresponding to a wavelength range of about 4084--8914\,\AA. The light curve is shown in Fig.\,\ref{fig:sa200_lightcurve}. The error bars denote the standard deviation of the data. At the end of the night one can see a subtle enhancement being in the range of 1$\sigma$ (fourth-to-last data point in Fig.\,\ref{fig:sa200_lightcurve}) coinciding in time with the start of the enhancement seen in H$\alpha$.

\begin{figure}
	\includegraphics[width=\columnwidth]{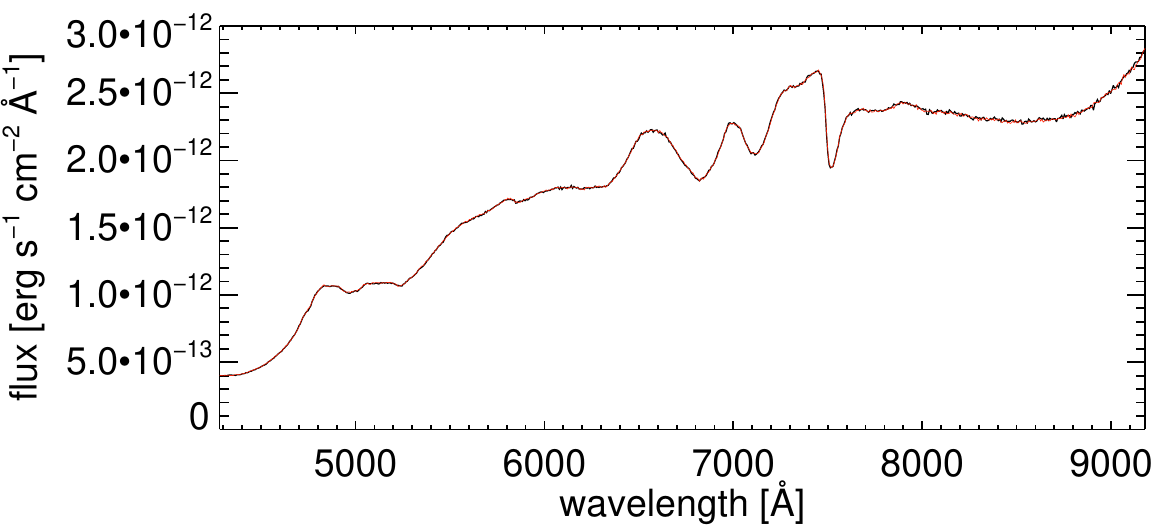}
    \includegraphics[width=\columnwidth]{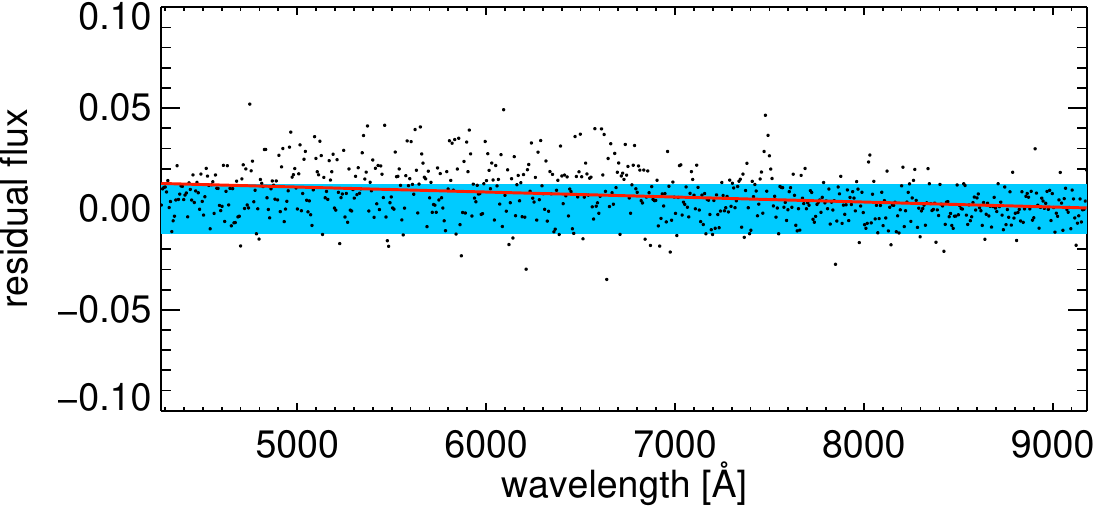}
    \caption{Upper panel: Flare spectrum (red solid line) overplotted with the pre-flare spectrum (black solid line) in the night of 2023-05-02. The spectra are not distinguishable to the eye. Lower panel: Residual flux, deduced as flare spectrum minus pre-flare spectrum. Overplotted is a line fit (red solid line) and the $\pm$1$\sigma$ error range (blue shaded area) of the residual.}
    \label{fig:sa200_spectrum}
\end{figure}

To test if this subtle increase could be due to a continuum enhancement in the blue indicative of a flare, we infer the residual flux of the SA200 spectra. We compile two representative spectra, a quiescent or pre-flare spectrum and a flare spectrum. The flare spectrum is comprised of the mean of the last four spectra of the time series (BJD-2460066=0.917--0.922), whereas the quiescent spectrum is comprised of the mean of three spectra before the flare spectra start (BJD-2460066=0.910--0.915). Both spectra are shown in the upper panel of Fig.\,\ref{fig:sa200_spectrum}. The red solid line represents the flare spectrum whereas the black solid line represents the quiescent spectrum. In the lower panel of Fig.\,\ref{fig:sa200_spectrum}, we show the residual (i.e. flare minus quiescent) spectrum. The blue-shaded area denotes the 1$\sigma$ error range. Additionally, we fit the residual data with a line. As one can see, the line shows a slight inclination towards the blue, as would be expected for a flare enhancement, but lying within the error range. Similarly to the subtle enhancement in the SA200 light curve, the continuum enhancement in the blue is therefore also not significant.

\subsection{Rotational modulation}
\begin{figure}
    \includegraphics[width=\columnwidth]{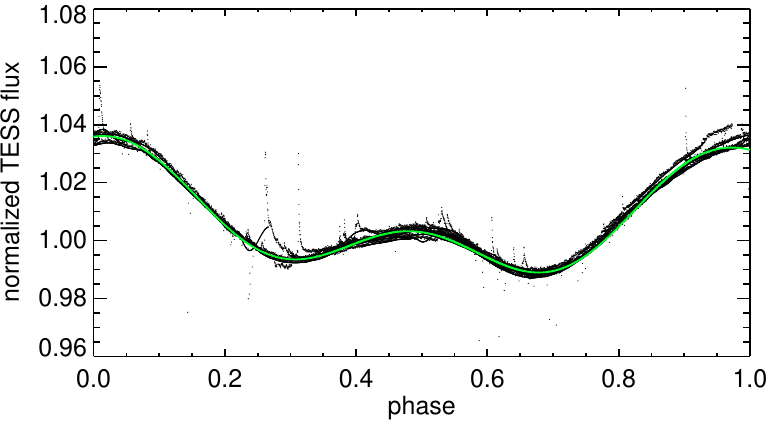}
    \includegraphics[width=\columnwidth]{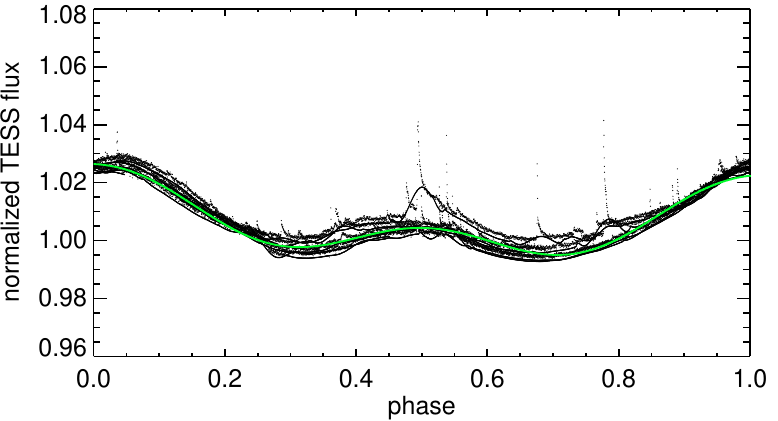}
    \includegraphics[width=\columnwidth]{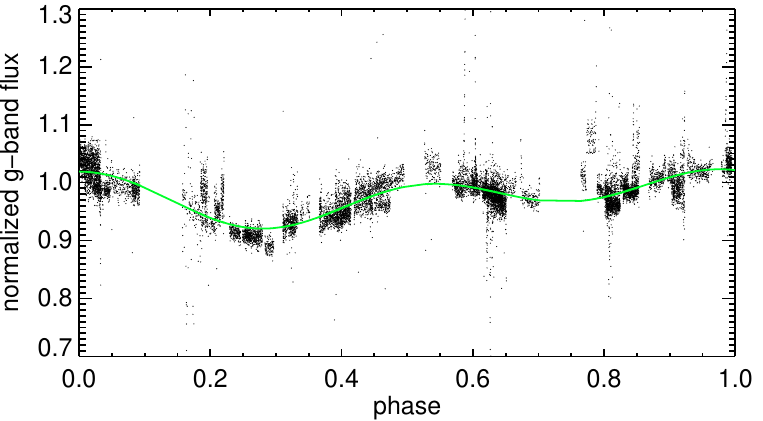}
    \caption{Upper and middle panels: Phase-folded light curves of AU~Mic obtained from \textit{TESS} data (sectors 1 and 27). Lower panel: Phase-folded light curve obtained from the coordinated photometry in the \textit{g'}-band taken with OndCam. From top to bottom, the displayed observations are from 2018, 2020, and 2023. In all panels, we overplotted sinusodial functions (green solid lines) representing the rotational modulation. The zero epochs are shifted to the respective maxima of the corresponding light curves.}
    \label{fig:rot}
\end{figure}

\begin{figure}
    \includegraphics[width=\columnwidth]{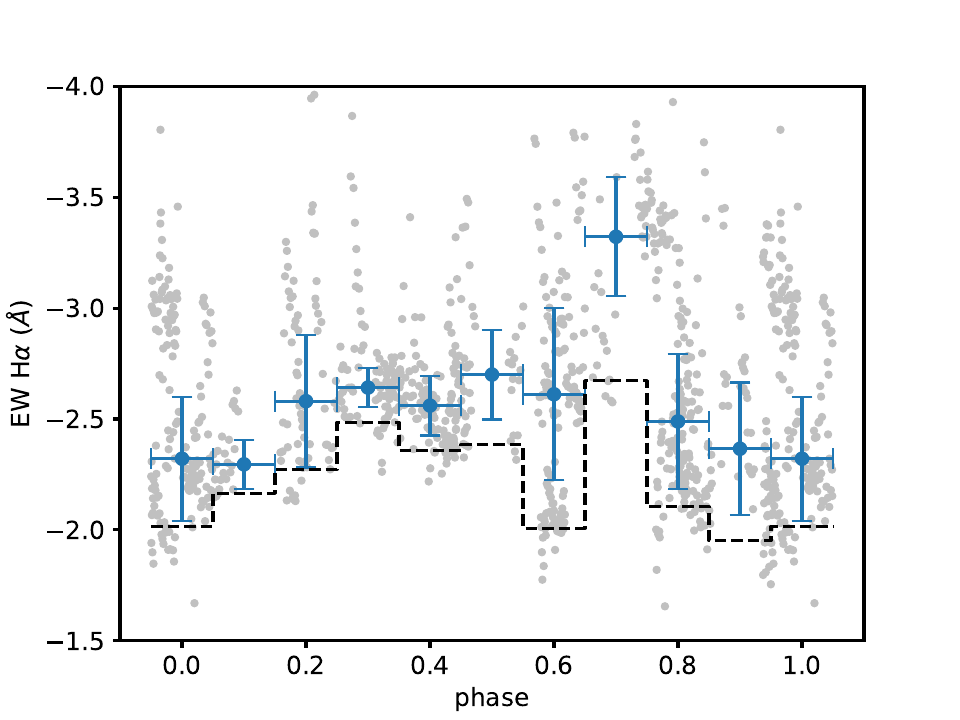}
    \includegraphics[width=\columnwidth]{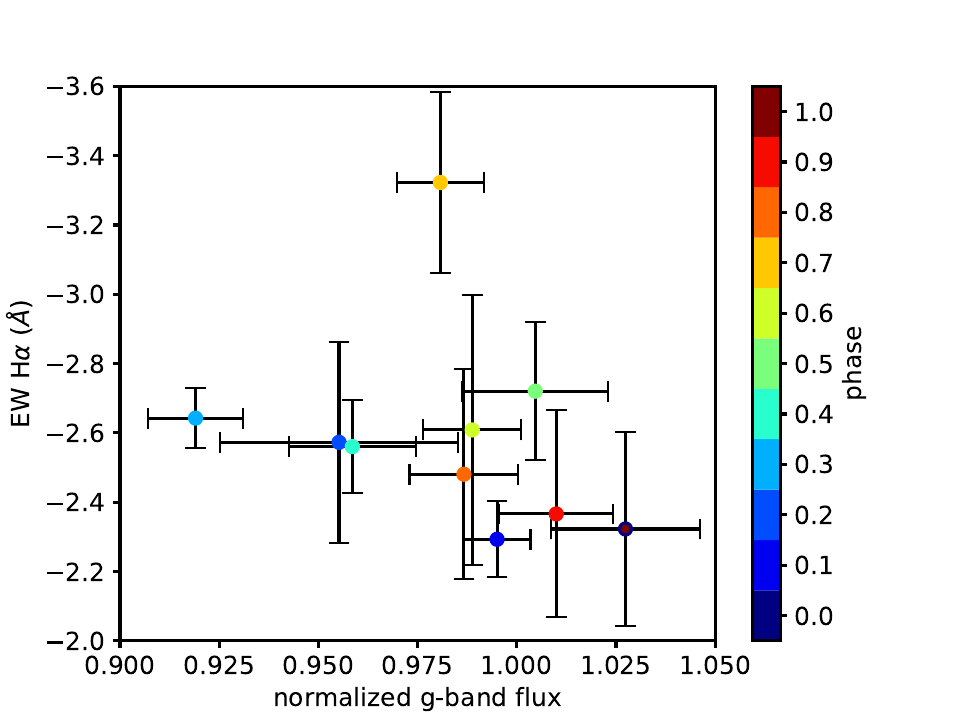}
    \caption{Upper panel: Phase-folded light curve of the H$\alpha$ EW (grey dots), with data outside the phase range between zero and one repeated for better visibility. The extreme flare on 2023-09-16 was omitted for clarity. The same zero epoch as for our \textit{g'}-band photometry (Fig.\,\ref{fig:rot}, lower panel) was used. Blue circles indicate the median value per phase bin, with error bars in x-direction indicating the bin width (0.1) and error bars in y-direction the median absolute deviation. The black dashed line shows the 90th percentile as an indicator for the lower envelope of the data. Lower panel: median of the normalized \textit{g'}-band flux vs. H$\alpha$ EW per phase bin (colour-coded). Error bars denote the respective median absolute deviations.}
    \label{fig:phased}
\end{figure}

We search for rotational modulation in the photometric data, as well as the H$\alpha$ EW (see Figs.\,\ref{fig:rot} and \ref{fig:phased}). From \textit{TESS} (Fig.\,\ref{fig:tess}), one can see that AU~Mic shows a prominent rotational modulation with a double-peaked morphology (one dominant main peak and a smaller secondary peak). From our photometric \textit{g'}-band observations, we also find a clear rotational pattern with a double-peaked structure which is similar to the one seen in \textit{TESS} data. In the lower panel of Fig.\,\ref{fig:rot}, we show the phase-folded light curve of AU~Mic obtained from \textit{g'}-band photometry. In the upper and middle panels, we show the same for the \textit{TESS} data from sectors~1 and 27. All phase-folded light curves were fitted by a sinusoidal function. Examining the amplitudes reveals that the \textit{g'}-band amplitudes from our observations are larger than from \textit{TESS}. The peak-to-peak amplitude of the main peak in the \textit{g'}-band is about 10\,per cent, and that of the secondary peak about 8\,per cent. This is larger than what was observed by \textit{TESS} in sector~1, 4\,per cent for the main peak and 1\,per cent for the secondary peak. The \textit{TESS} amplitude in sector~27 was slightly smaller, about 3.6\,per cent evolving to about 3\,per cent towards the end of the sector \citep{Szabo21, Gilbert22}. This can also be seen from the increased width of the phase-folded light curve in the middle panel of Fig.\,\ref{fig:rot}, indicating stronger amplitude variations throughout sector~27 compared to sector~1 (cf. also Fig.\,\ref{fig:tess}). Historic observations of AU~Mic's rotational modulation starting from the early seventies mention amplitudes ranging from $<$1 to 30\,per cent \citep[][and references therein]{Gilbert22}. Recently, \citet{Waalkes24} obtained semiamplitudes of 7.5\,per cent, 7.1--7.5\,per cent, and 4.1\,per cent in the \textit{g'}, \textit{r'}, and \textit{i'}-bands, respectively, in contemporary observations, indicating a colour-dependence. It is therefore clear that AU~Mic's spot properties and coverage are evolving over the years, despite the relatively stable rotational pattern seen on time-scales of months. Thus, the different amplitudes compared to \textit{TESS} which we find in our \textit{g'}-band data may be either due to the different observing epochs (\textit{TESS}: 2018+2020, here: 2023) and/or colour effects due to the spots' properties \citep[cf.][]{Waalkes24}.

We plot all H$\alpha$ EWs phase-folded with the rotation period in Fig.\,\ref{fig:phased}. From the \textit{TESS} light curves and our own photometry, it can be assumed that the rotational pattern seems to be rather stable on time-scales in the order of months. In contrast to photometric observations, the H$\alpha$ data show much more variability, and flare events can be clearly seen in many of the nights. However, the lower envelope of EWs encountered over the rotation period shows a clear rotational modulation and varies approximately between $-1.7$ and $-2.4$\,\AA. We used the same zero epoch as for the \textit{g'}-band curve (Fig.\,\ref{fig:rot}) and it can be seen that the H$\alpha$ EW behaves roughly inversely to the photometry, namely that photometric minima (i.e. maximum presence of spots) correspond to EW maxima (i.e. stronger chromospheric H$\alpha$ emission). This is consistent with the recent study of \citet{Tristan23}. As mentioned before, the rotational pattern is not as clearly visible as in the photometry, which is partly due to the large number of flares in the data, or may indicate that H$\alpha$ emitting regions evolve on different time-scales than the spots. For improved visibility, we bin the data in phase bins of width 0.1 and overplot in Fig.\,\ref{fig:phased} the median and the 90th percentile of each bin, where the latter gives a good approximation of the lower envelope of the phase curve. The median is quite robust against outlier data points (i.e. flares) and shows a clear rotational modulation in anti-phase to the photometry as well, although the uncertainty (here taken as the median absolute deviation) is large due to the frequent flares in the data. We cannot detect a counterpart to the secondary photometric minimum around phase 0.7, as this bin includes a gap in phase coverage and is only populated with data points of two nights with large flares, but very little non-flaring emission, which results in an artificially high median. To further assess their anti-phase behaviour, we plot in the lower panel of Fig.\,\ref{fig:phased} the median values in each phase bin of the normalized \textit{g'}-band flux against the H$\alpha$ EW. A slightly negative slope in the plot is indicative of the anti-phased behaviour described above. The data point around phase 0.7 is a clear outlier here as well.

\subsection{Flare luminosities and energies in spectral lines}
\label{sec:flareparams}
For all detected flares, we compute the parameters of peak flux, duration and energy for all spectral lines in which the flares can be identified. The flare energy is
\begin{equation}\label{eq:eflare}
E_\mathrm{f} = \int L_\mathrm{f}(t)\mathrm{d}t,
\end{equation}
i.e. the net flare luminosity of a spectral line integrated over the light curve. The net flare luminosity can be written as
\begin{equation}\label{eq:lflare}
L_\mathrm{f}(t) = F_\mathrm{f}(t)A_\mathrm{f}(t),
\end{equation}
where we generally assume that both the flare surface flux $F_\mathrm{f}$ and the flare area $A_\mathrm{f}$ evolve in time. What we measure from the observations is the total line flux $F(t)$ of the whole star which can be written as
\begin{equation}\label{eq:ftoton}
F(t) = F_\mathrm{nf}\left(1-\frac{A_\mathrm{f}(t)}{A_*}\right) + F_\mathrm{f}(t)\frac{A_\mathrm{f}(t)}{A_*},
\end{equation}
where $F_\mathrm{nf}$ is the non-flaring surface flux, $F_\mathrm{f}(t)$ the flare surface flux, and $A_*=\pi R_*^2$ the area of the stellar disc. We ignore here projection effects as we assume that the flare area corresponds to its disc-projected area, which may underestimate the computed flare parameters. Equation\,\ref{eq:ftoton} is for the case when the flare emission stems from the surface of the star. We discuss possible emission from flare loops in Section\,\ref{disc:loops}. Combining Eq.\,\ref{eq:lflare} with Eq.\,\ref{eq:ftoton}, we can express the net flare luminosity as
\begin{equation}\label{eq:lfon}
L_\mathrm{f}(t) = L_\mathrm{nf}\left(\frac{F(t)-F_\mathrm{nf}}{F_\mathrm{nf}}+\frac{A_\mathrm{f}(t)}{A_*}\right) = L_\mathrm{nf}\left(\bar{F}(t)+\frac{A_\mathrm{f}(t)}{A_*}\right),
\end{equation}
where we use $L_\mathrm{nf}=F_\mathrm{nf}A_*$, the non-flaring luminosity of the stellar disc. The normalized net flux
\begin{equation}\label{eq:fnorm}
\bar{F}(t)=[F(t)-F_\mathrm{nf}]/F_\mathrm{nf}
\end{equation}
can be directly measured from the data, noting that flux ratios of surface or observed fluxes are the same. From Eq.\,\ref{eq:lfon} one can compute the flare energy using Eq.\,\ref{eq:eflare}.

We note that Eq.\,\ref{eq:lfon} contains the unknown parameter $A_\mathrm{f}(t)/A_*$ which is mostly ignored in the literature, thereby implicitly assuming that the flare emission comes from very small areas on the stellar disc (e.g. footpoints of flare loops). However, areas of chromospheric flare emission may not necessarily be negligibly small \citep{Hawley92, Hawley2003}. Thus, assuming that the fractional flare area can reach at maximum one, a maximum flare luminosity (leading to a maximum flare energy) can be estimated as $L_\mathrm{f,max}(t) = L_\mathrm{nf}F(t)/F_\mathrm{nf}$
if assuming the limit $A_\mathrm{f}(t)/A_*{\sim}1$ during the whole flare.

The non-flare flux in each line is taken at the time with the highest EW (corresponding to the lowest flux) during the observation. However, this may not represent a ``true'' non-flaring state, as some nights are seemingly fully covered by an ongoing flare, which leads to an underestimation of all flare parameters (peak flux, energy, duration). Moreover, \citet{Guedel04} argued that the high temperature component seen in X-ray spectra is due to low-level flaring that is ongoing all the time. It has been estimated that 20--50\,per cent of the X-ray emission in active stars is due to low-level flaring \citep{Montmerle1983, Stern1992, Maggio2000}. This means that, strictly speaking, active stars always have some low-level flare activity, they are never really quiet. Line fluxes ($F(t)$, $F_\mathrm{nf}$) are computed from the EWs via
\begin{equation}\label{eq:fline}
    F_\mathrm{line}(t)=F_\mathrm{cont}(t)[\Delta\lambda-EW(t)],
\end{equation}
where $\Delta\lambda$ is the line window, which considers that the line fluxes are always positive independent if the EW values are positive (absorption) or negative (emission). Thus, at each point in the flare, we can compute the normalized net flux (Eq.\,\ref{eq:fnorm}) as
\begin{equation}\label{eq:fnetnorm}
\bar{F}(t) =  \bar{F}_\mathrm{cont}(t)\frac{\Delta\lambda-EW(t)}{\Delta\lambda-EW_\mathrm{nf}}-1,
\end{equation}
where we define
\begin{equation}\label{eq:fcontnorm}
\bar{F}_\mathrm{cont}(t) = \frac{F_\mathrm{cont}(t)}{F_\mathrm{cont,nf}},
\end{equation}
as the ratio of flaring and non-flaring continuum fluxes. We consider that the continuum flux $F_\mathrm{cont}(t)$ may also be affected by the flare and estimate the ratio $\bar{F}_\mathrm{cont}(t)$ from our normalized photometric light curves. As we do not detect any flare but \#13 in our photometry, we set $\bar{F}_\mathrm{cont}(t)=1$. Flare \#13 has a shorter duration (11\,min) than the exposure time of the spectra in this night (15\,min), and therefore a diluted amplitude of 4.5\,per cent when accounting for the exposure time of the peak spectrum during which it occurred. The uncertainty of $\bar{F}(t)$ is computed by standard error propagation accounting for uncertainties in $EW(t)$, $EW_\mathrm{nf}$ and $\bar{F}_\mathrm{cont}(t)$, where the latter is estimated from the standard deviation in the photometric light curves, corrected for the spectrum exposure times. Furthermore, for the uncertainty of $L_\mathrm{nf}$ (Eq.\,\ref{eq:lfon}), we include also the uncertainties of distance, calibration of the quiet flux (see below) and rotational modulation (semi-amplitude of 5\,per cent in the \textit{g'}-band). Combining Eqs.\,\ref{eq:lfon} with \ref{eq:fnetnorm}, the net flare luminosity can be written as
\begin{equation}\label{eq:lfonfinal}
L_\mathrm{f}(t) = L_\mathrm{nf}\left(\bar{F}_\mathrm{cont}(t)\frac{\Delta\lambda-EW(t)}{\Delta\lambda-EW_\mathrm{nf}}-1 + \frac{A_\mathrm{f}(t)}{A_*}\right).
\end{equation}

\begin{figure}
	\includegraphics[width=\columnwidth]{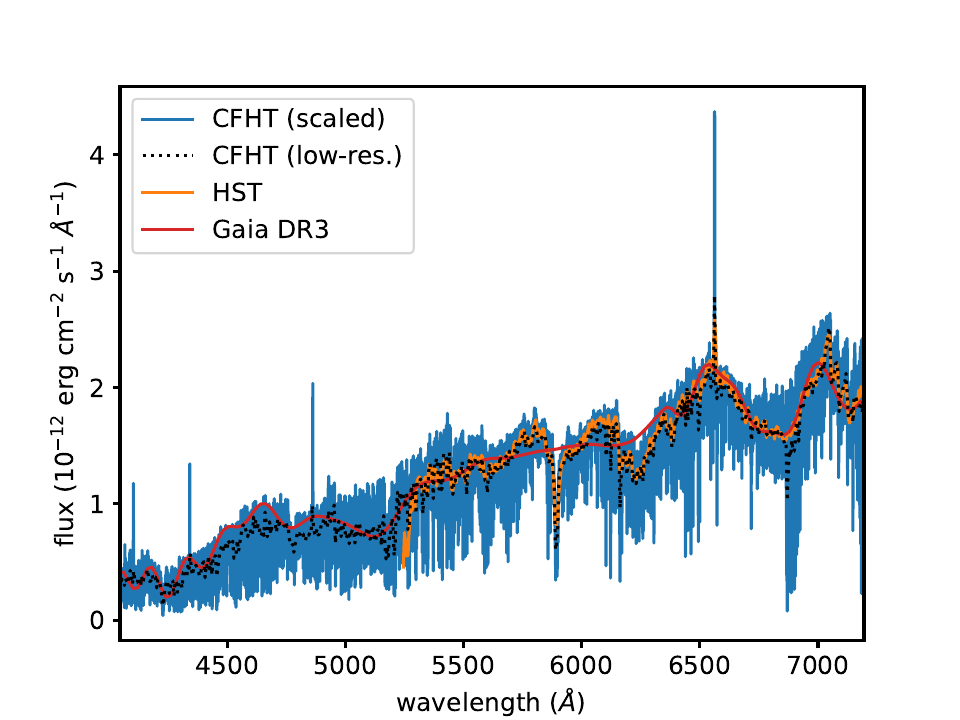}
    \caption{Flux-calibrated spectra of AU~Mic in the PUCHEROS+ wavelength range. The orange line shows the flux-calibrated \textit{HST} spectrum \citep{Lomax18}. The blue line shows the scaled CFHT spectrum in original resolution, the black dotted line the CFHT spectrum resampled to the lower resolution of the \textit{HST} data used for scaling. The red line gives the low-resolution \textit{Gaia} DR3 spectrum for comparison.}
    \label{fig:fluxcalib}
\end{figure}

To determine the non-flaring continuum fluxes around the spectral lines in physical units, we use a flux-calibrated \textit{HST}/STIS (G750L grating) spectrum \citep{Lomax18}. This spectrum covers the wavelength range 5240--10270\,\AA, but our PUCHEROS+ spectra reach farther into the blue (4045--7195\,\AA). Therefore, we additionally use unnormalized CFHT/ESPaDOnS spectra covering 3690--10481\,\AA. We select four subsequent 1200\,s CFHT spectra taken on 2006-08-03 (860776i, 860777i, 860778i, 860779i), because they seem to represent a minimally active state of AU~Mic (as seen from the low EWs of all chromospheric lines and negligible variability between these four spectra) and compute a high-SNR average. This mean spectrum is then resampled to the lower \textit{HST}/STIS resolution using \texttt{spectres} \citep{Carnall17} and shifted to the flux-calibrated \textit{HST} data (rms=0.08). Thus we obtain the continuum flux values in physical units around all considered spectral lines. The results are shown in Fig.\,\ref{fig:fluxcalib} and the extracted fluxes in the continuum windows, as computed from the calibrated average CFHT spectrum, are summarized in Table\,\ref{tab:lines}. AU~Mic's flux-calibrated low-resolution \textit{Gaia} DR3 spectrum \citep{GaiaCollaboration23} which we used in Section~\ref{sec:lrs} is overplotted, demonstrating a good match. 

Flare parameters are summarized in Table\,\ref{tab:flares}. The durations were computed as follows: first, we identify the flare peak as the minimum value of the EW in the light curve. The rise time is computed as the time between the maximum EW before the peak to the peak, the decay time as the time between the peak and the maximum EW after the peak. The maximum EWs are taken as those being closest to the chosen non-flaring value. In case of partially observed flares, the first or last data point of the observation is chosen instead. In some cases we need to select the start or end points manually if there are some clear outliers due to some noisier spectra. In the few cases with more than one flare in the observation, we choose the data points that are closest to the local flare peak, but fulfilling similar criteria. The total duration is then computed from the sum of rise and decay times. The peak luminosity is computed as the maximum of the net flare luminosity $L_\mathrm{f}$, and the energy as the integral of the net flare luminosity between the data points selected for computing the rise and decay times. For simplicity, and because there is no simple way of determining it without additional assumptions, we take $A_\mathrm{f}(t)/A_*{\sim}0$ as commonly assumed in the literature, but one needs to keep in mind that the computed flare parameters may thus be underestimated. However, as most of our flares are partial events not observed over their full duration and therefore their energies, and in some cases also the peak luminosities, are only lower limits anyway, we neglect this parameter as well.

\subsection{Balmer decrement during flares}
\label{sec:bd}
The Balmer decrement (BD) is the ratio between the fluxes in different Balmer lines, most often stated relative to H$\beta$ or H$\gamma$. During flares, it can give clues about the physical conditions in the flare. If plotting the BDs of several Balmer lines as a function of their upper levels, steeper BDs indicate smaller electron densities \citep{Garcia-Alvarez2002, Allred06, Kowalski2013}. We compute both the total, as well as the net flare BDs around the time of the peaks, where we use H$\beta$ as reference, as not all flares are convincingly detected in H$\gamma$ and the data there is typically noisier. The total BDs are computed as the ratio of the total observed line fluxes and are shown along with the light curves (Fig.\,\ref{fig:phbd2023-06-05} and Appendix~\ref{app:lc}). As a comparison, we also show the quiescent BDs as computed from the flux-calibrated spectrum shown in Section~\ref{sec:flareparams}, which gives H$\alpha$:H$\beta$:H$\gamma$:H$\delta$ of 4.05:1:0.44:0.27. This is in good agreement with the average quiescent BD of 17 flare stars determined by \citet{Pettersen89a}, namely (3.7$\pm$1.4):1:(0.5$\pm$0.1):(0.3$\pm$0.1). From Fig.\,\ref{fig:phbd2023-06-05} and Appendix~\ref{app:lc}, one can see that during the flares the BDs deviate from their quiescent values. During the flare peaks, the BDs of the different lines approach each other, i.e. becoming less steep if plotting the BDs vs. upper levels, which indicates higher electron densities compared to the quiescent state. The BDs thus also help to identify nights which are dominated by flares and no quiescent state is present. Additionally, we compute the net flare BDs as ratios of the net flare luminosities $L_\mathrm{f}$ relative to H$\beta$ around the peaks of the flares. As not all spectral lines peak at the same time in some of the flares, we compute these BDs at the time when H$\alpha$ peaks. These values are given in Table\,\ref{tab:flares}. The net flare BD gives clues about the electron densities in the flares \citep{Drake1980, Katsova90}.

\subsection{Relationships between flare parameters}
\label{sec:relations}
\begin{figure*}
	\includegraphics[width=\columnwidth]{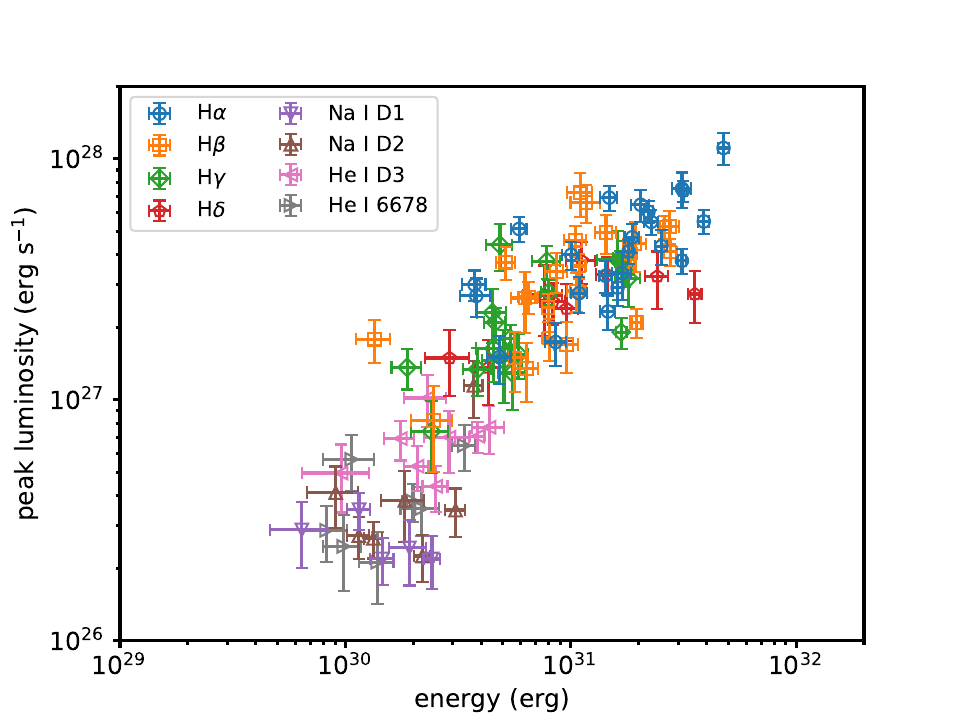}
    \includegraphics[width=\columnwidth]{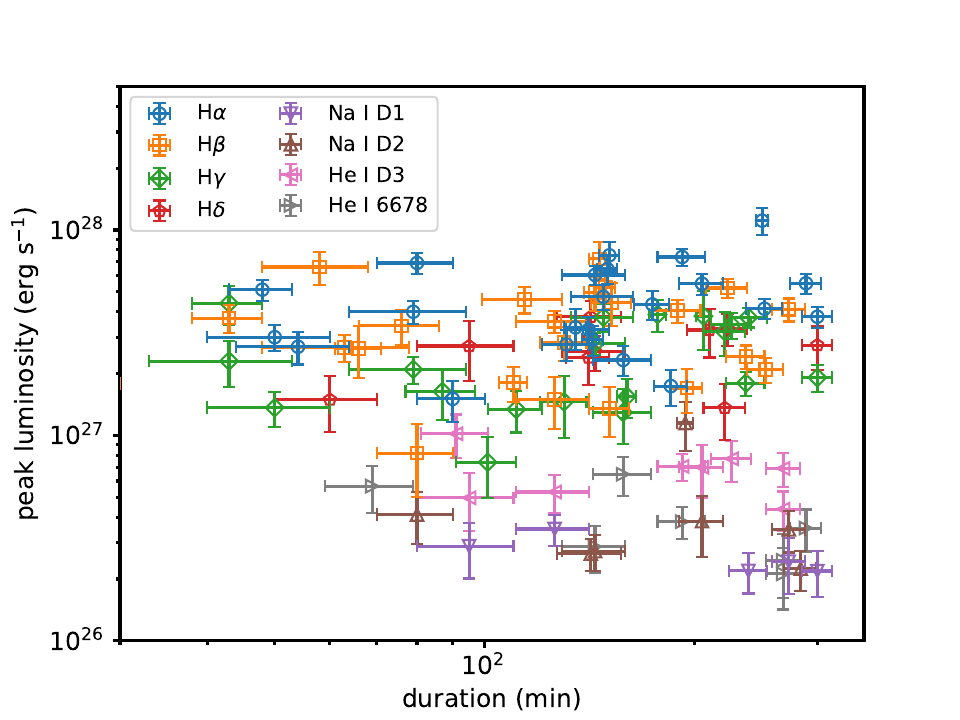}\\
    \includegraphics[width=\columnwidth]{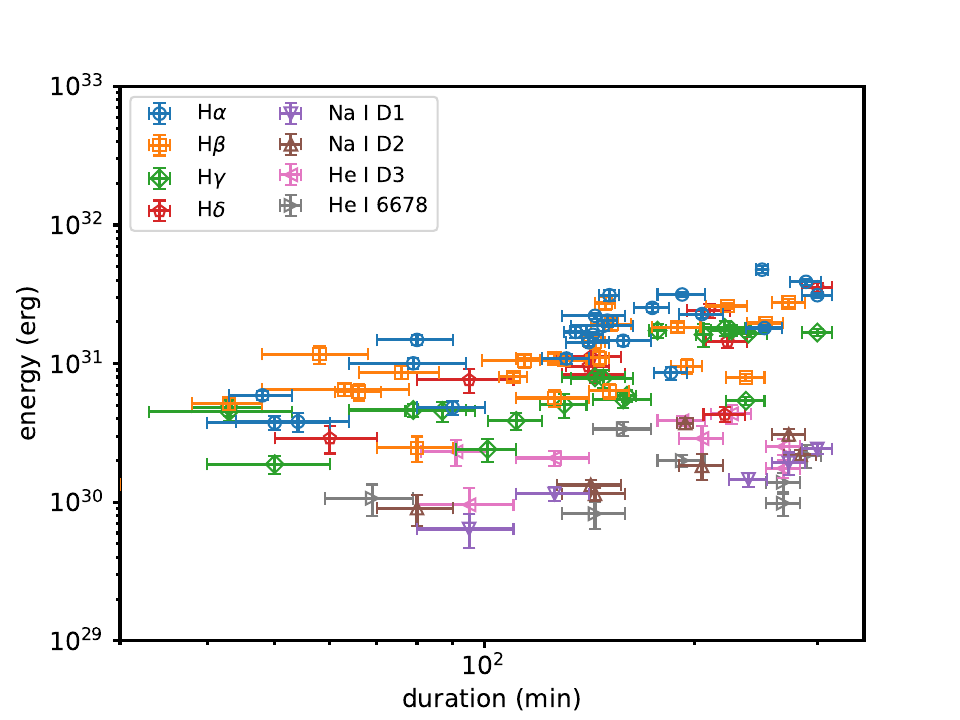}
    \includegraphics[width=\columnwidth]{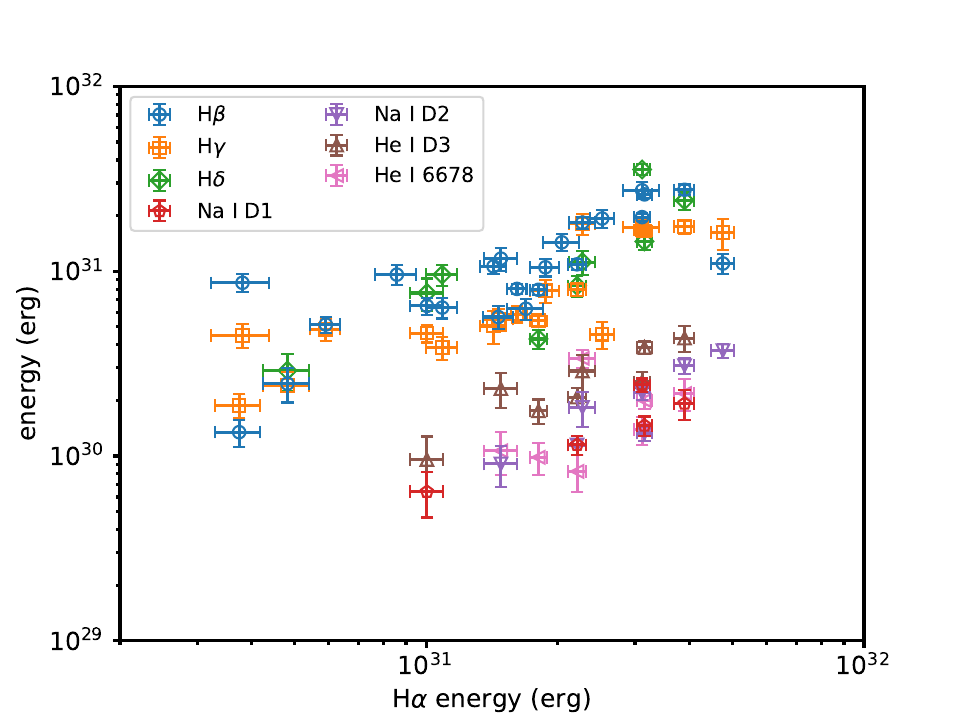}\\
    \caption{Relationships between flare parameters in different spectral lines. Upper left: flare energy vs. peak luminosity. Upper right: duration vs. peak luminosity. Lower left: duration vs. energy. Lower right: energy in H$\alpha$ vs. energy in other lines. The different spectral lines are colour-coded.}
    \label{fig:flareparams}
\end{figure*}

\begin{figure}
	\includegraphics[width=\columnwidth]{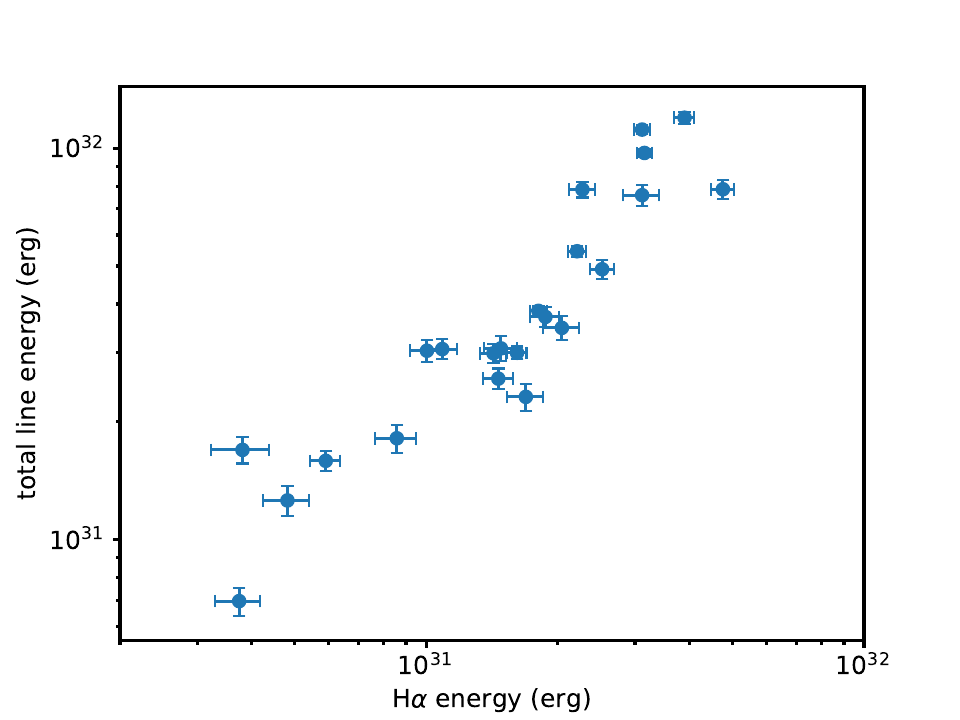}
    \caption{Relationship between flare energy in H$\alpha$ and the total energy in all considered spectral lines.}
    \label{fig:etot}
\end{figure}

We show relationships between flare parameters in the different spectral lines in Fig.\,\ref{fig:flareparams}. There is a clear relationship between peak luminosity and energy of the same line, although the spread is about a factor of a few. The increase in energy with peak luminosity is most obvious in the Balmer lines, but not so clear in the other chromospheric lines. However, there are fewer flares in which these lines are convincingly detected. We find no correlation between flare duration and peak luminosity. On the other hand, flare energy is correlated with duration, it increases with increasing duration. There are also rather tight relationships between the H$\alpha$ flare energy and the energy in other lines. Finally, we show the relation between H$\alpha$ energy and the total energy in the strong chromospheric lines in Fig.\,\ref{fig:etot}. One can see that H$\alpha$ is a good proxy for the total line energy. We note again that there are additional uncertainties in these figures, because many of our events were only partly observed, and thus their parameters are only lower limits to the true values.

To investigate the relationships between flare parameters further, we compute the Pearson correlation coefficient using \texttt{scipy.stats.pearsonr} and its p-value. We study linear relationships between the logarithms of flare energy, peak luminosity and duration. For significant correlations, which we define as $p<0.05$, we compute linear fits with Orthogonal Distance Regression using the \texttt{scipy.odr} package, which allows for uncertainties in both variables. The results are summarized in Table\,\ref{tab:relations}.

\begin{table}
	\centering
	\caption{Relationships between flare parameters (energy $E_\mathrm{f}$ in erg, peak luminosity $L_\mathrm{f,peak}$ in erg\,s$^{-1}$, duration $\Delta t$ in min) shown in Figs.\,\ref{fig:flareparams} and \ref{fig:etot}. For each relationship and line, we give the number of data points, the Pearson correlation coefficient (CC) and its p-value, as well as the parameters of the linear fit for correlations with $p<0.05$. The last column gives the residual variance of the fit.}
	\label{tab:relations}
	\begin{tabular}{lcccccc}
		\hline
		Line & n & CC & p-val. & $a_0$ & $a_1$ & res. var. \\
        \hline
        \multicolumn{7}{c}{$\log(L_\mathrm{f,peak}/10^{27}) = a_0\log(E_\mathrm{f}/10^{31}) + a_1$}\\
		\hline
        H$\alpha$ & 23 & 0.68 & 3.3e-4 & 0.41$\pm$0.11 & 0.57$\pm$0.04 & 5.80\\
        H$\beta$  & 23 & 0.65 & 8.6e-4 & 0.40$\pm$0.12 & 0.51$\pm$0.03 & 4.64\\
        H$\gamma$ & 19 & 0.71 & 6.3e-4 & 0.45$\pm$0.11 & 0.43$\pm$0.03 & 2.93\\
        H$\delta$ & 9  & 0.75 & 0.02   & 0.32$\pm$0.13 & 0.42$\pm$0.04 & 1.05\\
        \ion{Na}{i}\,D1 & 5  & -0.64 & 0.25 & -- & -- & --\\
        \ion{Na}{i}\,D2 & 7  & 0.51 & 0.25 & -- & -- & --\\
        \ion{He}{i}\,D3 & 8  & 0.41 & 0.31 & -- & -- & --\\
        \ion{He}{i}\,6678 & 7 & 0.53 & 0.22 & -- & -- & --\\
        \hline
        \multicolumn{7}{c}{$\log(L_\mathrm{f,peak}/10^{27}) = a_0\log(\Delta t/10^2) + a_1$}\\
		\hline
        H$\alpha$ & 23 & 0.26 & 0.22 & -- & -- & --\\
        H$\beta$  & 23 & 0.14 & 0.52 & -- & -- & --\\
        H$\gamma$ & 19 & 0.19 & 0.43 & -- & -- & --\\
        H$\delta$ & 9  & 0.32 & 0.40 & -- & -- & --\\
        \ion{Na}{i}\,D1 & 5  & -0.82 & 0.09 & -- & -- & --\\
        \ion{Na}{i}\,D2 & 7  & -0.07 & 0.88 & -- & -- & --\\
        \ion{He}{i}\,D3 & 8  & -0.17 & 0.68 & -- & -- & --\\
        \ion{He}{i}\,6678 & 7 & -0.66 & 0.11 & -- & -- & --\\
        \hline
        \multicolumn{7}{c}{$\log(E_\mathrm{f}/10^{31}) = a_0(\log\Delta t/10^2) + a_1$}\\
		\hline
        H$\alpha$ & 23 & 0.84 & 4.9e-7 & 1.44$\pm$0.24 & -0.023$\pm$0.067 & 11.54\\
        H$\beta$  & 23 & 0.76 & 2.6e-5 & 1.22$\pm$0.25 & -0.122$\pm$0.066 & 13.44\\
        H$\gamma$ & 19 & 0.75 & 2.4e-4 & 1.39$\pm$0.31 & -0.374$\pm$0.093 & 11.62\\
        H$\delta$ & 9  & 0.73 & 0.025  & 2.29$\pm$0.78 & -0.541$\pm$0.263 & 11.38\\
        \ion{Na}{i}\,D1 & 5  & 0.95 & 0.012  & 0.97$\pm$0.23 & -1.114$\pm$0.094 & 1.49\\
        \ion{Na}{i}\,D2 & 7  & 0.80 & 0.030  & 1.04$\pm$0.51 & -0.971$\pm$0.167 & 10.68\\
        \ion{He}{i}\,D3 & 8  & 0.50 & 0.20 & -- & -- & --\\
        \ion{He}{i}\,6678 & 7 & 0.23 & 0.62 & -- & -- & --\\
        \hline
        \multicolumn{7}{c}{$\log(E_\mathrm{f}/10^{31}) = a_0\log(E_\mathrm{f,H\alpha}/10^{31}) + a_1$}\\
		\hline
        H$\beta$  & 23 & 0.80 & 4.6e-6 & 1.07$\pm$0.14 & -0.22$\pm$0.06 & 10.66\\
        H$\gamma$ & 19 & 0.85 & 4.4e-6 & 1.02$\pm$0.14 & -0.36$\pm$0.05 & 7.97\\
        H$\delta$ & 9  & 0.80 & 9.3e-3 & 1.46$\pm$0.44 & -0.40$\pm$0.19 & 16.97\\
        \ion{Na}{i}\,D1 & 5  & 0.92 & 0.03   & 1.14$\pm$0.47 & -1.29$\pm$0.22 & 4.64\\
        \ion{Na}{i}\,D2 & 7  & 0.89 & 7.1e-3 & 1.51$\pm$0.37 & -1.46$\pm$0.19 & 4.06\\
        \ion{He}{i}\,D3 & 8  & 0.90 & 2.5e-3 & 1.06$\pm$0.24 & -0.99$\pm$0.11 & 1.93\\
        \ion{He}{i}\,6678 & 7 & 0.51 & 0.24 & -- & -- & --\\
		\hline
        \multicolumn{7}{c}{$\log(E_\mathrm{f,tot}/10^{31}) = a_0\log(E_\mathrm{f,H\alpha}/10^{31}) + a_1$}\\
        \hline
         & 23 & 0.92 & 5.2e-10 & 1.23$\pm$0.11 & 0.32$\pm$0.04 & 9.09\\
		\hline
	\end{tabular}
\end{table}

\begin{figure*}
	\includegraphics[width=0.65\columnwidth]{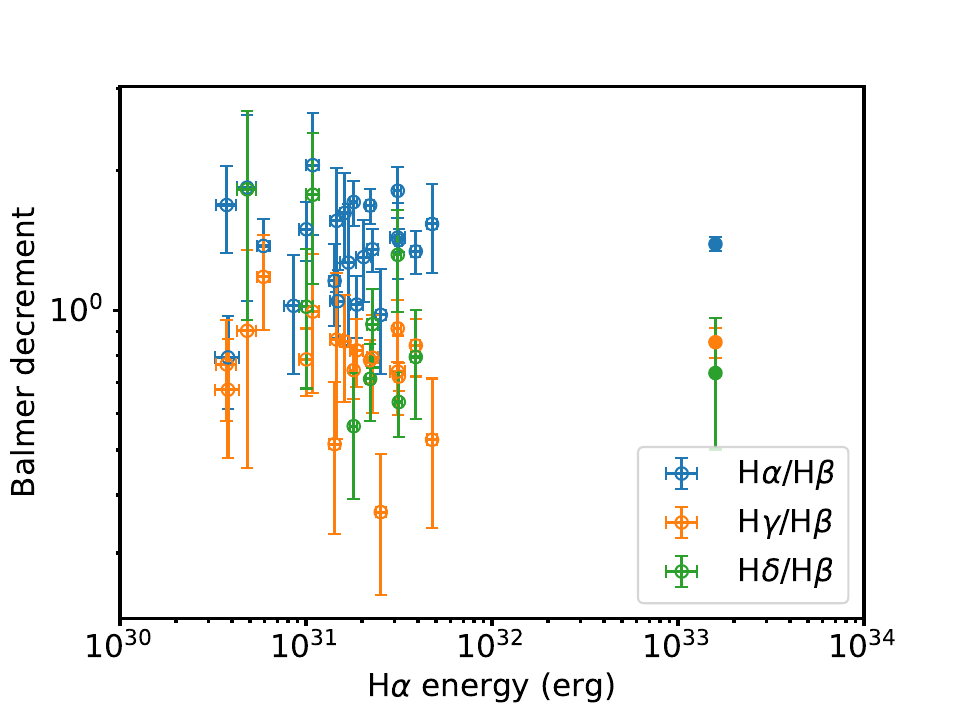}
    \includegraphics[width=0.65\columnwidth]{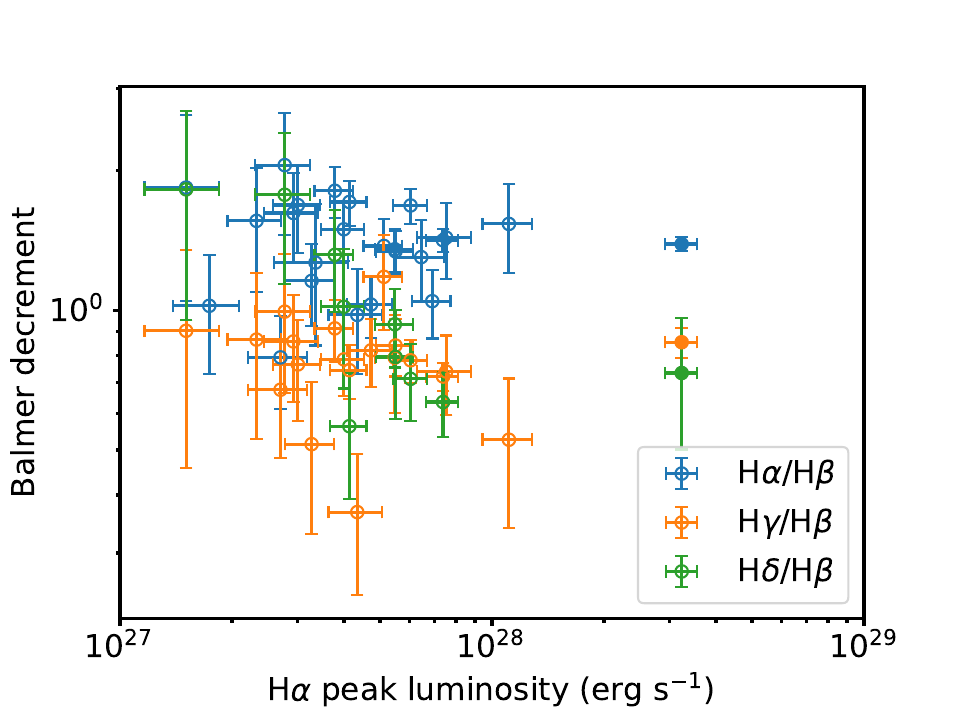}
    \includegraphics[width=0.65\columnwidth]{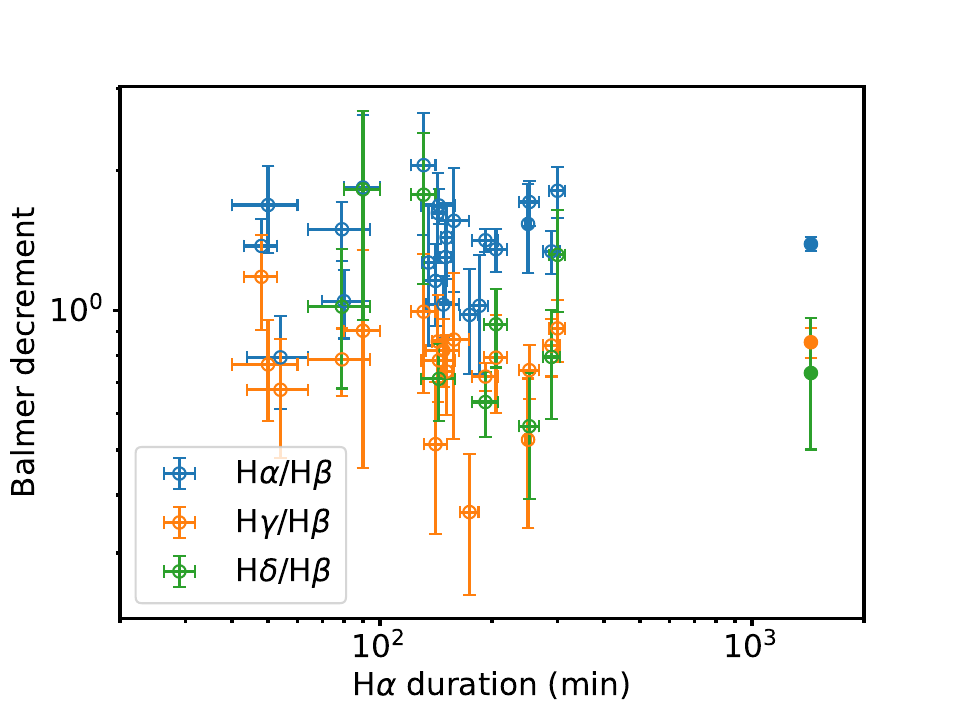}
    \caption{Net flare BD at the H$\alpha$ flare peaks vs. flare parameters taken from H$\alpha$ (left panel: energy, middle panel: peak luminosity, right panel: duration). Empty symbols denote all flares except for the superflare described in Section~\ref{sec:superflare}, whose parameters are shown as filled symbols. Its extrapolated values for energy and duration are taken from Section~\ref{disc:superflare}.}
    \label{fig:flare-bds}
\end{figure*}

We also test if there is a relationship between the peak BDs of the flares and other flare parameters to investigate some potential dependencies with the physical properties of the flare. Therefore, we plot the net flare BDs at the peak (Table\,\ref{tab:flares}) against flare energy, peak luminosity, and duration (all taken from H$\alpha$) in Fig.\,\ref{fig:flare-bds}. However, we do not find any correlations between the peak BD and these flare properties, indicating that there are no significant differences between the physical parameters of the weaker and stronger flares from our sample.

\subsection{The 2023-09-16 superflare}
\label{sec:superflare}
\begin{figure*}
	\includegraphics[width=\textwidth]{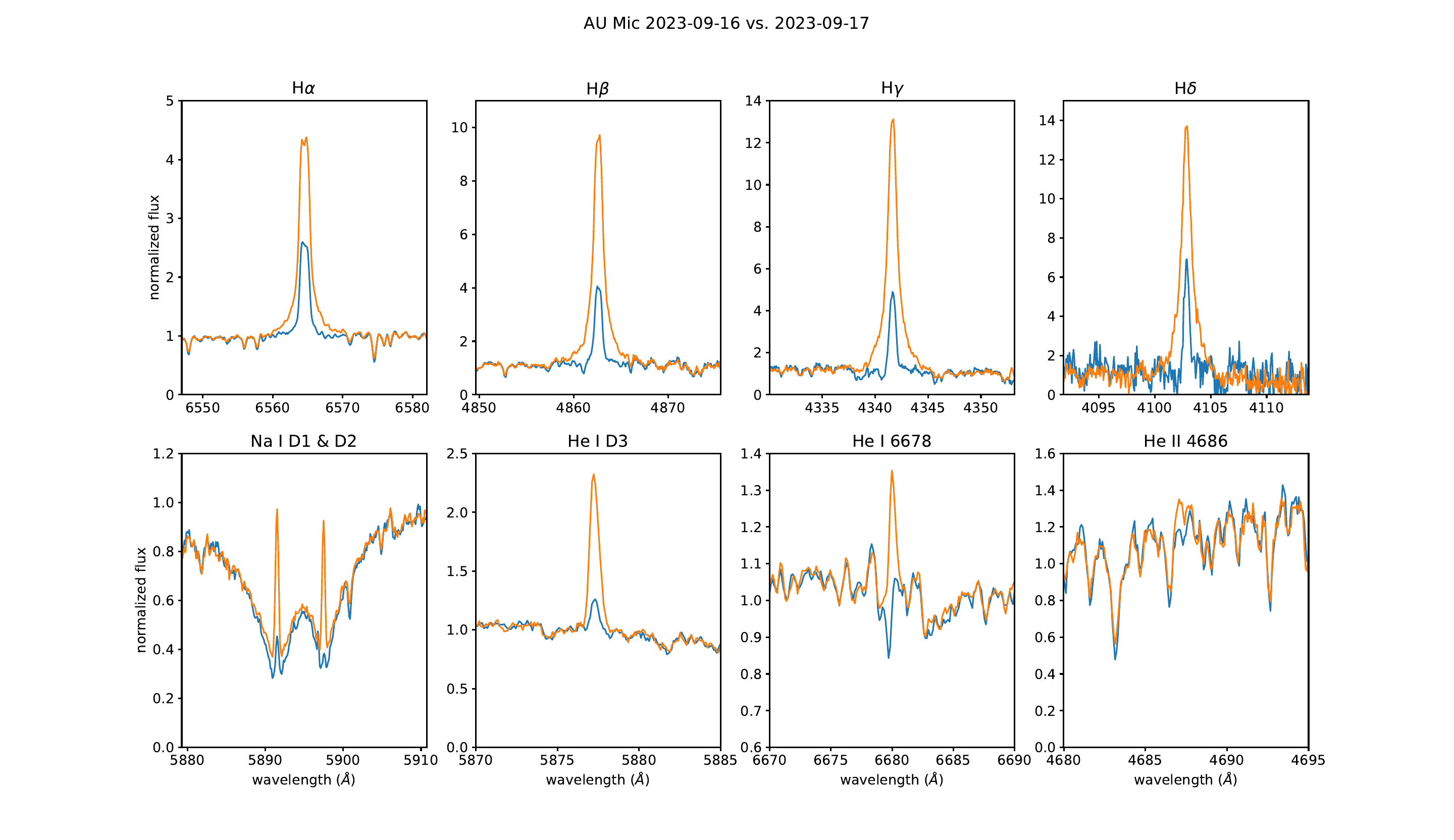}
    \caption{Comparison of the mean flare spectrum (orange) on 2023-09-16 with a mean of 23 spectra from the end of night 2023-09-17 (blue).}
    \label{fig:superflare_spectra}
\end{figure*}

During the night of 2023-09-16 an outstanding flare event happened, in which the EW of H$\alpha$ rose up to almost $-9$\,\AA, which is more than a factor of two above the more typical maximum values of all other strong flares we identified, which reach maximally up to about $-4$\,\AA. The strongest H$\alpha$ flare from the recent study of \citet{Tristan23}, who monitored AU~Mic for seven days, reached also about $-5$\,\AA\ only (however, they used slightly different wavelength windows). In the flare on 2023-09-16, all other prominent flare-affected emission lines rose to far higher than typical values as well. Unfortunately, only six spectra (of which one was too noisy) were recorded in this night. All usable spectra show these unusually high fluxes, meaning that our observations likely cover a time around the peak of this flare. The spectra cover about 36\,min, which is clearly a lower limit to the full duration of this exceptional flare, since no evolution to lower flux levels is observed during this time. For estimating some lower limits of flare parameters, we take an averaged spectrum from the following night as non-flaring comparison, although in this night most of the EWs were still higher than in a more typical non-flaring state. We average all five usable spectra from 2023-09-16 to create a ``flare'' spectrum, as we do not see large differences between them; for creating a ``non-flare'' spectrum, we average 23 spectra from the second half of the night 2023-09-17 after the main peak of flare \#24. In Fig.\,\ref{fig:superflare_spectra}, we compare the ``flare'' spectrum with the ``non-flare'' spectrum from the following night. All studied spectral lines were greatly enhanced compared to the following night. We also detect the \ion{He}{ii}\,4686 line, which was previously reported in some strong flares on M dwarfs \citep{Baranovskii2001, Paulson2006, Muheki2020a} and the Sun \citep{Zirin1985}. For the energies, we estimate lower limits by multiplying the peak luminosities with the duration of the whole observation on 2023-09-16. The estimated parameters are given in Table\,\ref{tab:flares}. However, it is very likely that such a strong flare had a duration far longer than our observations. We discuss estimates of its possible nature in Section~\ref{disc:superflare}.

The Balmer decrement during this flare (Fig.\,\ref{fig:bd2023-09-16}) shows large deviations from the quiescent values, much larger than for any other of the observed flares. Computing the net flare-only BD relative to H$\gamma$ in logarithmic units gives 0.21:0.07:0:4$-$0.07, which can be compared with fig.\,5 of \citet{Katsova90}. These numbers match closely the theoretical curve for $T=10000$\,K, $n_e=10^{13}$\,cm$^{-3}$ and $\tau(L\alpha)=3.3\times10^6$.

\section{Discussion}
In the following, we discuss parameters of the flares detected on AU~Mic, as well as the sources of flare emission. We focus in the present study on selected prominent flare-sensitive spectral lines (H$\alpha$, H$\beta$, H$\gamma$, H$\delta$, \ion{Na}{i}\,D1\&D2, \ion{He}{i}\,D3, and \ion{He}{i}\,6678). The spectral range of PUCHEROS+ covers many more interesting spectral lines. Other flare-affected spectral lines, as well as spectral line asymmetries, will be analysed in a separate study.

\subsection{Flare rates}
\label{disc:rates}
The AU~Mic campaign at E152 sums up to about 192\,h of spectroscopic monitoring. In this time, 24 flares in H$\alpha$ were identified. This results in a rate of $3.0_{-0.61}^{+0.74}$ H$\alpha$ flares per day \citep[1$\sigma$ error range after][]{Gehrels86}. However, the identification of flares in EW light curves is difficult because of several factors. First, the identified H$\alpha$ flares are typically long, in many cases even longer than the duration of one observing night, which leads not just to an underestimation of the computed flare parameters in the detected flares, but in nights with only a small number of spectra it cannot be unambiguously determined if the star is flaring, as one may see only elevated flux levels without clear flare shapes. Second, we find several flares with multiple peaks, which we assume to belong to the same event, but could alternatively be several overlapping but physically unassociated events. This ambiguity affects the flare counts, as well as the determined flare parameters. Third, our chosen exposure times of 5-15\,min limit our detection to longer events, as our observations cannot resolve short flares with durations of a few minutes. \citet{Tristan23}, who carried out a 7-day multiwavelength campaign targeting AU~Mic, detected 17 flares in H$\alpha$, yielding an H$\alpha$ flare rate of about 6.7\,d$^{-1}$, which is roughly a factor of two larger than we found here. This could be related to a possible change in flaring activity of AU~Mic (their observations were carried out in 2018) or also partly related to their shorter exposure time of 60\,s, which may have lead to detection of shorter flares which we would have missed.

\begin{figure}
	\includegraphics[width=\columnwidth]{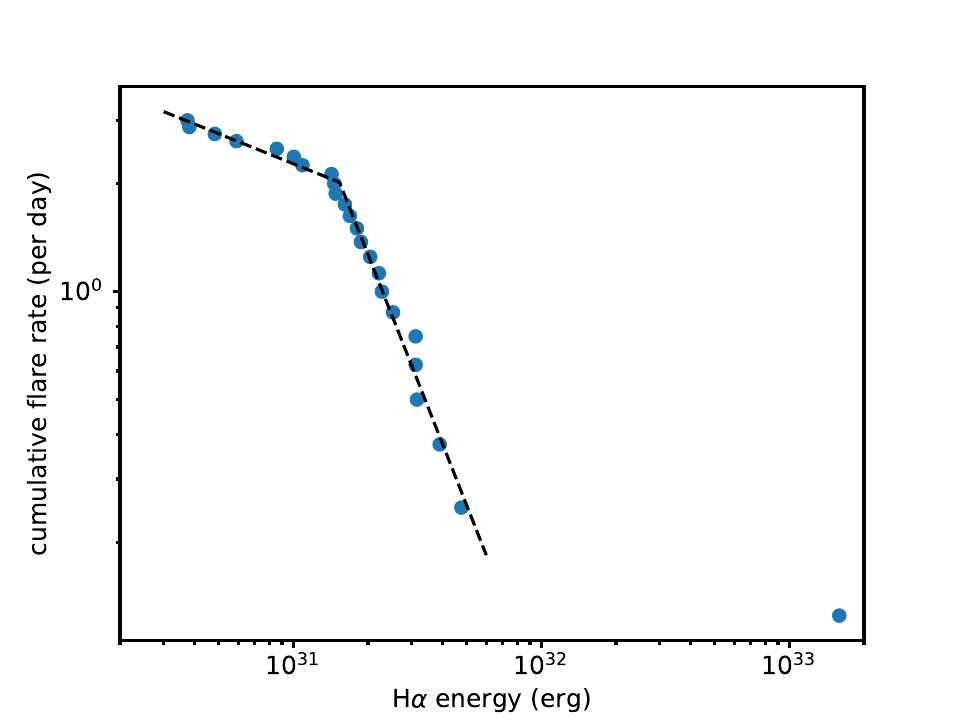}
    \caption{Cumulative FFD of H$\alpha$ flare energies. The black dashed line displays a broken power law fitted to the data. The fit excludes the outlier at the highest energy, the extreme superflare described in Section~\ref{sec:superflare}.}
    \label{fig:ffd}
\end{figure}

The flare frequency discussed above is related to the total number of flares detected in our observations. However, it is well established that the flare frequency increases with decreasing flare energy, and that the flare frequency distribution (FFD) follows a power law. We plot the cumulative FFD of H$\alpha$ flare energies in Fig.\,\ref{fig:ffd}. There are several apparent aspects of this FFD. First, there are two power law slopes, a more shallow one for H$\alpha$ flare energies below about $1.4\times10^{31}$\,erg and a steeper slope above. Second, we also plot the extrapolated energy of the 2023-09-16 superflare, which is a clear outlier from the distribution. We interpret the shallow slope at low energies as being due to incompleteness related to the detection limit of flares in our data, which is affected by our selection criteria, SNR, exposure time, and nightly coverage. On the other hand, the detection of the extreme superflare does likely not indicate a genuine transition to a shallower slope at high energies, but rather a by-chance detection of an event which is statistically rarer than what is expected to be found in the temporal coverage by our observations. We fit the distribution, excluding the obvious outlier, with a broken power law using least squares fitting, as implemented in \texttt{scipy.optimize.curve\_fit}. The steep part of the distribution between H$\alpha$ energies of about $1.4\times10^{31}$ and $5\times10^{31}$\,erg, i.e. excluding the likely incompletely sampled low-energy part and the outlier, is fitted with a power law slope of $-1.75\pm0.07$. This is similar to the slopes of the cumulative H$\alpha$ flare energy distributions of the active M dwarfs AD~Leo ($-1.56\pm0.13$) and EV~Lac ($-1.81\pm0.21$) obtained by \citet{Muheki2020b, Muheki2020a}. For the shallower, likely incomplete part, we find a slope of $-0.28\pm0.05$. Compared to \textit{TESS} which probes the optical broad-band continuum response of flares, the slope of the presumably complete part of the cumulative FFD of H$\alpha$ flare energies is steeper \citep[$-1.21\pm0.02$ for \textit{TESS} sector 1 and $-1.05\pm0.02$ for \textit{TESS} sector 27, based on the results from][]{Ikuta23}.

As we detected only one flare in our photometry, although AU~Mic is such a flare-active star, we can estimate if our photometric detection rate is consistent with \textit{TESS}. The average standard deviation of our \textit{g'}-band light curves (in units of normalized flux) is about 0.032. Assuming that for a significant detection a flare peak flux should be $>$3$\sigma$, we thus can only detect flares with peak amplitudes $>$0.1. According to \citet{Howard20}, peak amplitudes in the \textit{g'}-band are about a factor of 10 higher than in the \textit{TESS} band. Thus, we need to compare our flare rate with \textit{TESS} flares with peak amplitudes $\ge$0.01. This is around 0.31$^{+0.10}_{-0.08}$\,d$^{-1}$ \citep[0.24$^{+0.14}_{-0.09}$\,d$^{-1}$ for sector 1 and 0.39$^{+0.18}_{-0.13}$\,d$^{-1}$ for sector 27,][]{Ikuta23}. We find one significant event in 152.33\,hours of \textit{g'}-band photometry, which results in a flare rate of $0.16^{+0.36}_{-0.13}$\,d$^{-1}$, lower than, but consistent within the errors with the rough estimates based on \textit{TESS} data. However, as our \textit{g'}-band flare rate is lower than that from \textit{TESS} sector 1, similar to our lower H$\alpha$ flare rate compared to that from \citet{Tristan23}, this could indeed be a hint that AU~Mic showed reduced flare activity in 2023 compared to 2018. This is also consistent with the suggestions of \citet{Donati23} that AU~Mic was possibly moving towards an activity minimum in its potential cycle \citep{Ibanez2019} based on observation taken between 2019 and 2022.

The \textit{g'}-band flare is associated with flare \#13, which has the third-highest energy in H$\alpha$ (cf. Table\,\ref{tab:flares}). The only events with higher energies are the extreme event \#23, which has no usable accompanying photometry, as well as \#22, which occurred in the night with the highest photometric noise and suffers from several data gaps. This makes flare \#13 the most energetic flare with sufficient photometric coverage and quality to assess the presence of a white-light flare. Other flares with comparable, but slightly lower energies (\#2, \#9, \#11, \#12, \#15, \#19) had either no simultaneous photometry (\#2), data gaps (\#15, \#19), or the flares were already ongoing at the start of the observations (\#9, \#12), so the white-light flares could have been missed. In the case of flare \#11, which is a complex multi-peaked event in which the onset is likely covered, it is not clear if all peaks are indeed related to the same event or it represents a superposition of several unassociated, less energetic events. Comparing the white-light energy estimate of flare \#13 (${\sim}10^{33}$\,erg) with the energy emitted in H$\alpha$ ($3.89\times10^{31}$\,erg) reveals close agreement with the relationship determined by \citet{Namekata2024}, which predicts a white-light energy of $1.6\times10^{33}$\,erg for our measured H$\alpha$ energy.

We also searched for flare-related continuum enhancements in the flare spectra, but did not find any, not even in the extreme flare \#23, although this is hampered by low spectroscopic coverage. This is consistent with the low detection rate of flares in our photometry. There are several possible interpretations for this non-detection, including 1) there was no continuum enhancement associated with the chromospheric flares (i.e. non-white light flares); 2) the peak continuum enhancement was too weak and thus below our detection limit; 3) the integration times of our spectra were too long compared to the duration of the continuum enhancements (cf. flare \#13), so their signals were diluted; 4) continuum enhancements may have been missed in flares that were already ongoing at the start of the observations (9 out of 24) due to the Neupert effect. Regarding the latter two points, it has been often found in the literature that flare signatures in broad-band photometry seem not only to be shorter than the increase in H$\alpha$ line emission, but also peaking during the rise phase or peak of the Balmer line flare \citep[e.g.][]{Hawley2003, Notsu2024}, similar to the Neupert effect known from solar flares relating their hard and soft X-ray emissions.

\subsection{Properties of AU~Mic's flares}
In Table\,\ref{tab:flares}, all the deduced flare properties are listed. In Fig.\,\ref{fig:flareparams}, the scatter plots of flare energy versus flare peak luminosity (upper left panel), flare duration versus flare peak luminosity (upper right panel), flare duration versus flare energy (lower left panel), and H$\alpha$ flare energy versus flare energy in other lines (lower right panel) are shown. We see at least a trend for all of the scatter plots except the one for flare duration and flare peak luminosity. Apparently, the peak luminosity of the flares detected on AU~Mic are not dependent on the flare duration or vice versa. However, we remind that in most of our detected events we may have underestimated their properties because of either only partial coverage, multiple peaks which may or may not belong to the same event, or difficulties in determining a proper non-flaring spectrum.

The scatter plot of flare duration versus flare energy shows a correlation, that the longer the flare duration the larger the flare energy, which is reasonable as energy is determined by integrating the flare light curve over time. The duration of all flares varies from $\sim$0.5-5 hours and the flare energies from $\sim$10$^{29}$-10$^{32}$\,erg (for details see Table\,\ref{tab:flares}). \citet{CrespoChacon2006} monitored AD~Leo, which is a dM3.5Ve star with an age of 200\,Myr, so a bit later in terms of spectral type and a bit older than AU~Mic. These authors derive flare durations of $<$0.5\,h (in H$\beta$) and flare energies, derived from H$\alpha$, H$\beta$, H$\gamma$, and H$\delta$ in the range of 10$^{28}$-10$^{30}$\,erg. As we have integration times of 5-15\,min, we could not have resolved such short-lived flares. It is reasonable that the more short-lived flares detected in \citet{CrespoChacon2006} show lower energies than the flares we have detected on AU~Mic, as the AU~Mic flares have longer durations. \citet{Hawley2003} presented multi-wavelength observations of AD~Leo, including optical spectroscopy. For four flares in their study, durations and flare energies in the Balmer lines are given. The duration of the flares ranged from $\sim$0.5--1 hours with Balmer line flare energies in the range of 10$^{29}$--10$^{30}$\,erg, which are also comparable to the studies mentioned above. Also other dMe star (AD~Leo, YZ~CMi, AT~Mic) flare studies \citep{Doyle1988, Hawley1991, Gunn1994, Garcia-Alvarez2002} derived flare energies obtained from Balmer lines (H$\gamma$, H$\delta$) which agree ($\sim$10$^{29}$-10$^{31}$\,erg) with the ones given in Table\,\ref{tab:flares}. \citet{Koller2021} examined flares from SDSS DR14 spectra. Flares were found in this study mainly on dM stars, but also on few mid--late dK stars. The H$\alpha$ flare energy range from this study is $\sim$10$^{29}$--10$^{32}$\,erg (with very few data points reaching 10$^{33}$\,erg). Also here the range of the AU~Mic H$\alpha$ flares ($\sim$10$^{29}$--10$^{32}$\,erg) fits very well to their flare energies. The 41 flares on the active M dwarfs YZ~CMi, EV~Lac, and AD~Leo detected by \citet{Notsu2024} have H$\alpha$ parameter ranges (duration, peak flux, energy) very similar to those we determined for AU~Mic. Moreover, in their data also no correlation between peak luminosity and duration is apparent, but a trend very similar to ours between energy and duration (cf. their fig.\,37).

The relationship of flare duration with energy or peak luminosity, respectively, has been extensively studied on the Sun and other stars in different wavelength ranges. On the Sun, in soft X-ray observations a correlation of flare duration with fluence was found ($r=0.68$) with a slope of about 0.3, but only a weak correlation ($r=0.25$) between duration and peak flux \citep{Veronig2002}. A lack of correlation between duration and peak flux was also reported by \citet{Reep2019}. In hard X-ray observations of solar microflares, the correlation of duration with peak flux is stronger than in soft X-rays ($r=0.55$) and has a slope of about 0.2 \citep{Christe2008}, which is similar to observations of stellar X-ray flares \citep{Tsuboi2016}. In solar white-light flares, the slope of the relation between duration and energy is 0.38 \citep{Namekata2017}, similar to observations of stellar white-light flares, such as superflares on solar-type stars from \textit{Kepler} \citep[0.39;][]{Maehara2015} and \textit{TESS} \citep[0.42;][]{Tu2020}, flares and superflares on FGKM stars from \textit{TESS} \citep[0.46--0.5;][]{Pietras2022}, as well as superflares on a giant star from \textit{Kepler} \citep[0.325;][]{Kovari2020}. As the theory of magnetic reconnection predicts a slope of 1/3 for the relation of flare duration with total energy \citep{Maehara2015} and the observations of both solar and stellar (super)flares yield very similar values, this is indicative of a common generation mechanism. The slope we obtain for the duration and energy relationship in H$\alpha$ is significantly steeper ($\sim$0.7) than what other studies obtained for white-light emission. However, Balmer line emission is only a small fraction of the total flare emission and may thus not be a representative measure for the total flare energy. Similar as in solar soft X-ray observations, we find no correlation between duration and peak flux in the chromospheric lines. \citet{Reep2023} argued that existence of a relationship between flare duration and intensity should depend strongly on the wavelength range of the observations, as the flare duration is affected by the plasma temperature, the height of formation of the emission, as well as the relative importance of different cooling mechanisms. The relationships obtained in different wavelength ranges may thus also not be directly comparable.

For the scatter plot of flare energy and flare peak luminosity, a correlation is found for the Balmer lines, i.e. energetic flares have also large peaks. Flare peak luminosities are found for the flares on AU~Mic in the range of $\sim$2--31$\times$10$^{27}$\,erg\,s$^{-1}$. Comparing this to H$\alpha$ flare peak luminosities determined from SDSS data of M~dwarfs \citep{Koller2021}, we find that our values lie in the middle of the flare peak luminosity distribution from \citet{Koller2021}.

Finally, we investigated the dependence of H$\alpha$ flare energy on the flare energies determined from the other lines. Also here a trend is evident. The larger the H$\alpha$ flare energy, the larger also the flare energies of the other spectral lines. In Fig.\,\ref{fig:etot}, the H$\alpha$ flare energy is plotted versus the total flare energy in chromospheric lines. Here, we see a correlation between both quantities, following the trend seen in the lower right panel of Fig.\,\ref{fig:flareparams}.

Regarding flare energies, for two events we can confirm with our data that they are superflares, namely the 2023-06-05 event with simultaneous detection in the \textit{g'}-band (Figs.\,\ref{fig:2023-06-05} and \ref{fig:phbd2023-06-05}) and the extreme event on 2023-09-16 (Figs.\,\ref{fig:superflare_spectra} and \ref{fig:superflare}). For the rest of the flares, we may estimate their bolometric output using relationships from the literature. From fig.\,13 of \citet{Hawley1991}, partially based on \citet{Butler1988}, we can estimate that both the \textit{U}-band and soft X-ray flare energies are typically a factor of about 30 higher than the H$\gamma$ flare energy. \citet{Osten15} found that the \textit{U}-band emission is about 10\,per cent of the bolometric emission of a flare. Thus, flares with H$\gamma$ energies above ${\sim}3\times10^{30}$\,erg could already be superflares. We detected 18 flares in H$\gamma$ apart from the two confirmed superflares, from which 16 have energies higher than this threshold, which could thus be superflares as well. If further assuming the typical Balmer decrement of flares \citep{Butler1988}, the energy in H$\alpha$ should be about 3 times the energy in H$\gamma$, which would mean that flares with H$\alpha$ energies greater than ${\sim}10^{31}$\,erg could be superflares. From the 22 flares apart from the two confirmed superflares, 17 have higher energies and could thus be superflares. Recently, \citet{Namekata2024} also presented relationships between H$\alpha$, white-light, and X-ray flare energies. From those, we can infer that flares with H$\alpha$ energies greater than ${\sim}2\times10^{31}$\,erg could be superflares. From the same 22 flares as before, 8 have higher energies. Taking all these different estimates together, this means that in total 40--90\,per cent of the flares we detected may be superflares.

\subsection{The 2023-09-16 extreme event}
\label{disc:superflare}
Interestingly, during the beginning of the night following the 2023-09-16 event (\#23), we observe a gradual decay from about $-3.4$\,\AA\ to $-2.8$\,\AA\ values in the H$\alpha$ EW, interrupted by a short flare (\#24) at BJD-2460304=0.62 (cf. Fig.\,\ref{fig:2023-09-17}). This decay is also apparent in the H$\beta$ and H$\gamma$ lines. If we assume that this decay still belongs to the extreme flare \#23 of the previous night, we can fit the data by a simple flare model, which we show in Fig.\,\ref{fig:superflare}. The flare model is taken to be
\begin{equation}\label{eq:flaremodel}
    y=
    \begin{cases}
        a_0+a_1 e^{-(t-t_0)^2/(2a_3^2)} \dots t\le t_0\\
        a_0+a_1 e^{-(t-t_0)/a_2} \dots t>t_0,
    \end{cases}
\end{equation}
i.e. a Gaussian rise and exponential decay \citep{Pitkin14}. The parameter $y$ is the EW, $t$ the time after JD=2460203 in days, $t_0$ is the time of the peak, and $a_0$, $a_1$, $a_2$, $a_3$ the fit parameters. As in Section\,\ref{sec:relations}, we use the \texttt{scipy.odr} package for fitting. We fix the peak time $t_0$ to 0.54, as this aids fitting the light curves of the weaker lines with larger EW errors, and it is the value obtained for H$\alpha$ if it is taken as a fit parameter as well. We omit several apparent outlier data points in the 2023-09-17 data, which are likely affected by additional flares (incl. flare \#24) in the fitting procedure and are marked as gray symbols in Fig.\,\ref{fig:superflare}. The best-fitting parameters are summarized in Table\,\ref{tab:flarefit}. We do not obtain satisfactory fits for the H$\delta$ and \ion{He}{i}\,6678 lines, so we fix their decay time parameters to values similar to H$\gamma$ and \ion{He}{i}\,D3, respectively.

The flare energies obtained by integrating over the fits are also given, along with the flare energy estimates from the extrapolation of the relation between flare energy and peak luminosity obtained in Section~\ref{sec:relations}. The values obtained with both methods agree to within a factor of two, although there are some more discrepant lines (e.g. H$\delta$). This is likely because the relations in some lines are based on only a small number of data points. Due to the consistency between both estimation methods, it is indeed likely that we observed a small portion of a rare extreme event that reached superflare energies (${\sim}10^{33}$\,erg) in H$\alpha$ alone and had a duration of $>$24\,hours.

The flare energy in chromospheric lines is typically just a small part of the total flare energy budget \citep[e.g.][]{Hawley1991}. These authors discuss a relation based on \citet{Butler1988} showing a proportionality between H$\gamma$ and soft X-rays, as well as \textit{U}-band flare energies. According to that, a flare with $E_\mathrm{H\gamma}{\sim}5\times10^{32}$\,erg can have both \textit{U}-band and soft X-ray energies ${\gtrsim}10^{34}$\,erg. \cite{Osten15} estimated the relative contributions of flares in different energy bands to their bolometric emission, finding that the \textit{U}-band contributes typically about 11\,per cent and soft X-rays about 30\,per cent. Thus, the total bolometric energy of this flare could have been as large as ${>}10^{35}$\,erg. The flare energy relationships from \citet{Namekata2024} predict a total bolometric energy of ${\gtrsim}10^{35}$\,erg for the H$\alpha$ energy estimates in Table\,\ref{tab:flarefit} as well. As a comparison, the strongest flare found in the \textit{TESS} data of AU~Mic had a total white-light energy of about $8\times10^{33}$\,erg, estimated assuming a 10000\,K blackbody by \citet{Ikuta23}. This means that the extreme event presented here was likely much rarer than what is currently known from this star. However, it could be in the same total energy range as the event observed by the \textit{EUVE} satellite in 1992 \citep{Cully94, Katsova99} which has an estimated radiated energy in the 1--2000\,\AA\ range of ${\sim}10^{35}$\,erg.

Unfortunately, the photometry taken during the spectroscopic observations in this night was not usable, so we cannot constrain the presence or magnitude of a possible white-light component. We also attempted to identify a possible continuum enhancement in the blue range of the spectra following the method of \citet{Muheki2020b}. We first generated full-range low-resolution spectra by taking the fluxes in the centres of the Echelle orders, normalized them to one in the reddest order, and compared spectra taken at similar airmass of nights 2023-09-16 and 2023-09-17 to identify any potential excess flux in the blue for flare \#23. As we could not identify such excess flux, we possibly did not catch the continuum enhancement in the limited observing time during that night. As described in Section~4.1, the main continuum response of this flare may have occurred before the start of the spectroscopic observations, since the temporal behaviour of the optical continuum flux and H$\alpha$ line flux of typical flares shows a relationship similar to the Neupert effect \citep[e.g.][]{Hawley2003, Notsu2024}. This is consistent with \citet{Muheki2020b}, who identified three white-light flares on EV~Lac using this method and found that their durations were not only more than a factor of 10 shorter than the durations of the associated H$\alpha$ flares, but they also occurred during the impulsive and peak phases of the H$\alpha$ flares. As we were unable to identify a possibly associated white-light component of flare \#23, we computed the energies from the flare light curve fits (Table\,\ref{tab:flarefit}) assuming $\bar{F}_\mathrm{cont}(t)=1$ as for the other flares, thus yielding lower limits.

\begin{table*}
	\centering
	\caption{Fit parameters and energies of the 2023-09-16 superflare. Energies (in erg) were determined by integrating the fit of the EW light curves ($E_\mathrm{f,fit}$), as well as extrapolating the peak luminosity-flare energy relations to the observed peak fluxes of this flare ($E_\mathrm{f,rel}$).}
	\label{tab:flarefit}
	\begin{tabular}{lrrrrrr}
		\hline
		Line & $a_0$ & $a_1$ & $a_2$ & $a_3$ & $E_\mathrm{f,fit}$ & $E_\mathrm{f,rel}$ \\
		\hline
        H$\alpha$ & -2.47032802 & -6.19575854 & 0.45616892 & 0.03889632 & $1.59\times10^{33}$ & $1.87\times10^{33}$ \\
        H$\beta$  & -2.9827632 & -11.50032822 & 0.31680448 & 0.05863422 & $8.58\times10^{32}$ & $1.47\times10^{33}$ \\
        H$\gamma$ & -1.58070253 & -15.29770129 & 0.31964602 & 0.06201179 & $7.14\times10^{32}$ & $8.11\times10^{32}$ \\
        H$\delta$ & -1.95939438 & -19.31412056 & 0.32 (fixed) & 0.06 (fixed) & $5.77\times10^{32}$ & $3.76\times10^{33}$ \\
        Na I D1   & 1.12613693 & -0.43107111 & 0.38836716 & 0.03427485 & $5.79\times10^{31}$ & -- \\
        Na I D2   & 1.16284777 & -0.48048704 & 0.38565761 & 0.03432492 & $6.40\times10^{31}$ & -- \\
        He I D3   & -0.14366357 & -0.90602625 & 0.43020309 & 0.04568159 & $1.49\times10^{32}$ & -- \\
        He I 6678 & 0.11400856 & -0.30711339 & 0.43 (fixed) & 0.045 (fixed) & $4.88\times10^{31}$ & -- \\
		\hline
        total     &&&&& $4.06\times10^{33}$ & $7.91\times10^{33}$ \\
        \hline
	\end{tabular}
\end{table*}

\subsection{Sources of flare emission}
\label{disc:loops}
In Section~\ref{sec:flareparams}, we computed flare energies in the chromospheric spectral lines assuming they emerge from surface regions of the chromosphere, either by heating the footpoint regions (flare ribbons) of flare loops by energetic electron beams \citep[e.g.][]{Druett17} or backwarming from the X-ray/EUV emitting coronal flare plasma \citep{Hawley2003}. However, the emission may also come from cool flare loops overlying the chromosphere, such as H$\alpha$ post-flare loops, which are frequently observed on the Sun. These loops evolve from the initially hotter flare loops due to cooling. On the Sun, however, these are typically observed in absorption in the H$\alpha$ line if viewed against the solar disc \citep{Heinzel92}. \citet{Heinzel2018} demonstrated that even white-light emission from superflares may stem from loops instead of footpoints, provided that the electron densities are sufficiently high. They also found that for M dwarfs, this would be possible for lower electron densities compared to Sun-like stars. Thus, also stellar Balmer line flare emission may stem, at least partly, from loop emission. This concept was recently addressed by \citet{Wollmann2023} to explain the frequently observed red asymmetries in the Balmer lines of M dwarf flares. They presented a model of flare loops in which the cooling plasma moves downwards during the gradual phase of the flare, i.e. the concept of coronal rain. The results were consistent with observations of the dMe star AD~Leo. However, as AU~Mic is an earlier and thus hotter M dwarf than AD~Leo, we need to estimate which loop parameters would be required to observe Balmer line emission from such loops on AU~Mic.

\begin{figure}
	\includegraphics[width=\columnwidth]{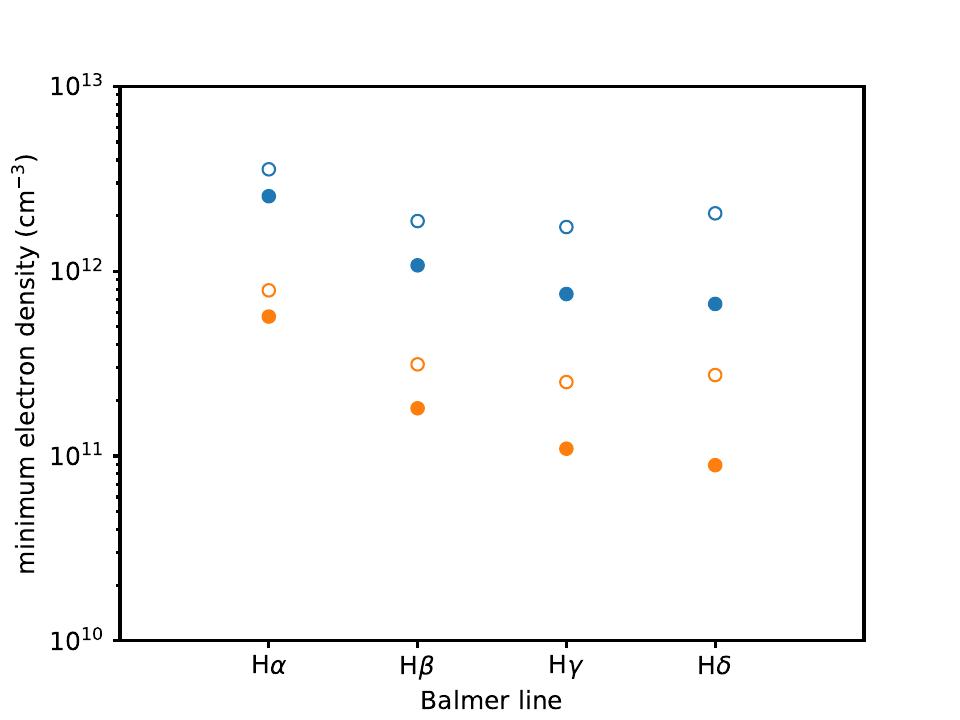}
    \caption{Minimum electron density required that a flare loop may be seen in emission against the stellar disc for the first four Balmer lines. Filled symbols are for AU~Mic and empty symbols for the Sun. Blue symbols use a loop temperature of 10000\,K, orange symbols 20000\,K. A dilution factor of $W=0.5$ corresponding to $h=0$ was adopted. For $h>0$, the minimum electron densities can become at maximum a factor of two higher than the corresponding $h=0$ values.}
    \label{fig:nemin}
\end{figure}

We can estimate the electron densities that would be required to observe excess emission if cool flare loops are observed against the stellar disc on an early M dwarf like AU~Mic. The emergent intensity $I$ of a 1D slab of plasma can be written as
\begin{equation}\label{eq:iout}
I = I_\mathrm{bg}e^{-\tau}+S(1-e^{-\tau}),
\end{equation}
where $I_\mathrm{bg}$ is the background intensity, which we assume to correspond to the stellar quiescent intensity $I_*$ in the line centre, $\tau$ the optical thickness, and $S$ the source function \citep{Heinzel15}. Emission would be observed if $I/I_*>1$, which leads to the constraint $S/I_*>1$. In a simple two-level atom model, the source function in the Balmer lines can be approximated by
\begin{equation}\label{eq:s}
S=(1-\epsilon)\bar{J}+\epsilon B(T)
\end{equation}
\citep[e.g.][]{Heinzel15}, where the first term describes scattering of the incident radiation and the second term the thermal emission of the plasma at temperature $T$, where $B(T)$ is the Planck function. The parameter $\bar{J}$ is the mean intensity inside the slab, which is the sum of direct external illumination and diffuse intensity field. For simplicity, we approximate it here as $WI_*$ \citep[cf.][]{Wollmann2023}, where $W$ is the geometrical dilution factor which depends on the height of the slab above the stellar surface \citep{Heinzel15}. Furthermore, $\epsilon=C_{ji}/(C_{ji}+A_{ji})$, where $C_{ji}=n_e\Omega_{ji}(T)$ is the collisional transition rate \citep{Johnson72} and $A_{ji}$ the Einstein coefficient for spontaneous emission of the line, with lower and upper levels $i$ and $j$. Since $C_{ji}$ depends on the electron density $n_e$, we can find the minimum electron density as a function of $T$, $W$, and $I_*$ for the Balmer lines. The minimum electron density can be then written as
\begin{equation}\label{eq:nemin}
n_e > \frac{A_{ji}}{\Omega_{ji}(T)}\frac{1-W}{\frac{B(T)}{I_*}-1}.
\end{equation}
The results of this estimate are shown in Fig.\,\ref{fig:nemin}. Considering that the temperature of H$\alpha$-emitting plasma is typically in the order of 10000\,K, we show a range of 10000--20000\,K. For the geometrical dilution factor, we adopt 0.5 corresponding to a loop height $h=0$. We note that $W$ decreases with height and reaches zero for $h\rightarrow\infty$, which means that higher loop heights increase $n_\mathrm{e,min}$, but at maximum up to a factor of two compared to the value for $h=0$, so we do not show these results in Fig.\,\ref{fig:nemin}. For $I_*$, we adopt the line centre values from the quiescent flux-calibrated CFHT spectrum (Fig.\,\ref{fig:fluxcalib}), and further use the relation between surface flux and intensity, $F_*=\pi I_*$.

There are some general trends seen in Fig.\,\ref{fig:nemin}. First, the minimum electron density $n_\mathrm{e,min}$ decreases slightly from lower to higher Balmer lines, meaning that if seeing H$\alpha$ in emission against the disc all other lines would be in emission as well; second, larger loop temperatures move $n_\mathrm{e,min}$ to lower values. But most importantly, our estimates show that for an M dwarf like AU~Mic, flare loops are expected to be seen in emission against the disc at lower electron densities compared the the Sun, and the difference in $n_\mathrm{e,min}$ becomes more pronounced for the higher Balmer lines for otherwise same loop parameters.

If the flare emission we observe may indeed come from loops instead of ribbons, it would affect the flare luminosities and energies which we computed in Section~\ref{sec:flareparams}. The flare energy is related to the source function of the flaring plasma, not to the emergent intensity from the flare region which can be, as in the case of on-disc loops, not exclusively related to the flare, but containing transmitted radiation. The relationship between emergent intensity and source function is given by Eq.\,\ref{eq:iout}, where $I_\mathrm{bg}=0$ for off-disc loops. We note that in reality, the flare emission may be a mixture of both footpoint and loop emission, with their relative contributions changing over the evolution of the flare.

Generally, the contributions of footpoints and loops can only be reliably disentangled using spatially resolved observations, like on the Sun. Without spatial resolution, one may in specific cases be able to estimate if emission from loops contributes to the overall flare emission. One possibility is the detection of red asymmetries in the Balmer lines during the decay phase of a flare, indicative of plasma downflows in cool loops \citep{Wollmann2023}, as this necessarily requires the existence of Balmer line-emitting loops. For very specific geometries, a cool loop system can lead to periodic occultations of the flare's footpoint regions, allowing a determination of the physical parameters (and their evolution) of footpoints and loop system, as succeeded with observations of an exceptional long-duration ($>$24\,h) flare by \textit{TESS} \citep{Bicz2024}. Another hint could be the detection of a secondary peak in a flare, as observed in some white-light flares from \textit{TESS} data, which has been attributed to loop emission \citep{Yang2023}.

Sun-as-a-star observations can give clues about the underlying causes of spatially integrated flare light curve shapes and line asymmetries \citep{Namekata2022, Otsu2022, Ma2024, Leitzinger2024, Otsu2024} and may be helpful to identify characteristic signatures from footpoints and loops. Recently, \citet{Otsu2024a} analysed H$\alpha$ observations of a solar flare and identified distinct signatures related to footpoint and loop emission. They found that the emergence of cool loops caused a secondary peak in the Sun-as-a-star flare light curve, similar to what has been suggested for white-light flares \citep{Yang2023}. Moreover, they detected both blue and red absorption signatures from downflows in the cool loops, as well as a flattening of the flare decay due to the off-disc emission of growing post-flare loops, as the flare was located close to the solar limb. These hints from solar observations are useful to interpret spatially unresolved stellar data. However, they are currently based on a very small number of solar flare events and more statistics are needed to better understand both possible variations of these signatures between different events, as well as potential alternative causes of these signatures. Moreover, directly translating these solar signatures to stars with different spectral types may not be trivial, as we discussed above. For instance, loops with physical parameters creating absorption signatures if viewed against the solar disc could as well produce emission signatures if viewed against an M dwarf's disc. Despite these limitations, we find that some of the flare light curves of AU~Mic do indeed show secondary peaks (e.g. Figs.\,\ref{fig:2023-06-05} and \ref{fig:2022-11-17}), but these could also be due to a secondary energy release in the flaring region, or the superposition of a flare occurring at another location on this active star. However, combination with an investigation of potentially associated line asymmetries, as we plan for a future study, may provide more clues about their nature.


\section{Conclusions}
We have conducted an optical spectroscopic and coordinated photometric monitoring campaign of the southern, young and active exoplanet host star AU~Mic with the E152 telescope at La Silla, Chile, operated by the PLATOSpec consortium. Within 56 nights, we detected 24 flares on AU~Mic from which two are confirmed superflares, but up to 90\,per cent of the sample may potentially be superflares based on flare energy relationships from the literature. For 15 flares we have coordinated photometry in the \textit{g'}-band. For only one flare we find also its counterpart in the photometry, likely due to the small size of the telescope. Accordingly, we find flare rates from H$\alpha$ spectroscopy of $3.0_{-0.61}^{+0.74}$\,d$^{-1}$ and from \textit{g'}-band photometry of $0.16^{+0.36}_{-0.13}$\,d$^{-1}$.

We investigate the flares in eight different spectral lines (H$\alpha$, H$\beta$, H$\gamma$, H$\delta$, \ion{Na}{i}\,D1\&D2, \ion{He}{i}\,D3, \ion{He}{i}\,6678) and in five flares all spectral lines are affected and show significant flare signatures. The flare parameter analysis reveals correlations between flare energy and flare peak luminosity, flare duration and flare energy, and H$\alpha$ flare energy and total flare energy in the chromospheric lines. We find no correlation between flare duration and flare peak luminosity. 

We detect one exceptional flare event for which only the possible flare peak phase has been captured. Fitting the flare peak data and data from the following night, which show a decay at the beginning of the night, with a flare model reveals the actual duration and energy of that event, as confirmed by independent estimates utilizing our peak luminosity-energy relation. From the energy measured in H$\alpha$ alone the flare reaches superflare energies and represents therefore a rare and highly energetic extreme event.

We detect rotational modulation in our \textit{g'}-band photometry, as well as in the H$\alpha$ EWs. In the \textit{TESS} phase-folded light curve, two local maxima are identified which we see also in our \textit{g'}-band phase-folded light curve. In the \textit{g'}-band, which is bluer than the \textit{TESS} wavelength band, the amplitude of rotational modulation is found to be larger (10\,per cent) than in the \textit{TESS} band (4\,per cent).

We follow the approach in \citet{Heinzel2018} and \citet{Wollmann2023} and consider a possible contribution of flare loop emission, in addition to footpoint emission, producing the chromospheric flare emission. We show that flare loops on AU~Mic may appear in emission when being located in front of the stellar disc already for slightly lower electron densities compared to the Sun.

The exceptional activity level of AU~Mic allowed detailed investigations of prominent chromospheric spectral lines during flares. Additional flare-affected spectral lines, as well as investigation of line broadening and line asymmetries during flares, will be the subject of a future study.

\section*{Acknowledgements}
We thank the referee for valuable comments that helped to improve the paper. This research was funded in whole, or in part, by the Austrian Science Fund (FWF) [10.55776/I5711]. For the purpose of open access, the author has applied a CC BY public copyright licence to any Author Accepted Manuscript version arising from this submission. PK, JL, RK, PH, and JW acknowledge GACR grant 22-30516K. The EXOWORLD project is supported by the European Union under the Horizon Europe Programme Marie Sk{\l}odowska-Curie Actions Staff Exchanges. PK and JL acknowledge travel funding for mobility to Chile from Horizon 2020 Project ID: 101086149 EXOWORLD. PH was supported by the program 'Excellence Initiative - Research University' for years 2020–2026 at University of Wroc{\l}aw, project No. BPIDUB.4610.96.2021.KG. This work was generously supported by the Th\"uringer Ministerium f\"ur Wirtschaft, Wissenschaft und Digitale Gesellschaft and the Th\"uringer Aufbaubank. LV acknowledges projects ANID Fondecyt n. 1211162, ANID Quimal ASTRO20-0025 and ANID BASAL FB210003.
We thank the observers who contributed to collecting the data presented in this paper: Eva \v{Z}\v{d}\'arsk\'a, Lud\v{e}k \v{R}ezba, Zuzana Balk\'oov\'a, Michaela V\'itkov\'a, Veronika Schaffenroth and Joanne Yoshua. We also thank Jan Fuchs for software support.
This paper includes data collected by the \textit{TESS} mission, which are publicly available from the Mikulski Archive for Space Telescopes (MAST). Funding for the \textit{TESS} mission is provided by the NASA's Science Mission Directorate. This work has made use of data from the European Space Agency (ESA) mission {\it Gaia} (\url{https://www.cosmos.esa.int/gaia}), processed by the {\it Gaia} Data Processing and Analysis Consortium (DPAC, \url{https://www.cosmos.esa.int/web/gaia/dpac/consortium}). Funding for the DPAC has been provided by national institutions, in particular the institutions participating in the {\it Gaia} Multilateral Agreement. This research used observations obtained at the Canada-France-Hawaii Telescope (CFHT) which is operated by the National Research Council of Canada, the Institut National des Sciences de l'Univers of the Centre National de la Recherche Scientique of France, and the University of Hawaii.

\section*{Data Availability}
The data collected for the work presented in this article are available upon a request in their raw, but also in their CERES+ reduced versions. We can provide photometric and also spectroscopic data if requested. Later in 2025, we will be feeding the data sets into the ESO archive. The \textit{TESS} light curve data are available at the Mikulski Archive for Space Telescopes (MAST; \url{https://mast.stsci.edu}). The \textit{Gaia} DR3 spectrum is available at VizieR (\url{https://vizier.cds.unistra.fr}) in Table I/355/spectra. The CFHT spectra are available at the Canadian Astronomy Data Centre (\url{https://www.cadc-ccda.hia-iha.nrc-cnrc.gc.ca}).
 



\bibliographystyle{mnras}
\bibliography{aumicbib} 



\appendix

\section{Log of observations}
\label{app:obslog}

\begin{table*}
	\centering
	\caption{Observing log. The first two columns give the start times of the first and last spectra of the night, respectively. Columns 3-6 give the total observing time during this night in hours, the number of spectra per night, the exposure time in seconds, and the net on-source time (number of spectra multiplied by exposure time, i.e. excluding data gaps) in hours. The last column summarizes details about the simultaneous photometry if available (filter/exposure time in seconds). In three nights only photometry was available, thus columns 1-2 give the times of the first and last photometric image instead.}
	\label{tab:obslog}
	\begin{tabular}{llccccc}
		\hline
		Start (UT) & End (UT) & $T_\mathrm{obs}$ (h) & $N_\mathrm{spec}$ & $T_\mathrm{exp}$ (s) & $T_\mathrm{net}$ (h) & phot\\
		\hline
		2022-10-31T00:34:17 & 2022-10-31T02:31:12 & 2.03 & 20 & 300 & 1.67 & -- \\
        2022-11-02T00:51:30 & 2022-11-02T02:07:29 & 1.43 & 8  & 600 & 1.33 & -- \\
        2022-11-05T00:02:13 & 2022-11-05T03:17:35 & 3.42 & 17 & 600 & 2.83 & -- \\
        2022-11-05T23:49:32 & 2022-11-06T02:43:10 & 3.06 & 17 & 600 & 2.83 & -- \\
        2022-11-06T23:57:10 & 2022-11-07T02:12:47 & 2.43 & 10 & 600 & 1.67 & -- \\
        2022-11-08T00:28:08 & 2022-11-08T03:26:58 & 3.15 & 14 & 600 & 2.33 & -- \\
        2022-11-09T01:57:46 & 2022-11-09T02:52:01 & 1.07 & 6  & 600 & 1.0 & -- \\
        2022-11-10T02:25:31 & 2022-11-10T03:05:56 & 0.84 & 5  & 600 & 0.83 & -- \\
        2022-11-14T01:00:18 & 2022-11-14T03:57:53 & 3.13 & 15 & 600 & 2.5 & -- \\
        2022-11-15T01:59:10 & 2022-11-15T03:47:27 & 1.97 & 11 & 600 & 1.83 & -- \\
        2022-11-17T00:04:51 & 2022-11-17T02:36:33 & 2.69 & 15 & 600 & 2.5 & -- \\
        2023-04-26T07:04:25 & 2023-04-26T09:45:14 & 2.85 & 17 & 600 & 2.83 & \textit{g'}/15 \\
        2023-04-27T08:13:28 & 2023-04-27T08:33:33 & 0.50 & 2  & 600 & 0.33 & \textit{g'}/10 \\
        2023-04-28T07:02:08 & 2023-04-28T08:12:28 & 1.34 & 8  & 600 & 1.33 & \textit{g'}/10 \\
        2023-04-29T07:08:26 & 2023-04-29T09:59:16 & 3.01 & 18 & 600 & 3.0 & \textit{g'}/10 \\
        2023-05-01T07:08:04 & 2023-05-01T09:58:53 & 3.01 & 18 & 600 & 3.0 & \textit{g'}/20+SA200 \\
        2023-05-02T06:31:52 & 2023-05-02T10:03:03 & 3.69 & 22 & 600 & 3.67 & \textit{g'}/10+SA200 \\
        2023-05-06T06:25:56 & 2023-05-06T10:06:45 & 3.93 & 15 & 900 & 3.75 & \textit{g'}/20 \\
        2023-05-08T08:35:05 & 2023-05-08T09:06:38 & 0.78 & 3  & 900 & 0.75 & \textit{g'}/8 \\
        2023-05-09T09:08:56 & 2023-05-09T10:12:01 & 1.30 & 5  & 900 & 1.25 & \textit{g'}/5 \\
        2023-05-10T07:23:20 & 2023-05-10T10:03:02 & 2.91 & 11 & 900 & 2.75 & \textit{g'}/8 \\
        2023-05-13T08:43:22 & 2023-05-13T10:02:14 & 1.56 & 6  & 900 & 1.5 & \textit{r'}/5 \\
        2023-05-26T08:03:48 & 2023-05-26T10:32:04 & 2.72 & 10 & 900 & 2.5 & \textit{g'}/10 \\
        2023-05-29T07:59:07 & 2023-05-29T10:29:31 & 2.76 & 10 & 900 & 2.5 & \textit{g'}/5 \\
        2023-05-30T05:13:41 & 2023-05-30T10:29:11 & 5.51 & 21 & 900 & 5.25 & \textit{g'}/10 \\
        2023-05-31T05:29:34 & 2023-05-31T10:29:25 & 5.25 & 20 & 900 & 5.0 & \textit{g'}/10 \\
        2023-06-01T06:16:25 & 2023-06-01T10:15:31 & 4.23 & 16 & 900 & 4.0 & \textit{g'}/3 \\
        2023-06-02T05:29:05 & 2023-06-02T10:44:45 & 5.51 & 20 & 900 & 5.0 & \textit{g'}/2 \\
        2023-06-05T05:06:32 & 2023-06-05T10:11:32 & 5.33 & 18 & 900 & 4.5 & \textit{g'}/7 \\
        2023-06-07T05:55:20 & 2023-06-07T10:33:26 & 4.88 & 18 & 900 & 4.5 & \textit{g'}/7 \\
        2023-06-08T06:04:32 & 2023-06-08T10:22:41 & 4.55 & 17 & 900 & 4.25 & \textit{g'}/5 \\
        2023-06-09T07:21:50 & 2023-06-09T10:15:24 & 3.14 & 12 & 900 & 3.0 & \textit{g'}/5 \\
        2023-06-10T05:30:50 & 2023-06-10T07:52:50 & 2.62 & 10 & 900 & 2.5 & \textit{g'}/5 \\
        2023-06-12T05:27:02 & 2023-06-12T10:26:47 & 5.25 & 20 & 900 & 5.0 & \textit{g'}/5 \\
        2023-06-13T06:21:54 & 2023-06-13T10:18:33 & 4.19 & 16 & 900 & 4.0 & \textit{g'}/7 \\
        2023-06-14T05:23:49 & 2023-06-14T06:42:43 & 1.56 & 6  & 900 & 1.5 & \textit{g'}/5 \\
        2023-06-15T05:52:55 & 2023-06-15T10:21:15 & 4.72 & 17 & 900 & 4.25 & \textit{g'}/5 \\
        2023-06-16T06:03:04 & 2023-06-16T10:31:20 & 4.72 & 12 & 900 & 3.0 & \textit{g'}/3 \\
        2023-06-17T07:28:33 & 2023-06-17T10:40:04 & -- & -- & --  & --  & \textit{r'}/12 \\
        2023-06-21T07:41:39 & 2023-06-21T08:54:23 & 1.38 & 7  & 600 & 1.17 & \textit{g'}/10 \\
        2023-06-23T06:25:06 & 2023-06-23T10:10:27 & 4.01 & 11 & 900 & 2.75 & \textit{g'}/5 \\
        2023-06-24T07:09:26 & 2023-06-24T10:11:24 & 3.28 & 11 & 900 & 2.75 & \textit{r'}/8 \\
        2023-06-28T04:17:34 & 2023-06-28T09:35:37 & 5.55 & 20 & 900 & 5.0 & \textit{g'}/10+5 \\
        2023-06-29T04:09:25 & 2023-06-29T05:28:18 & 1.56 & 6  & 900 & 1.5 & \textit{g'}/5 \\
        2023-06-30T03:51:58 & 2023-06-30T09:59:13 & 6.20 & 49 & 300 & 4.08 & -- \\
        2023-07-01T04:16:27 & 2023-07-01T09:33:38 & 5.37 & 49 & 300 & 4.08 & \textit{g'}/5 \\
        2023-07-08T07:52:27 & 2023-07-08T09:50:30 & -- & -- & --  & --  & \textit{g'}/5 \\
        2023-07-13T03:02:01 & 2023-07-13T10:08:15 & 7.19 & 35 & 300 & 2.92 & \textit{g'}/15 \\
        2023-07-31T03:31:33 & 2023-07-31T10:38:34 & -- & -- & --  & --   & \textit{g'}/5 \\
        2023-08-01T04:04:18 & 2023-08-01T08:46:28 & 4.79 & 37 & 300 & 3.08 & \textit{g'}/4 \\
        2023-08-11T08:01:07 & 2023-08-11T09:49:29 & 1.89 & 15 & 300 & 1.25 & -- \\
        2023-08-12T04:36:27 & 2023-08-12T09:19:29 & 4.80 & 47 & 300 & 3.92 & -- \\
        2023-08-13T05:05:52 & 2023-08-13T08:19:11 & 3.30 & 29 & 300 & 2.42 & -- \\
        2023-08-14T07:39:28 & 2023-08-14T08:07:39 & 0.55 & 5  & 300 & 0.42 & -- \\
        2023-08-15T00:18:10 & 2023-08-15T07:34:53 & 7.36 & 50 & 300 & 4.17 & \textit{g'}/4 \\
        2023-08-17T23:54:34 & 2023-08-18T09:30:13 & 9.68 & 89 & 300 & 7.42 & \textit{g'}/15 \\
        2023-09-16T00:29:08 & 2023-09-16T01:05:04 & 0.68 & 6  & 300 & 0.5 & \textit{g'}/5 \\
        2023-09-16T23:27:25 & 2023-09-17T05:59:42 & 6.62 & 55 & 300 & 4.58 & \textit{g'}/10 \\
        2023-09-21T00:03:17 & 2023-09-21T00:55:12 & 0.95 & 10 & 300 & 0.83 & \textit{g'}/5 \\
		\hline
        Total: && 192.24 & 1037 && 159.08 & \\
        \hline
	\end{tabular}
\end{table*}

\clearpage

\section{Flare light curves}
\label{app:lc}

\begin{figure*}
	\includegraphics[width=\textwidth]{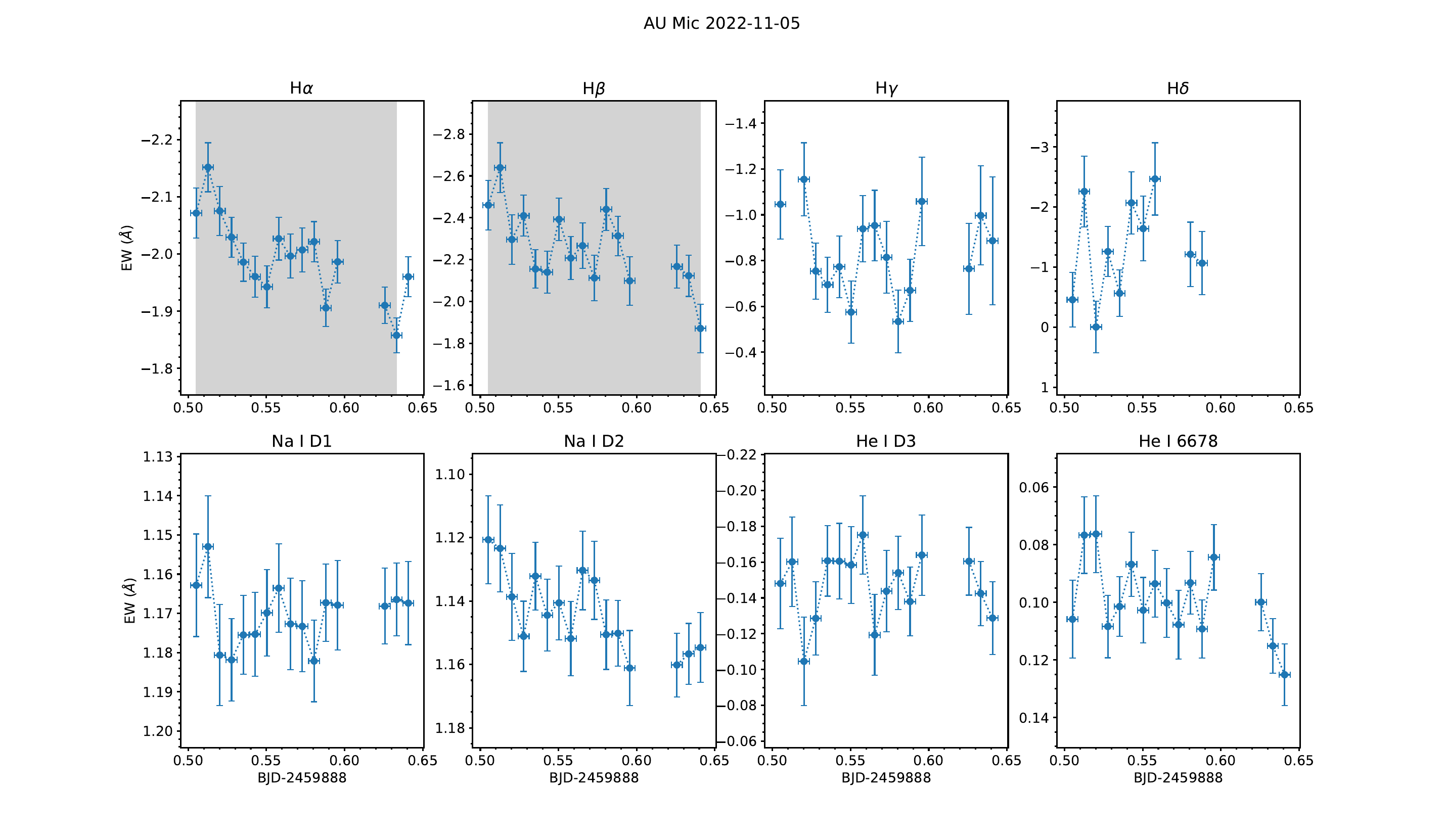}
    \caption{Flare \#1 on 2022-11-05 in all studied spectral lines. Significant detection in H$\alpha$ and H$\beta$.}
    \label{fig:2022-11-05}
\end{figure*}

\begin{figure*}
 \includegraphics[width=\columnwidth]{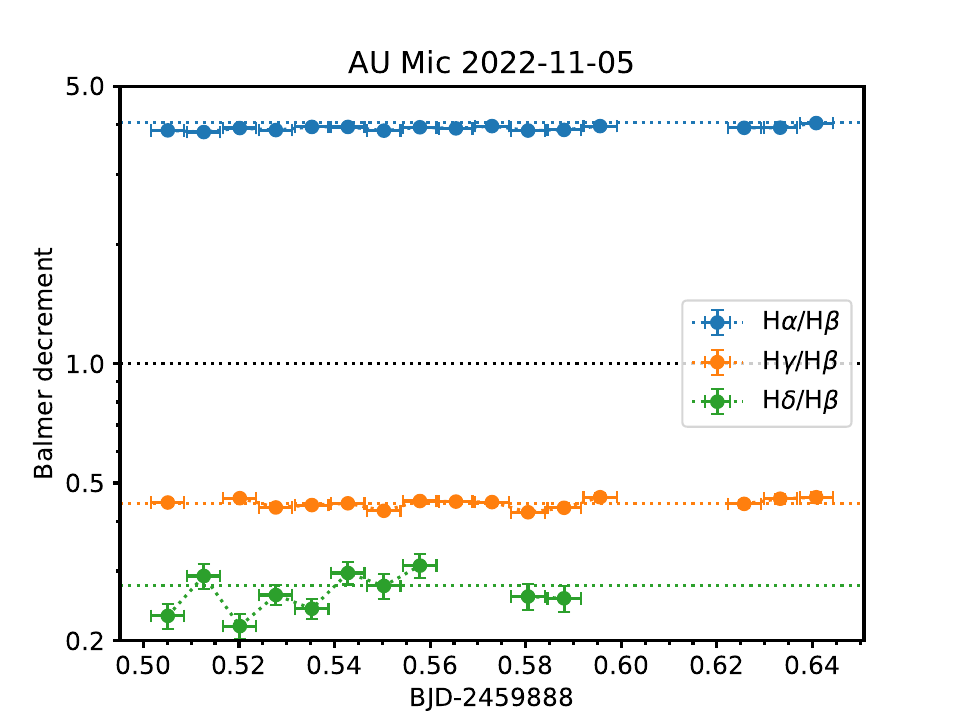}
    \caption{Balmer decrements relative to H$\beta$ along the flare.}
    \label{fig:bd2022-11-05}
\end{figure*}

\begin{figure*}
	\includegraphics[width=\textwidth]{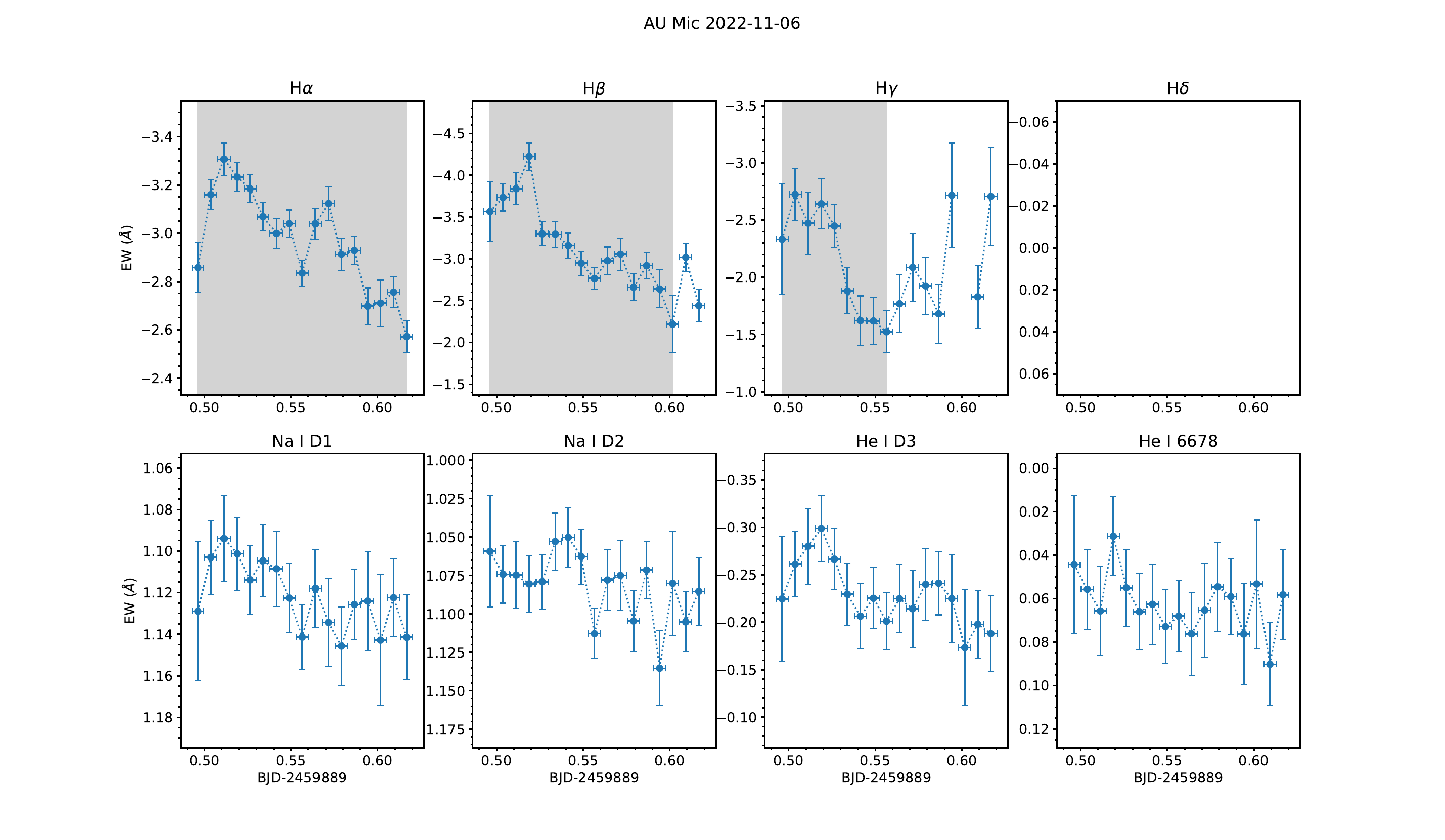}
    \caption{Flare \#2 on 2022-11-06 in all studied spectral lines. Significant detection in H$\alpha$, H$\beta$ and H$\gamma$.}
    \label{fig:2022-11-06}
\end{figure*}

\begin{figure*}
 \includegraphics[width=\columnwidth]{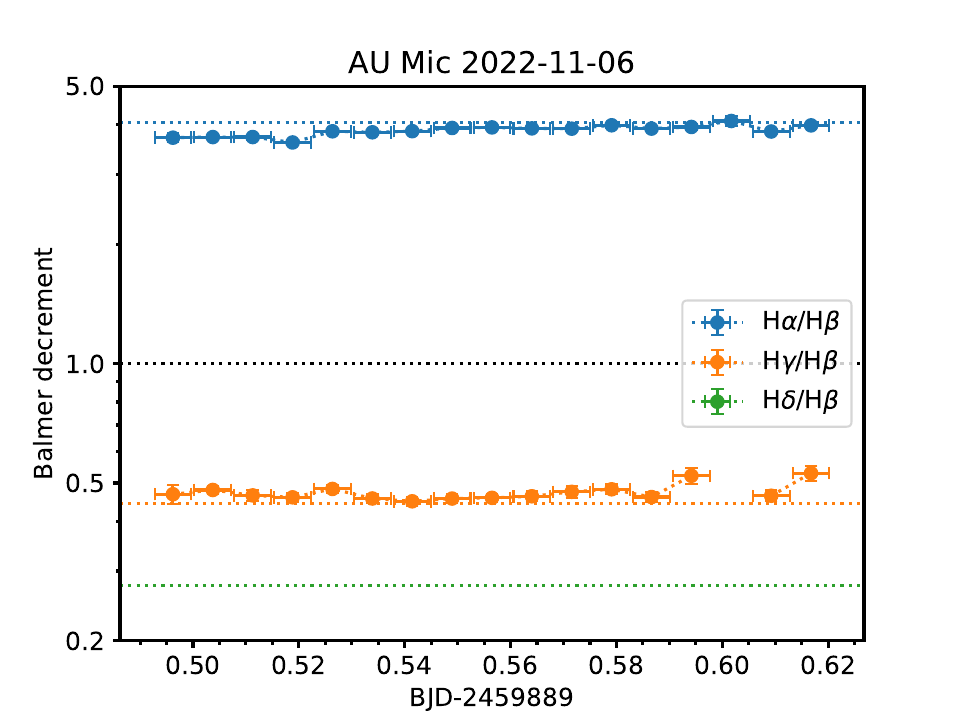}
    \caption{Balmer decrements relative to H$\beta$ along the flare.}
    \label{fig:bd2022-11-06}
\end{figure*}

\begin{figure*}
	\includegraphics[width=\textwidth]{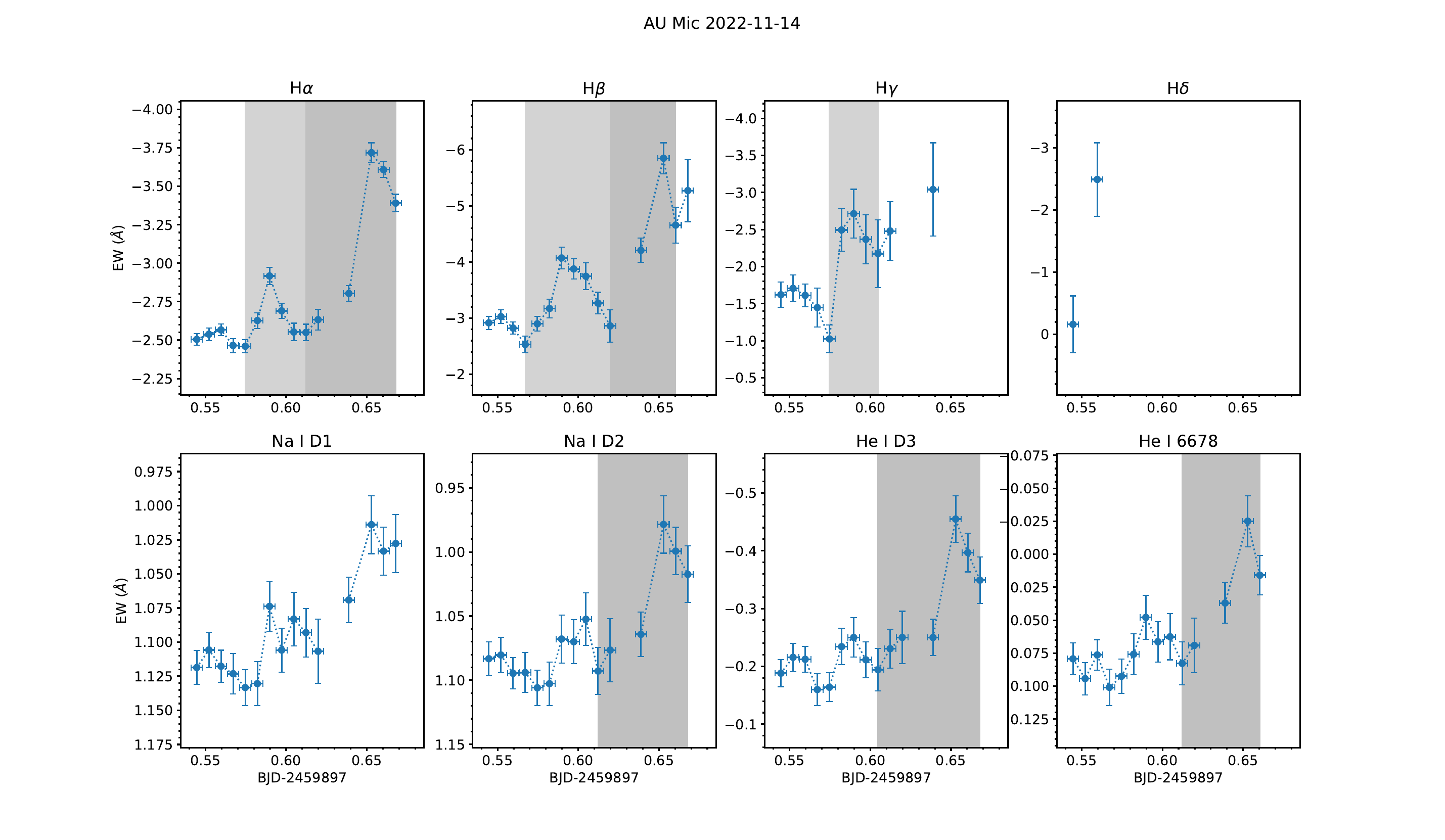}
    \caption{Flares \#3 and \#4 on 2022-11-14 in all studied spectral lines. Significant detection of flare \#3 in H$\alpha$, H$\beta$ and H$\gamma$. Significant detection of flare \#4 in H$\alpha$, H$\beta$, \ion{Na}{i}\,D2, \ion{He}{i}\,D3 and \ion{He}{i}\,6678.}
    \label{fig:2022-11-14}
\end{figure*}

\begin{figure*}
 \includegraphics[width=\columnwidth]{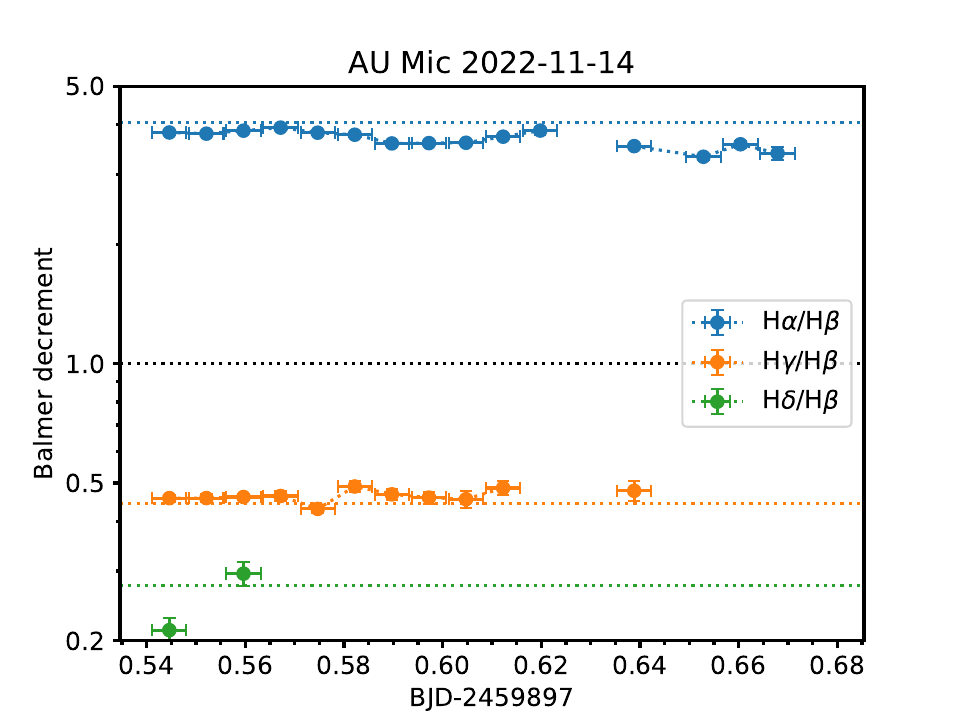}
    \caption{Balmer decrements relative to H$\beta$ along the flare.}
    \label{fig:bd2022-11-14}
\end{figure*}

\begin{figure*}
	\includegraphics[width=\textwidth]{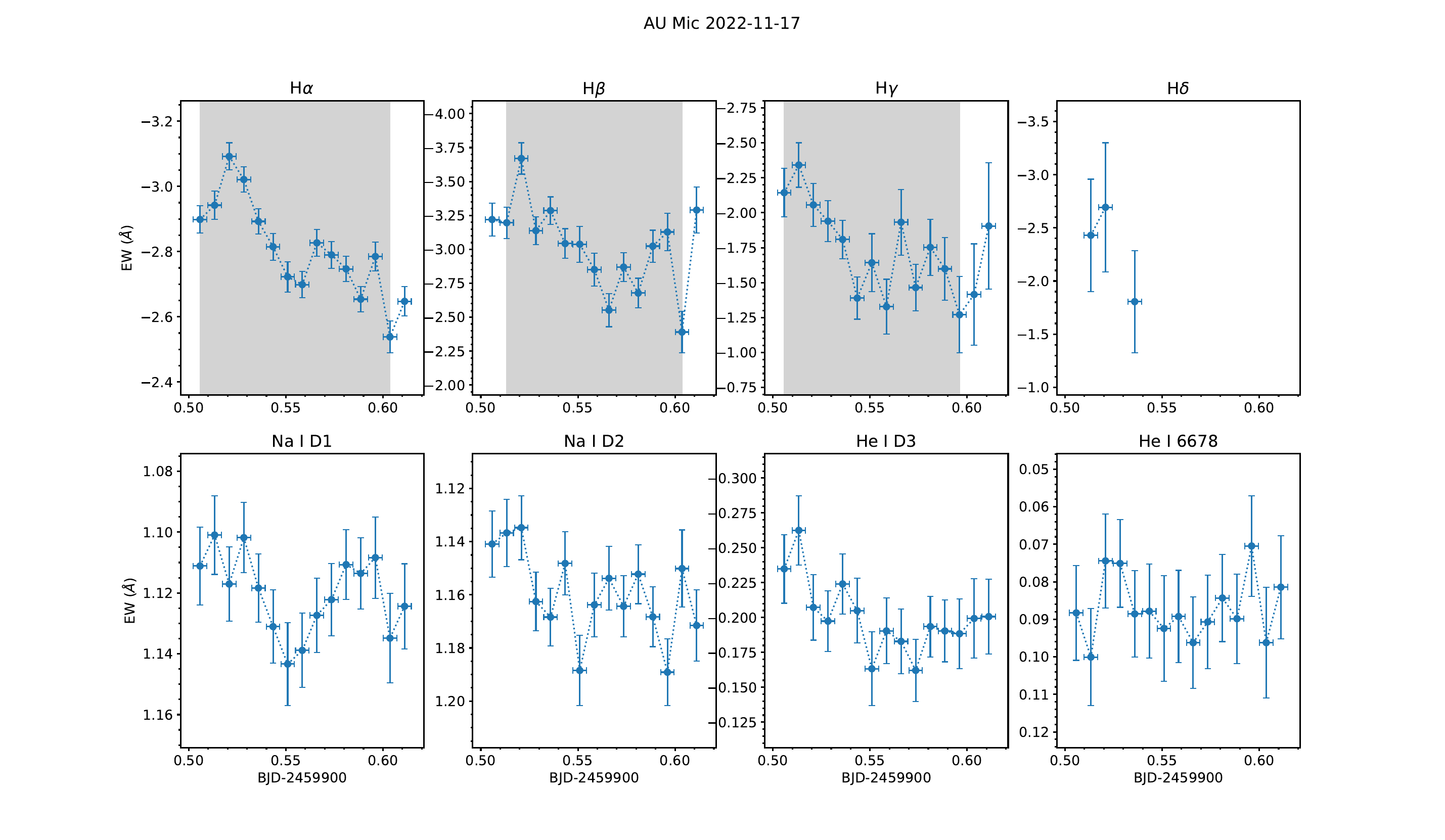}
    \caption{Flare \#5 on 2022-11-17 in all studied spectral lines. Significant detection in H$\alpha$, H$\beta$ and H$\gamma$.}
    \label{fig:2022-11-17}
\end{figure*}

\begin{figure*}
 \includegraphics[width=\columnwidth]{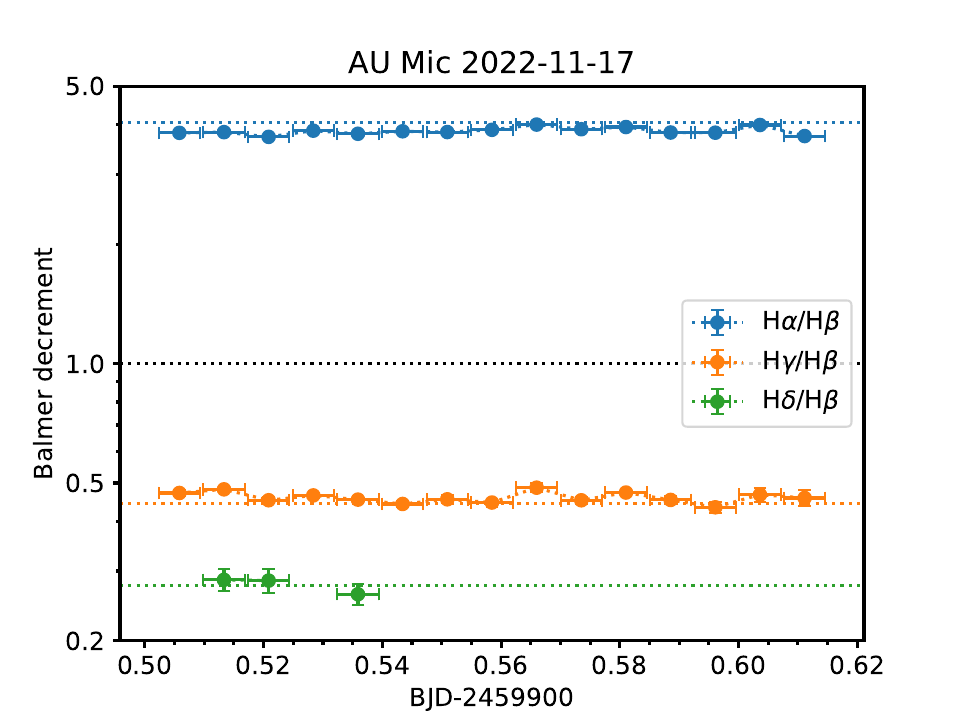}
    \caption{Balmer decrements relative to H$\beta$ along the flare.}
    \label{fig:bd2022-11-17}
\end{figure*}

\begin{figure*}
	\includegraphics[width=\textwidth]{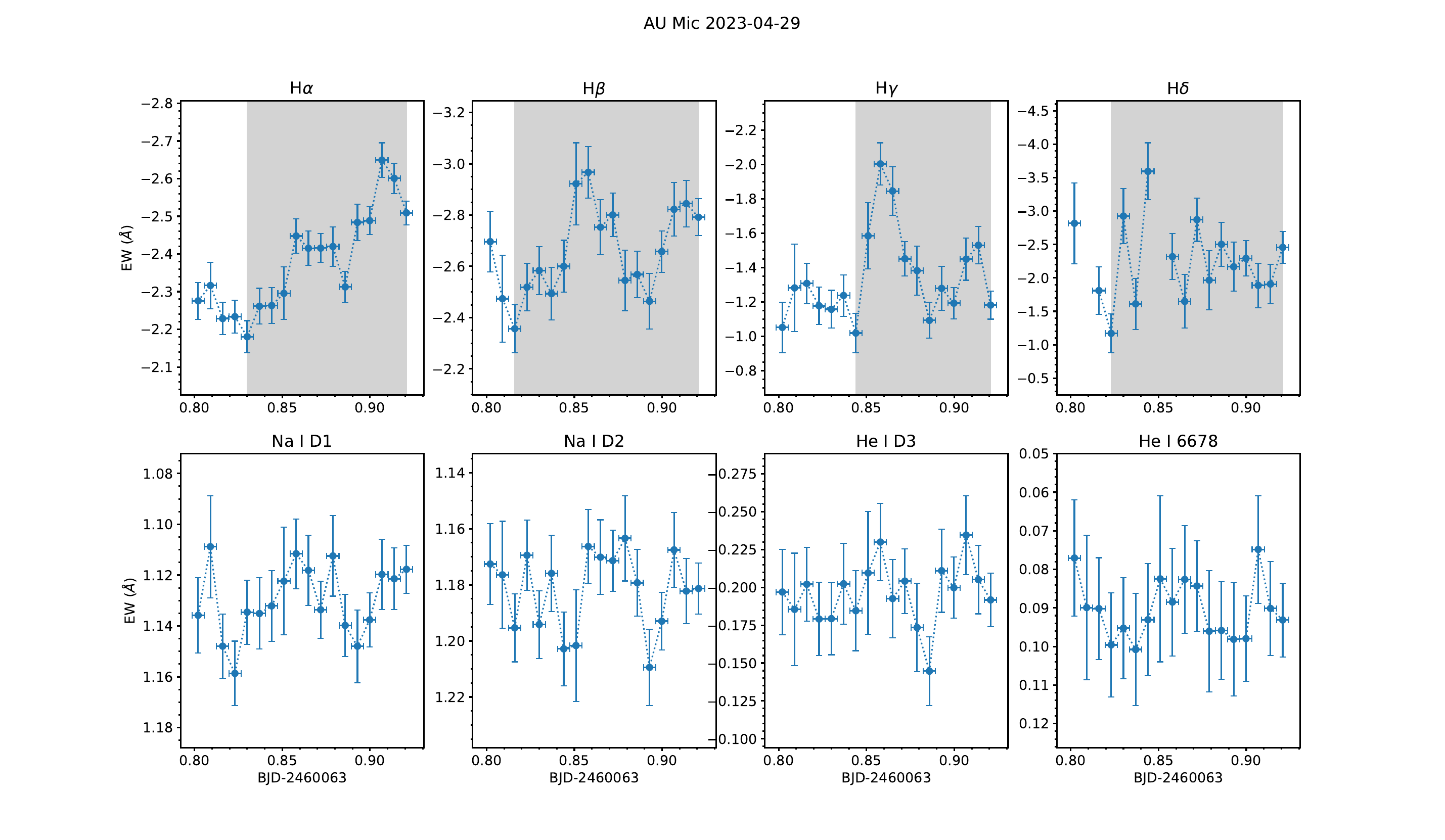}
    \caption{Flare \#6 on 2023-04-29 in all studied spectral lines. Significant detection in H$\alpha$, H$\beta$, H$\gamma$ and H$\delta$.}
    \label{fig:2023-04-29}
\end{figure*}

\begin{figure*}
	\includegraphics[width=\columnwidth]{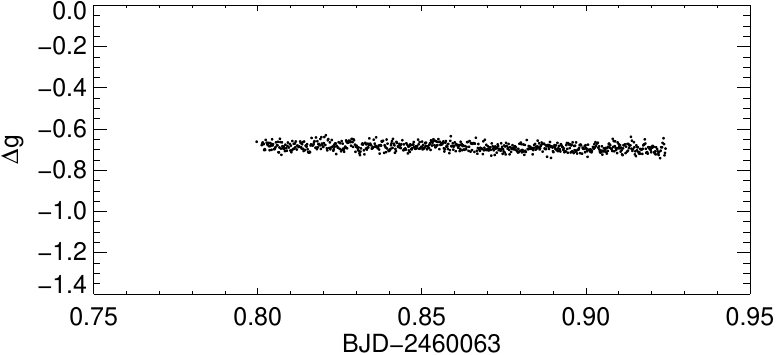}
 \includegraphics[width=0.7\columnwidth]{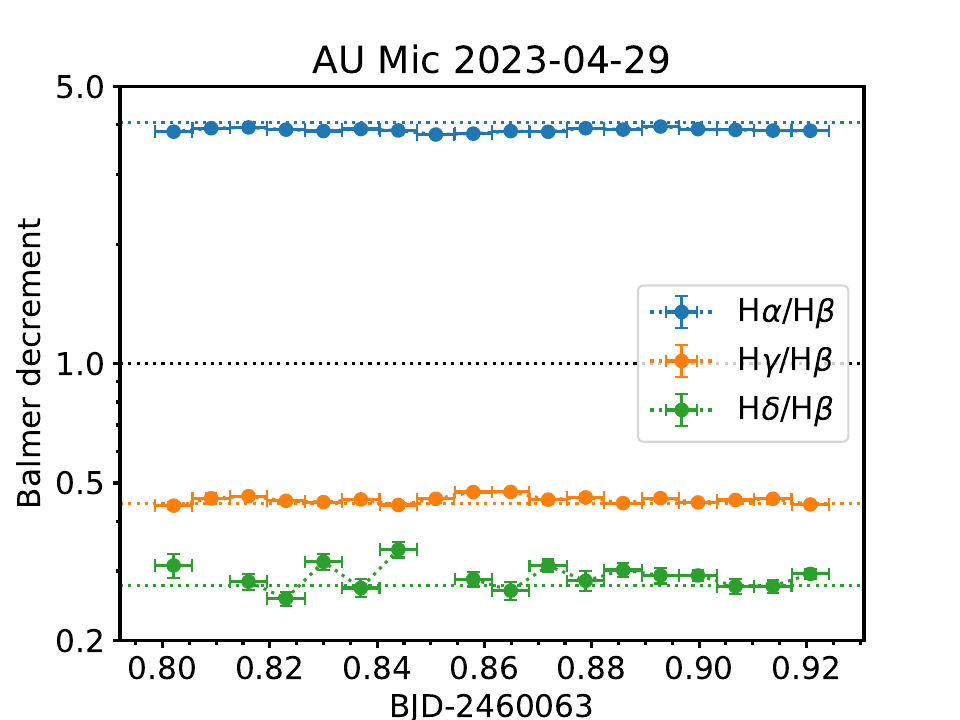}
    \caption{Left panel: Coordinated \textit{g'}-band photometry on 2023-04-29. Right panel: Balmer decrements relative to H$\beta$ along the flare.}
    \label{fig:phbd2023-04-29}
\end{figure*}

\begin{figure*}
	\includegraphics[width=\textwidth]{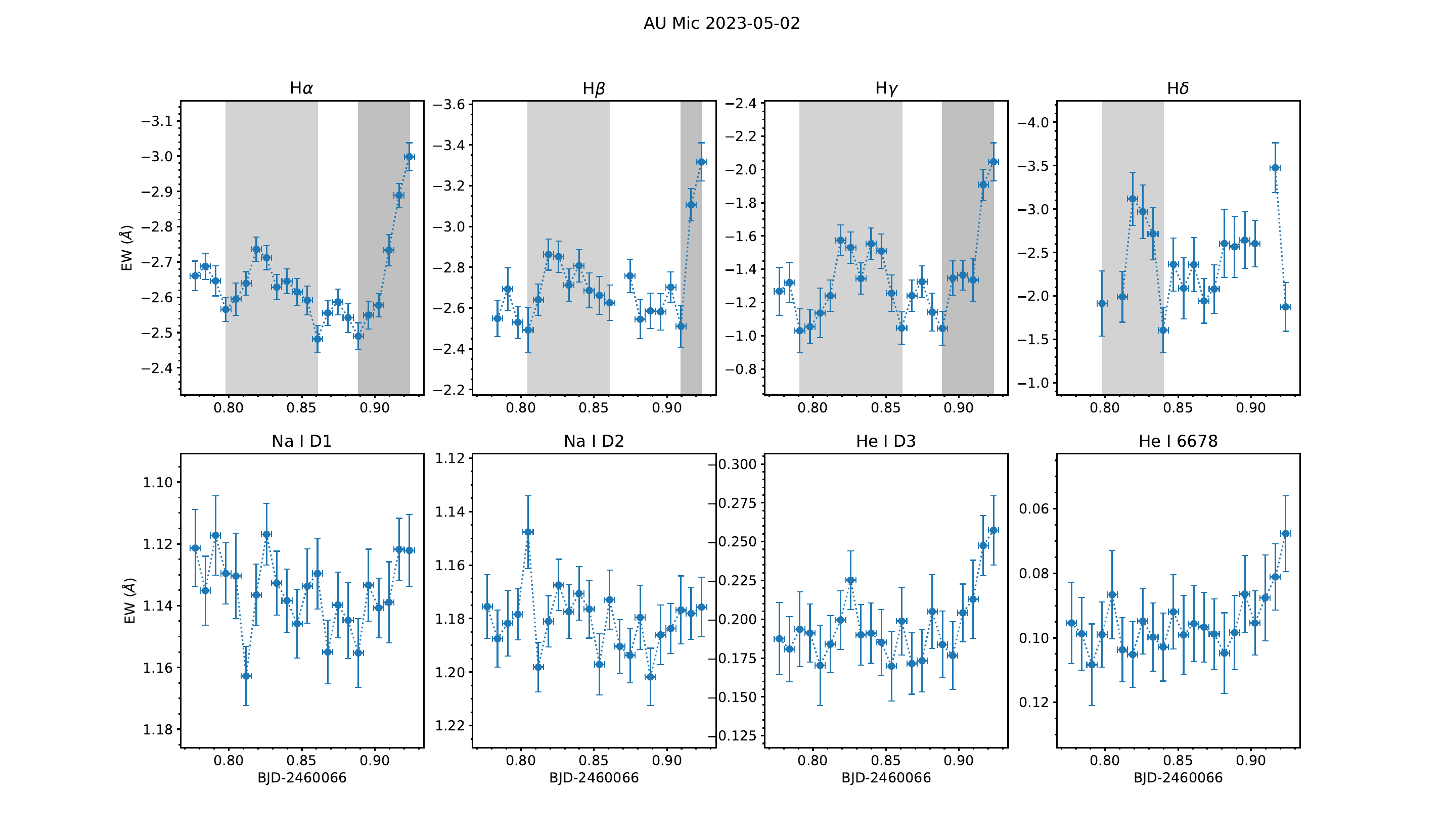}
    \caption{Flares \#7 and \#8 on 2023-05-02 in all studied spectral lines. Significant detection of flare \#7 in H$\alpha$, H$\gamma$ and H$\delta$. The flare was not significantly detected in H$\beta$, but added after visual inspection. Significant detection of flare \#8 in H$\alpha$, H$\beta$ and H$\gamma$.}
    \label{fig:2023-05-02}
\end{figure*}

\begin{figure*}
	\includegraphics[width=\columnwidth]{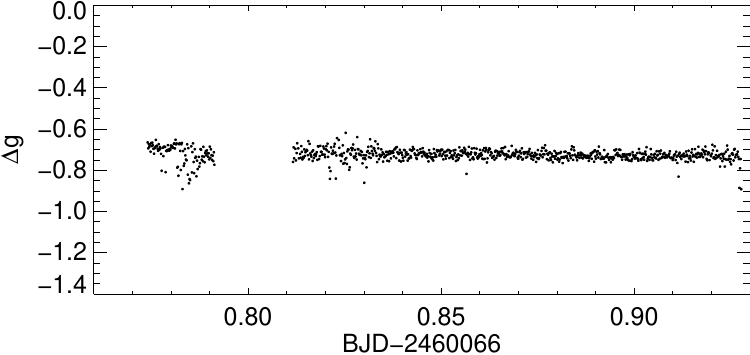}
 \includegraphics[width=0.7\columnwidth]{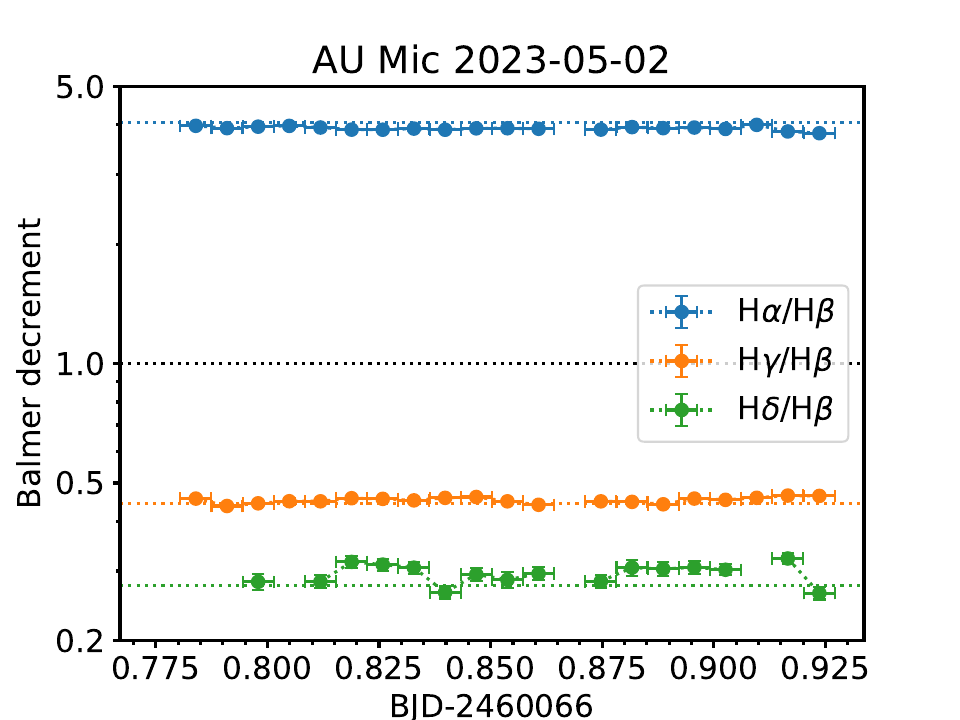}
    \caption{Left panel: Coordinated \textit{g'}-band photometry on 2023-05-02. Right panel: Balmer decrements relative to H$\beta$ along the flare.}
    \label{fig:phbd2023-05-02}
\end{figure*}

\begin{figure*}
	\includegraphics[width=\textwidth]{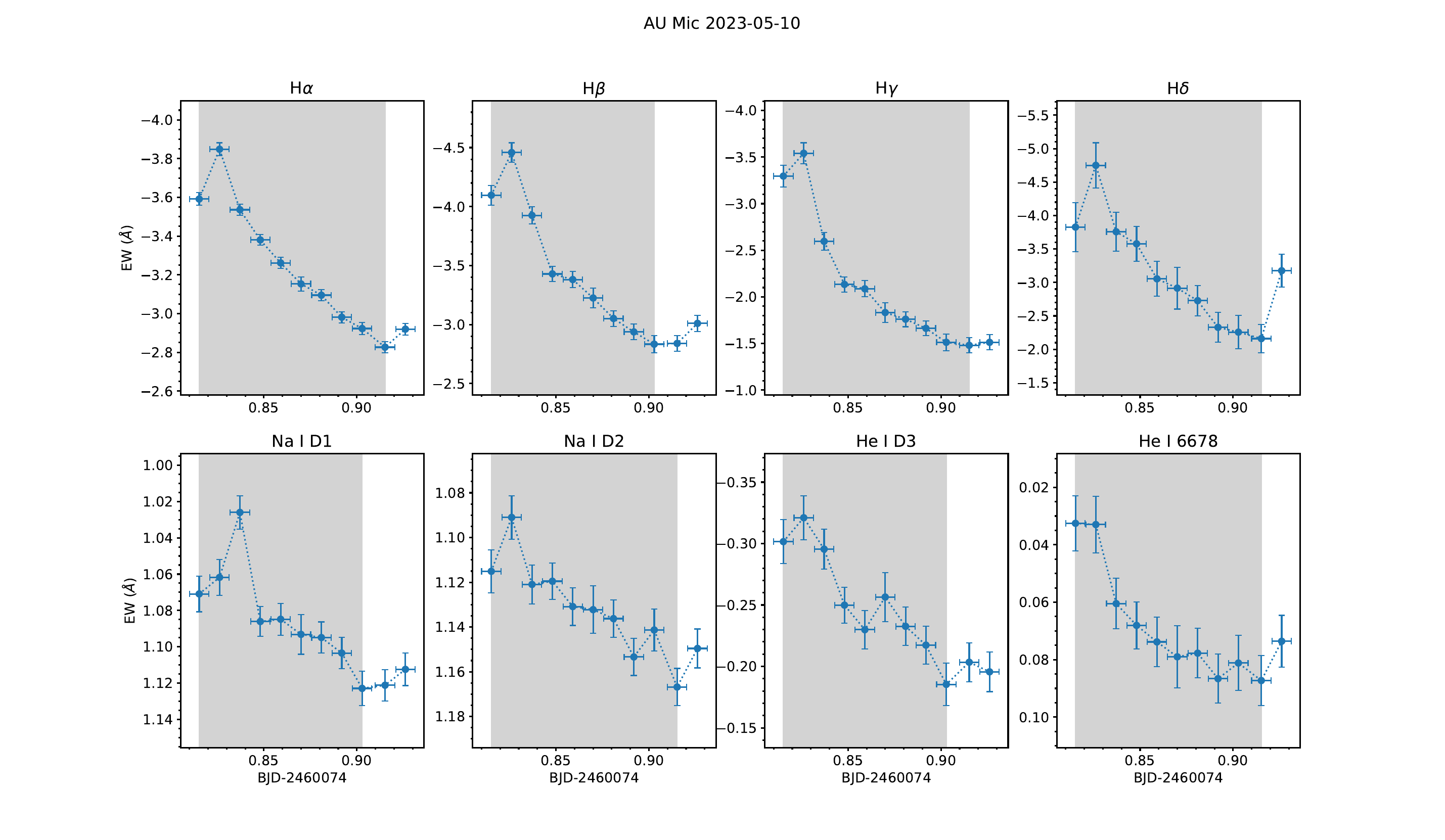}
    \caption{Flare \#9 on 2023-05-10 in all studied spectral lines. Significant detection in all spectral lines.}
    \label{fig:2023-05-10}
\end{figure*}

\begin{figure*}
	\includegraphics[width=\columnwidth]{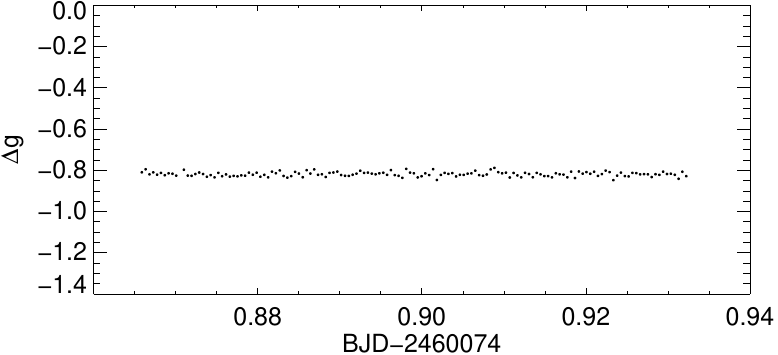}
 \includegraphics[width=0.7\columnwidth]{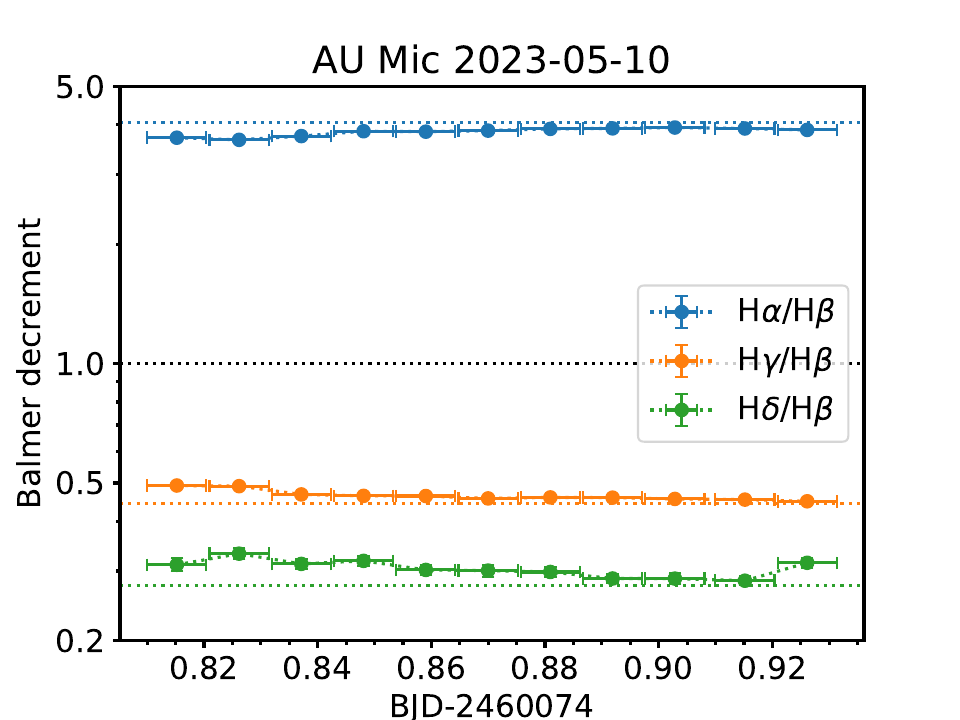}
    \caption{Left panel: Coordinated \textit{g'}-band photometry on 2023-05-10. The first half of the night was not usable. Right panel: Balmer decrements relative to H$\beta$ along the flare.}
    \label{fig:phbd2023-05-10}
\end{figure*}

\begin{figure*}
	\includegraphics[width=\textwidth]{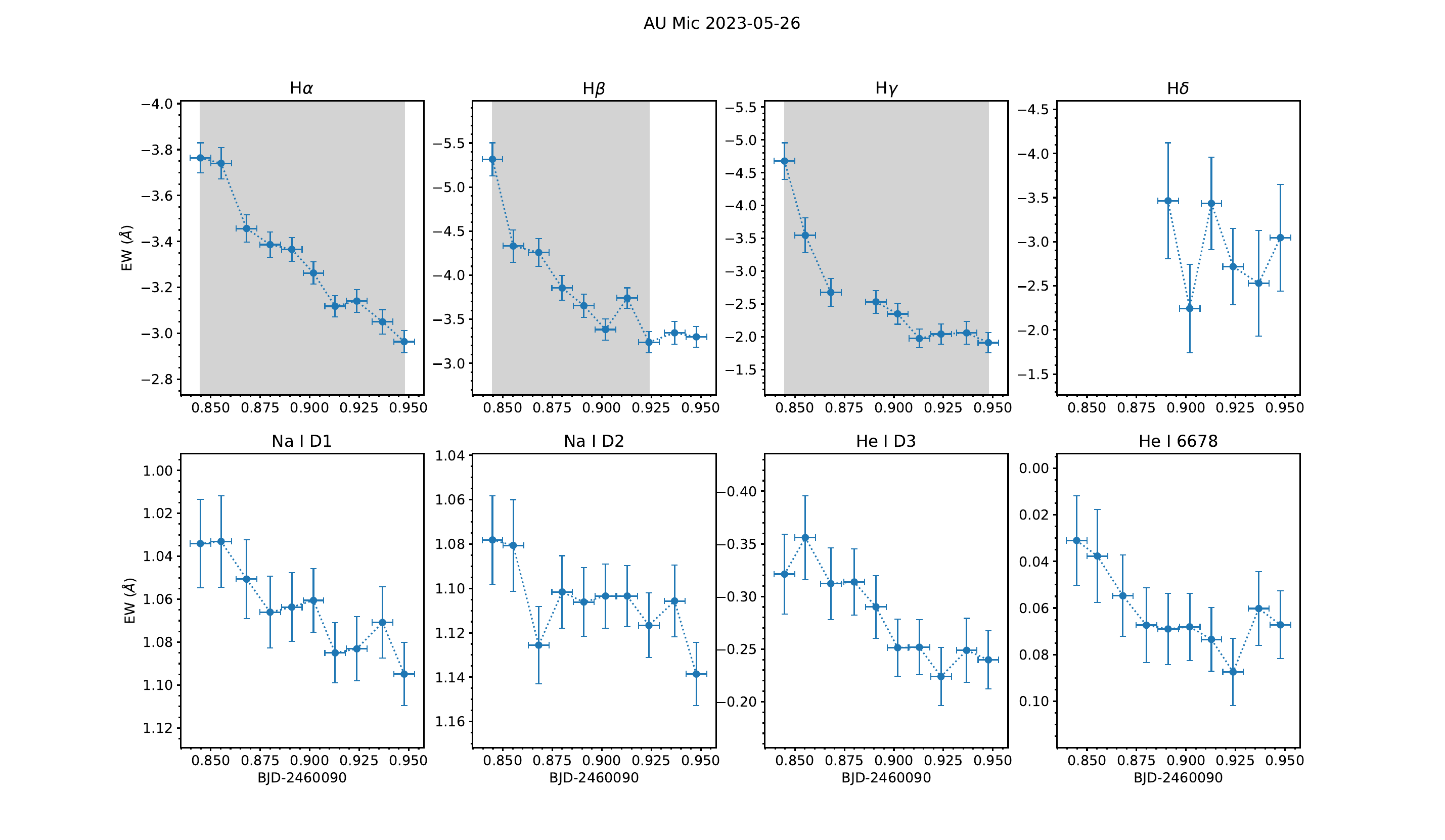}
    \caption{Flare \#10 on 2023-05-26 in all studied spectral lines. Significant detection in H$\alpha$, H$\beta$ and H$\gamma$.}
    \label{fig:2023-05-26}
\end{figure*}

\begin{figure*}
	\includegraphics[width=\columnwidth]{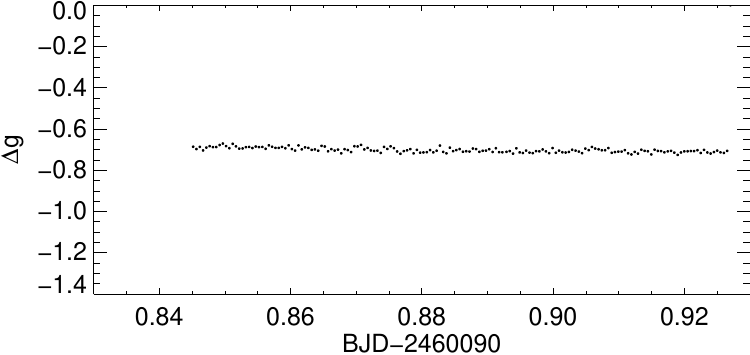}
  \includegraphics[width=0.7\columnwidth]{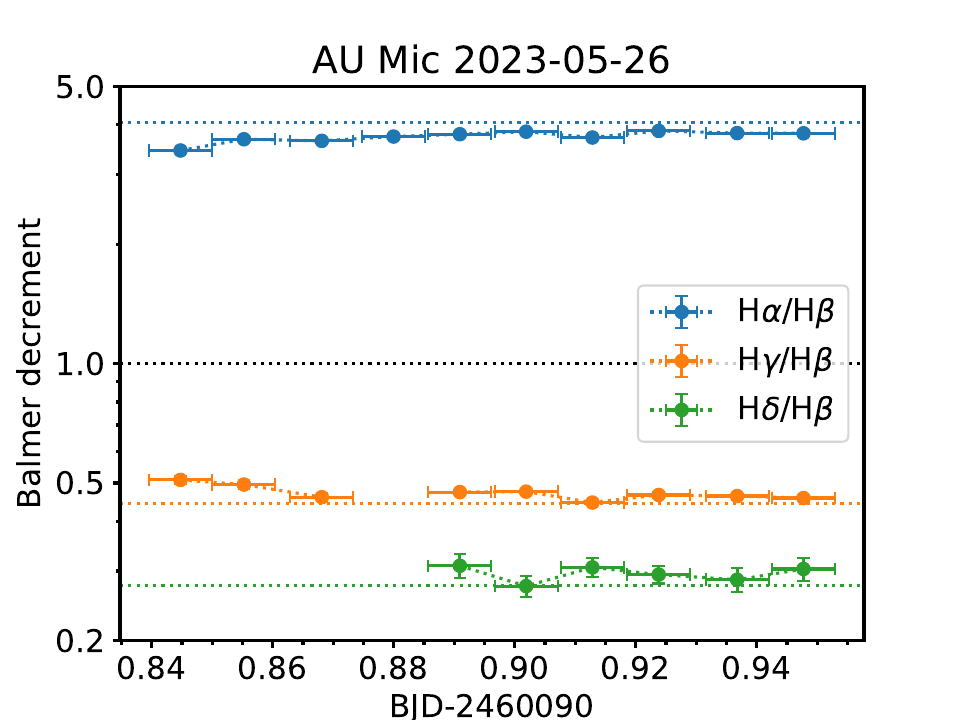}
    \caption{Left panel: Coordinated \textit{g'}-band photometry on 2023-05-26. Right panel: Balmer decrements relative to H$\beta$ along the flare.}
    \label{fig:phbd2023-05-26}
\end{figure*}

\begin{figure*}
	\includegraphics[width=\textwidth]{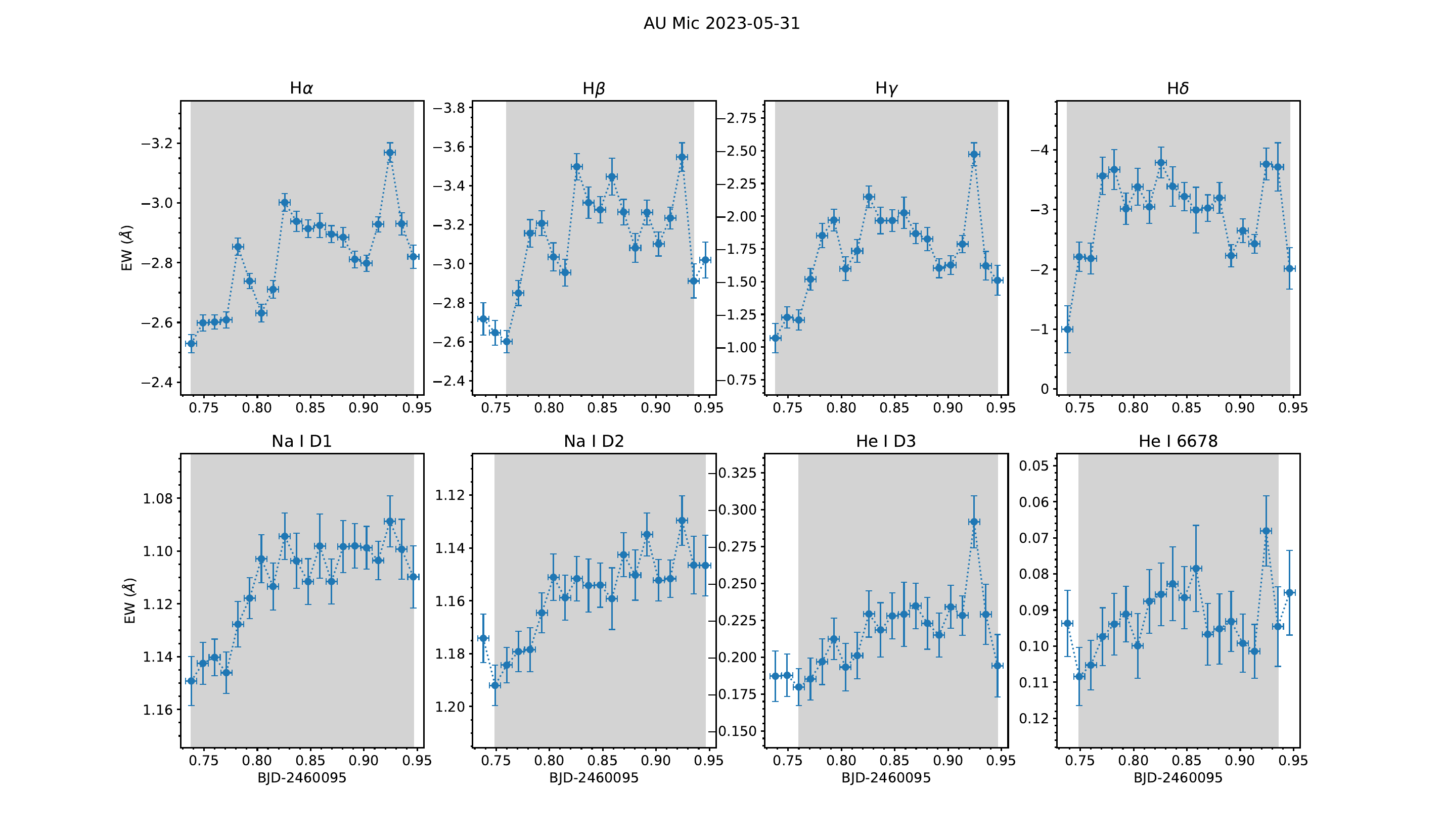}
    \caption{Flare \#11 on 2023-05-31 in all studied spectral lines. Significant detection in all spectral lines.}
    \label{fig:2023-05-31}
\end{figure*}

\begin{figure*}
	\includegraphics[width=\columnwidth]{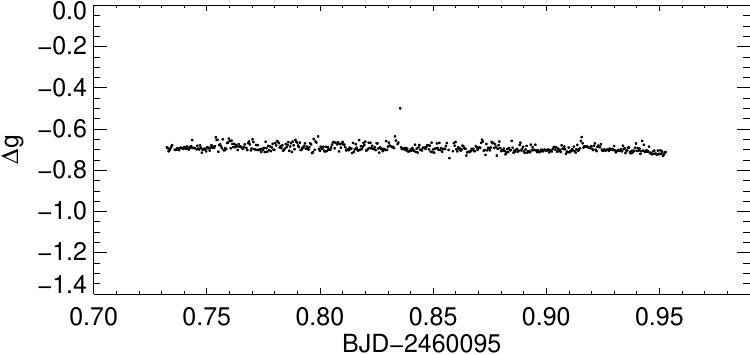}
 \includegraphics[width=0.7\columnwidth]{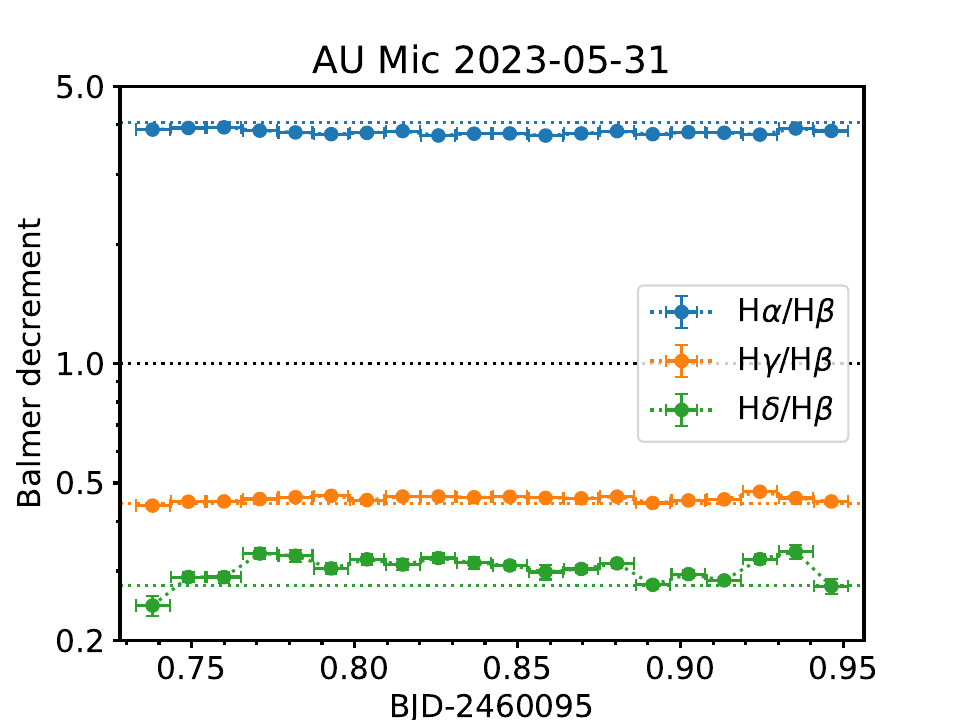}
    \caption{Left panel: Coordinated \textit{g'}-band photometry on 2023-05-31. Right panel: Balmer decrements relative to H$\beta$ along the flare.}
    \label{fig:phbd2023-05-31}
\end{figure*}

\begin{figure*}
	\includegraphics[width=\textwidth]{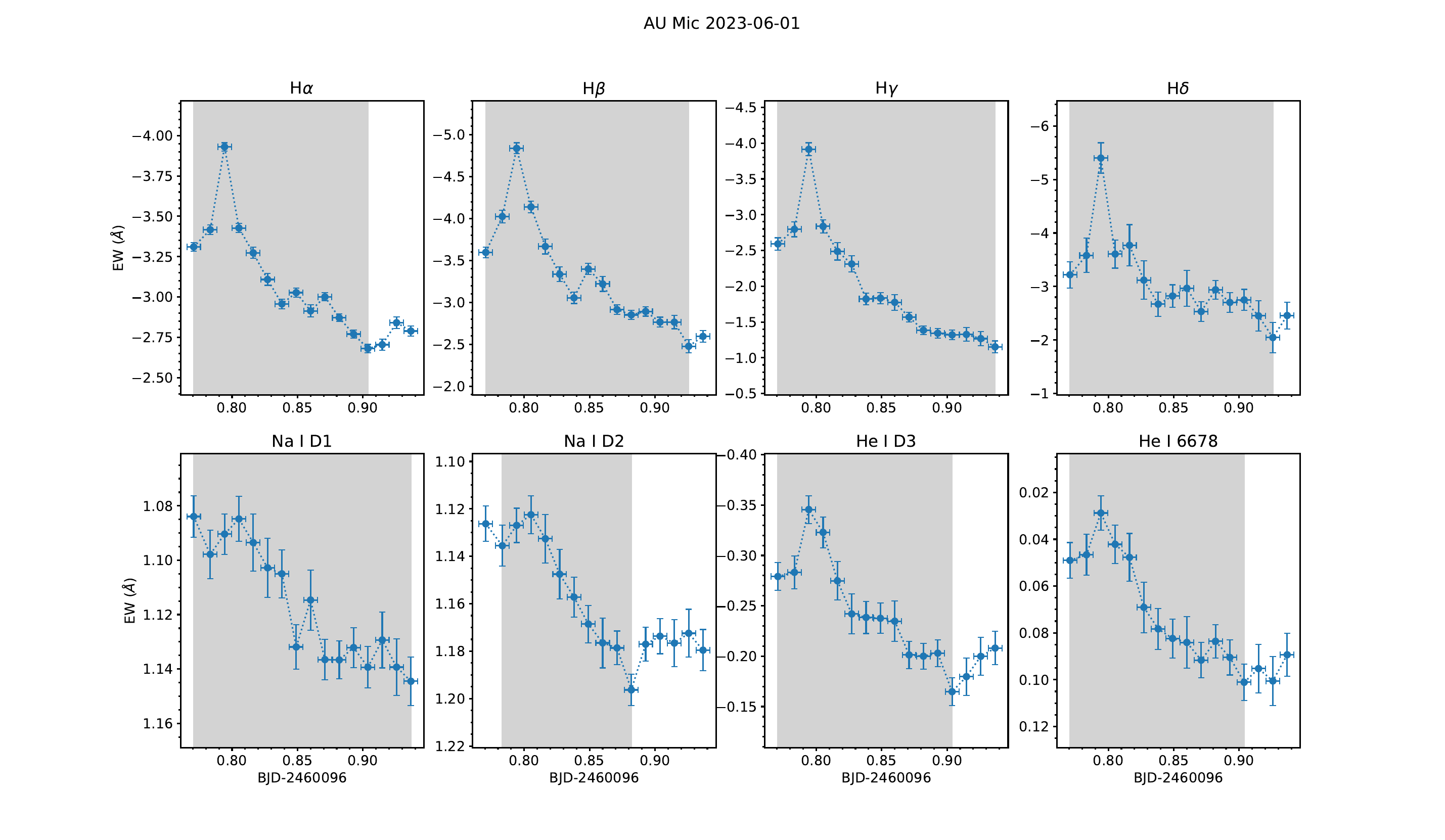}
    \caption{Flare \#12 on 2023-06-01 in all studied spectral lines. Significant detection in all spectral lines.}
    \label{fig:2023-06-01}
\end{figure*}

\begin{figure*}
	\includegraphics[width=\columnwidth]{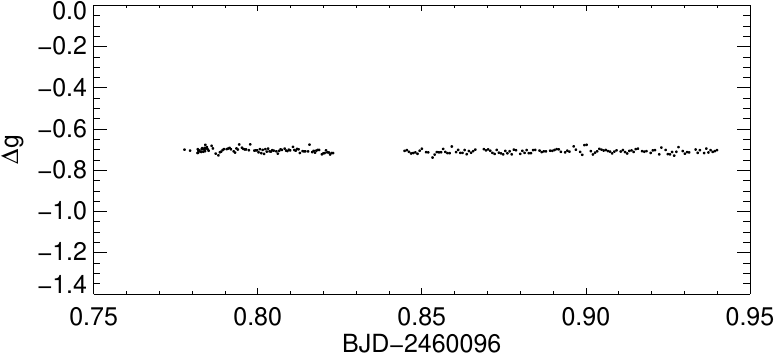}
 \includegraphics[width=0.7\columnwidth]{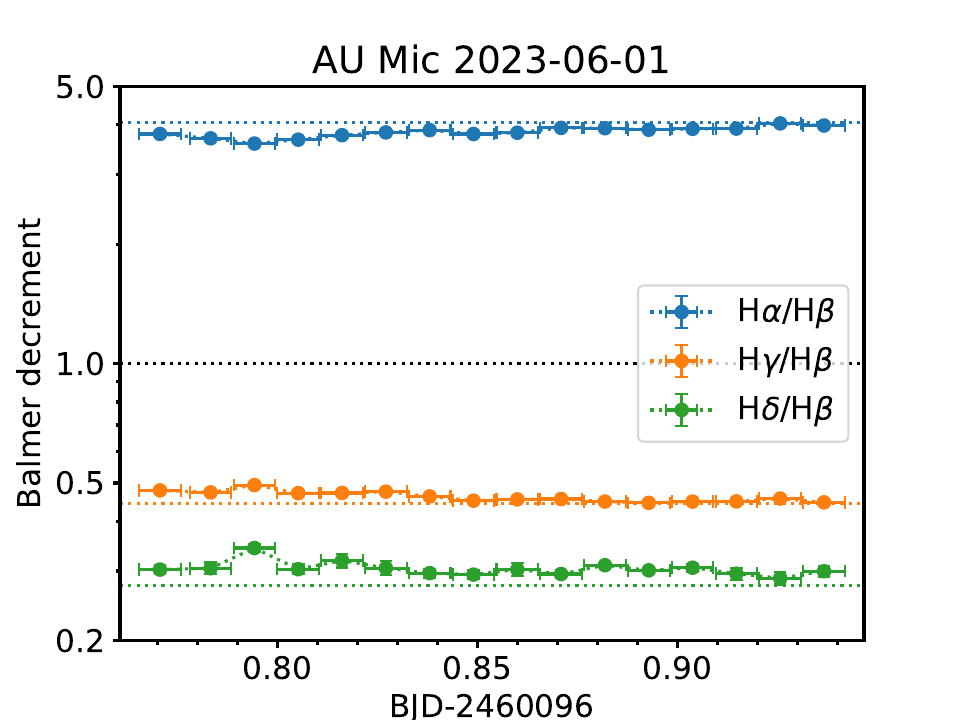}
    \caption{Left panel: Coordinated \textit{g'}-band photometry on 2023-06-01. Right panel: Balmer decrements relative to H$\beta$ along the flare.}
    \label{fig:phbd2023-06-01}
\end{figure*}

\begin{figure*}
	\includegraphics[width=\textwidth]{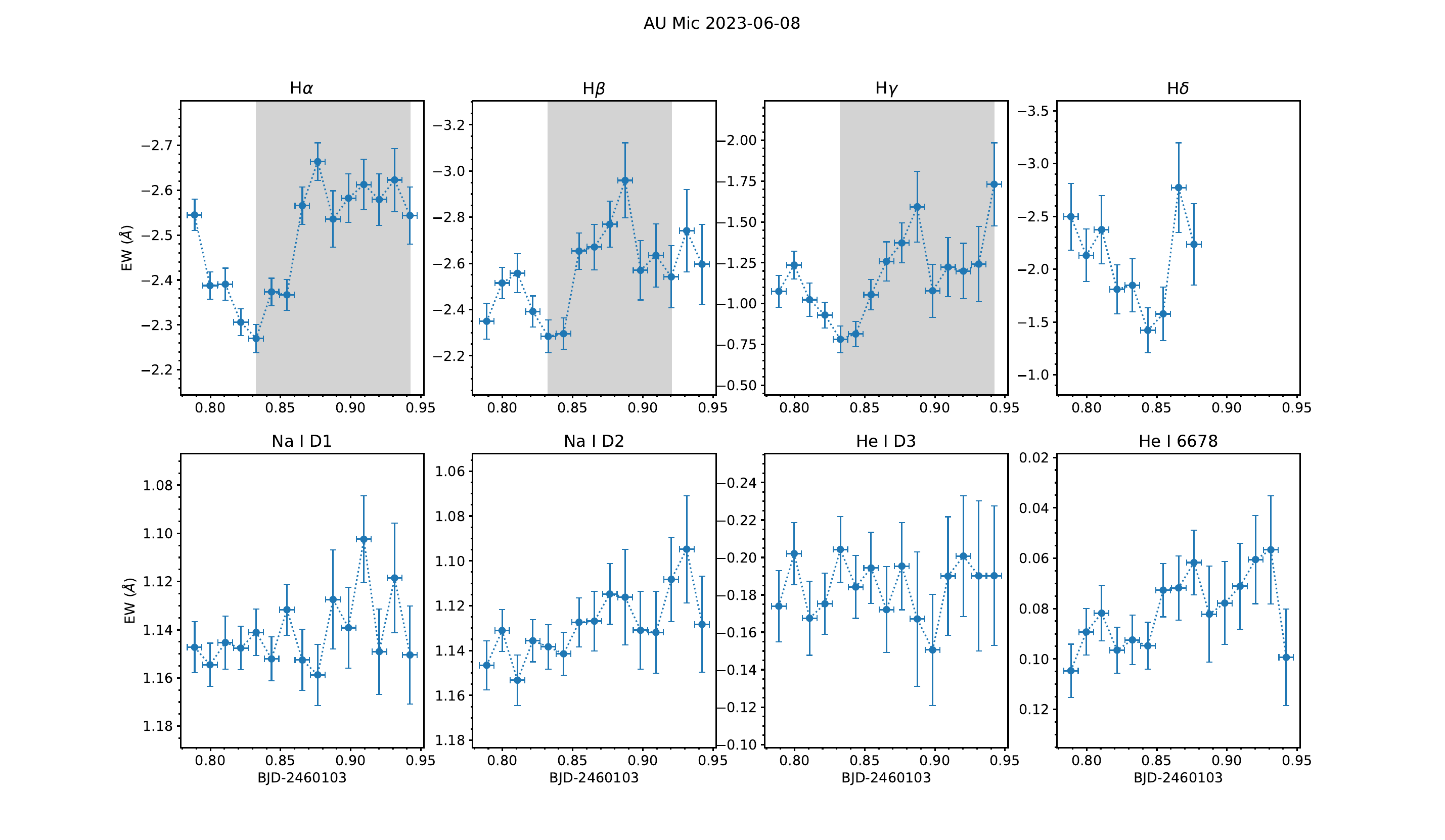}
    \caption{Flare \#14 on 2023-06-08 in all studied spectral lines. Significant detection in H$\alpha$, H$\beta$ and H$\gamma$.}
    \label{fig:2023-06-08}
\end{figure*}

\begin{figure*}
	\includegraphics[width=\columnwidth]{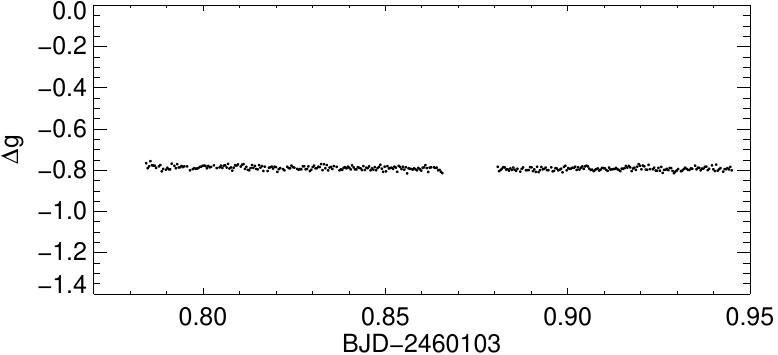}
 \includegraphics[width=0.7\columnwidth]{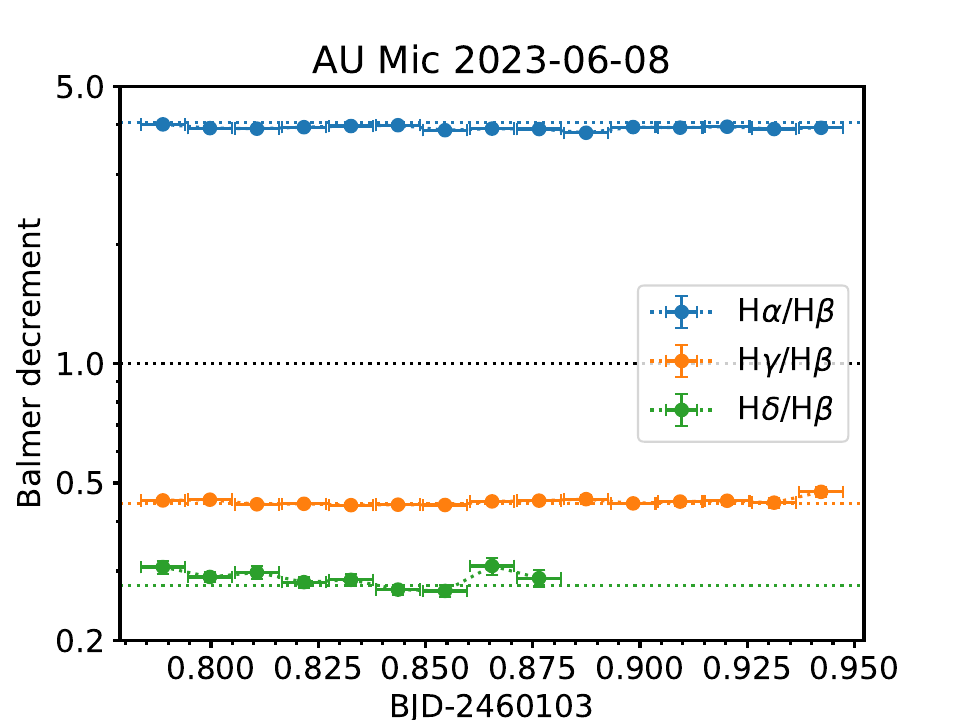}
    \caption{Left panel: Coordinated \textit{g'}-band photometry on 2023-06-08. The first part of the night was not usable. Right panel: Balmer decrements relative to H$\beta$ along the flare.}
    \label{fig:phbd2023-06-08}
\end{figure*}

\begin{figure*}
	\includegraphics[width=\textwidth]{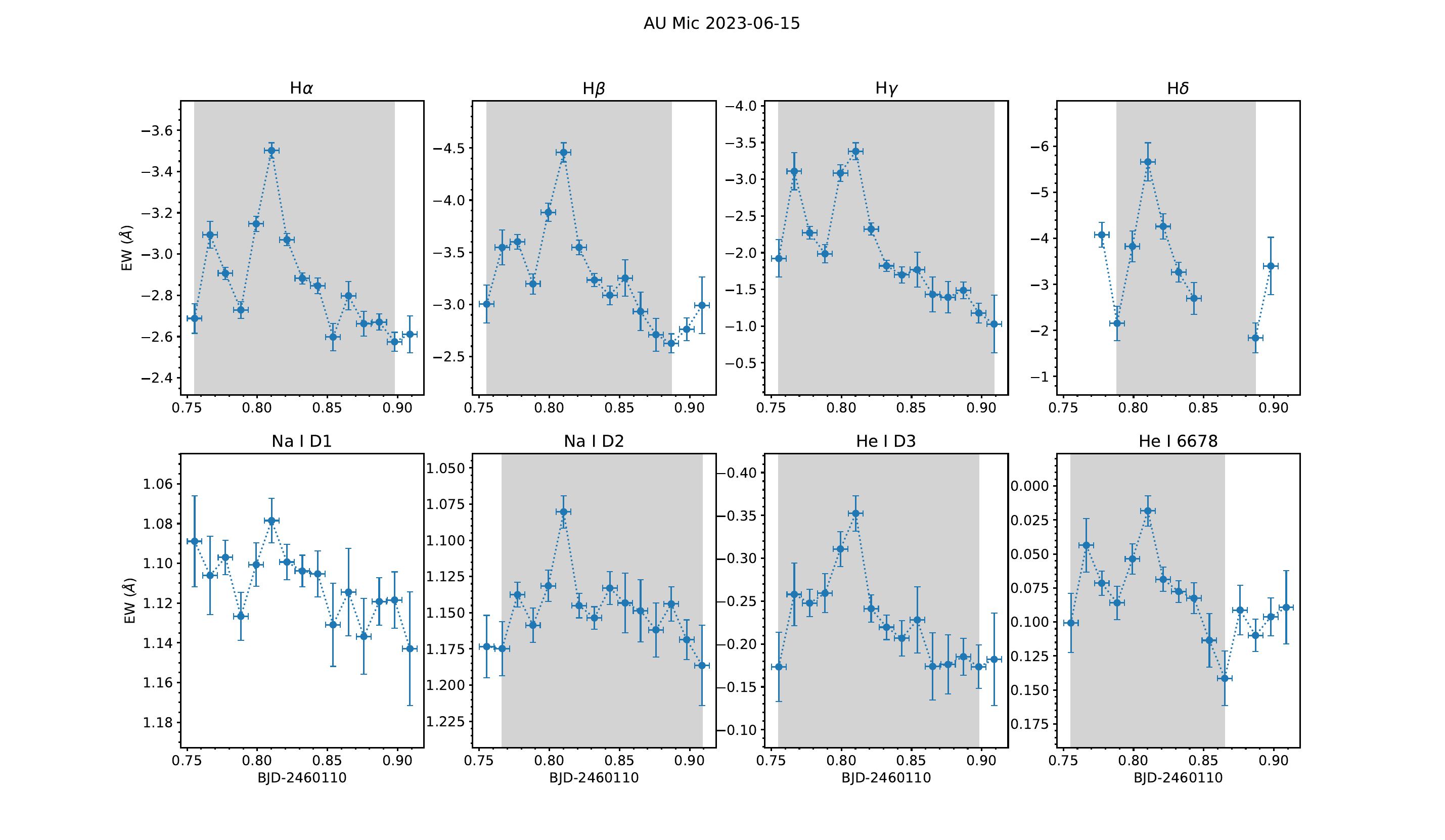}
    \caption{Flare \#15 on 2023-06-15 in all studied spectral lines. Significant detection in H$\alpha$, H$\beta$, H$\gamma$ and H$\delta$, \ion{Na}{i}\,D2, \ion{He}{i}\,D3 and \ion{He}{i}\,6678.}
    \label{fig:2023-06-15}
\end{figure*}

\begin{figure*}
	\includegraphics[width=\columnwidth]{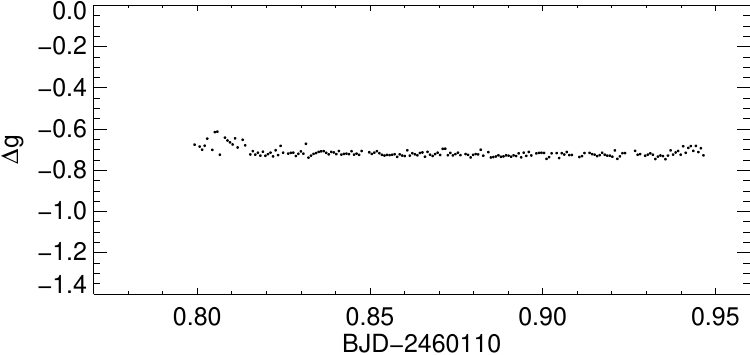}
 \includegraphics[width=0.7\columnwidth]{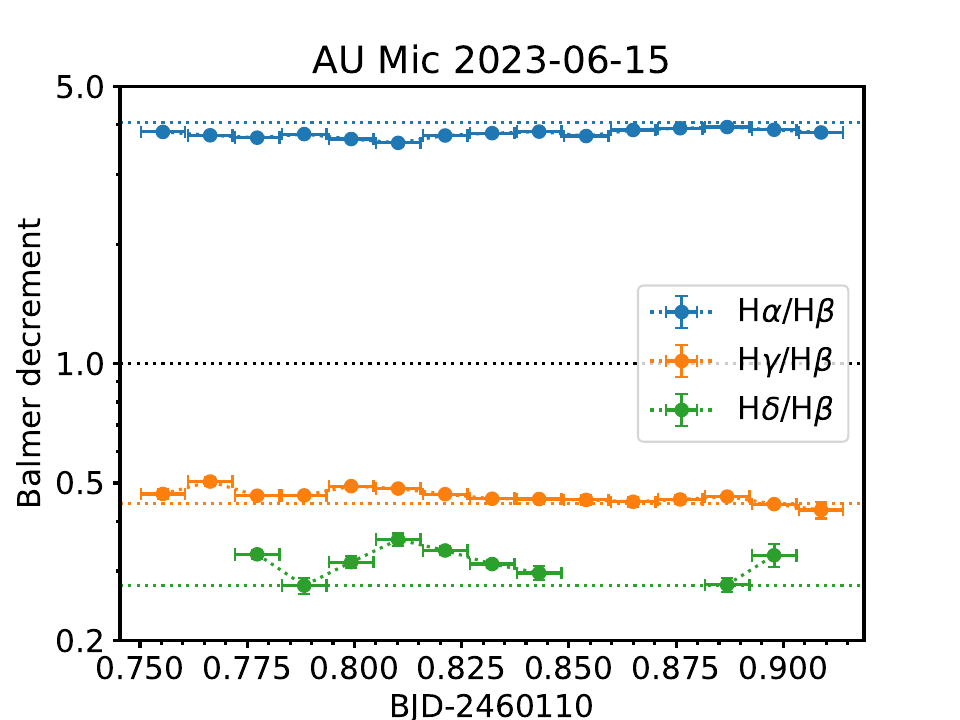}
    \caption{Left panel: Coordinated \textit{g'}-band photometry on 2023-06-15. The first part of the night was not usable. Right panel: Balmer decrements relative to H$\beta$ along the flare.}
    \label{fig:phbd2023-06-15}
\end{figure*}

\begin{figure*}
	\includegraphics[width=\textwidth]{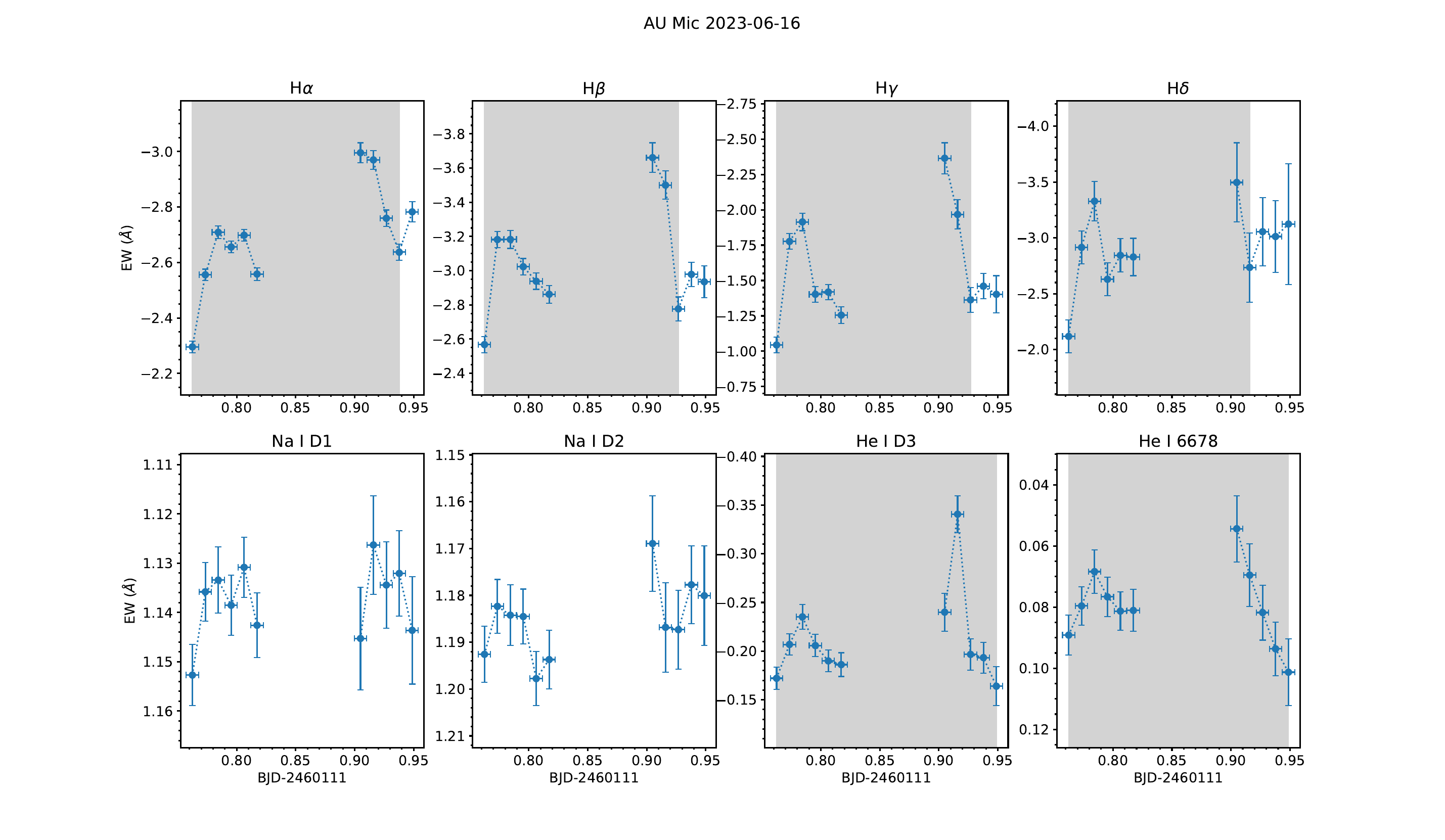}
    \caption{Flare \#16 on 2023-06-16 in all studied spectral lines. Significant detection in H$\alpha$, H$\beta$, H$\gamma$, H$\delta$, \ion{He}{i}\,D3 and \ion{He}{i}\,6678.}
    \label{fig:2023-06-16}
\end{figure*}

\begin{figure*}
	\includegraphics[width=\columnwidth]{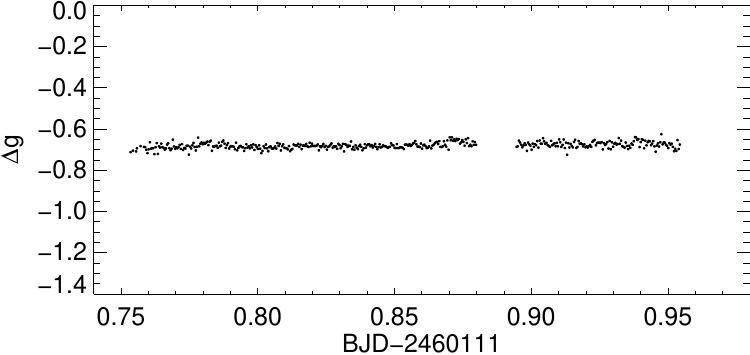}
 \includegraphics[width=0.7\columnwidth]{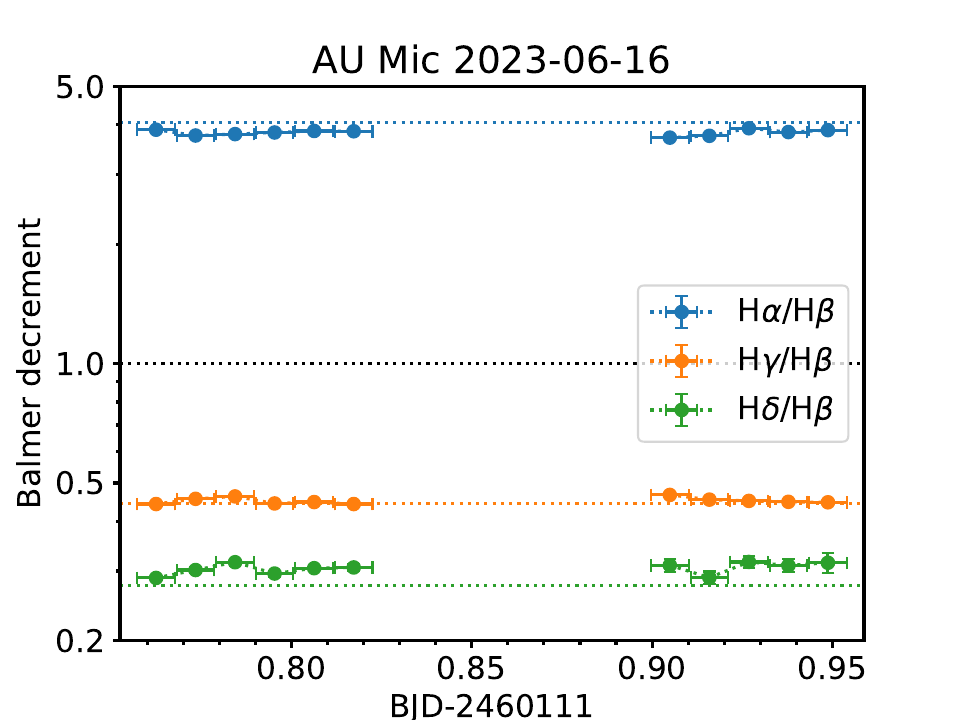}
    \caption{Left panel: Coordinated \textit{g'}-band photometry on 2023-06-16. Right panel: Balmer decrements relative to H$\beta$ along the flare.}
    \label{fig:phbd2023-06-16}
\end{figure*}

\begin{figure*}
	\includegraphics[width=\textwidth]{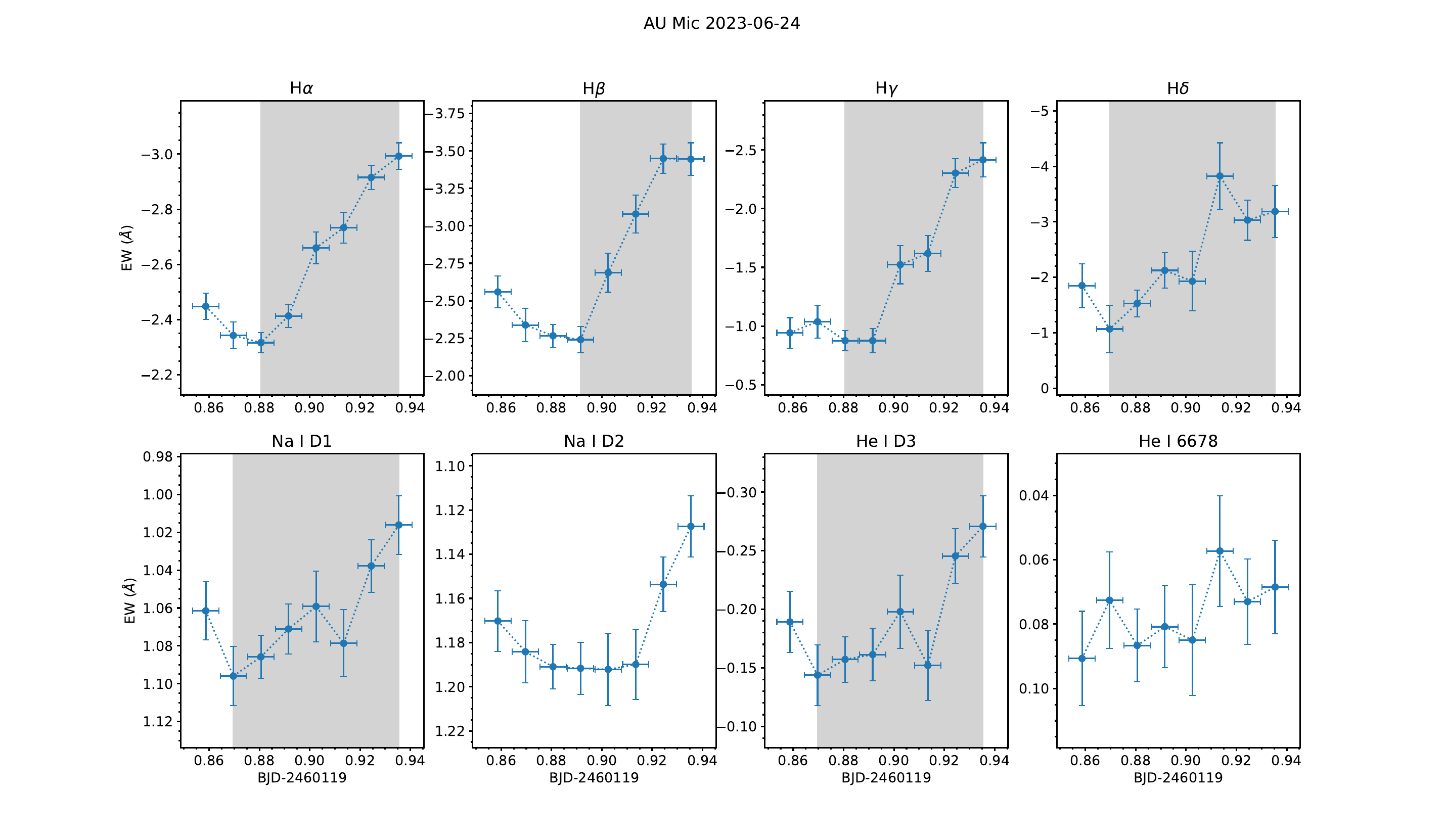}
    \caption{Flare \#17 on 2023-06-24 in all studied spectral lines. Significant detection in H$\alpha$, H$\beta$, H$\gamma$, H$\delta$, \ion{Na}{i}\,D1 and \ion{He}{i}\,D3.}
    \label{fig:2023-06-24}
\end{figure*}

\begin{figure*}
	\includegraphics[width=\columnwidth]{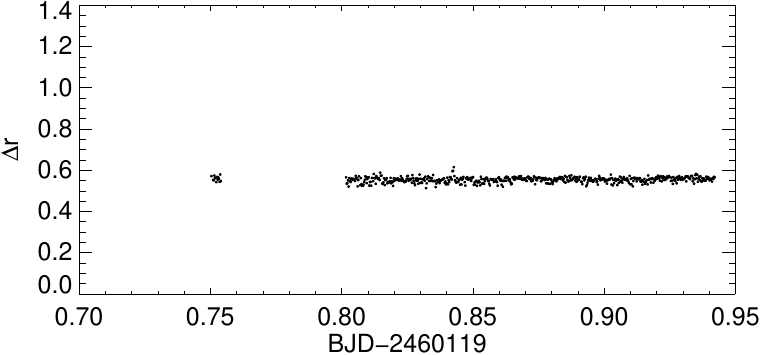}
 \includegraphics[width=0.7\columnwidth]{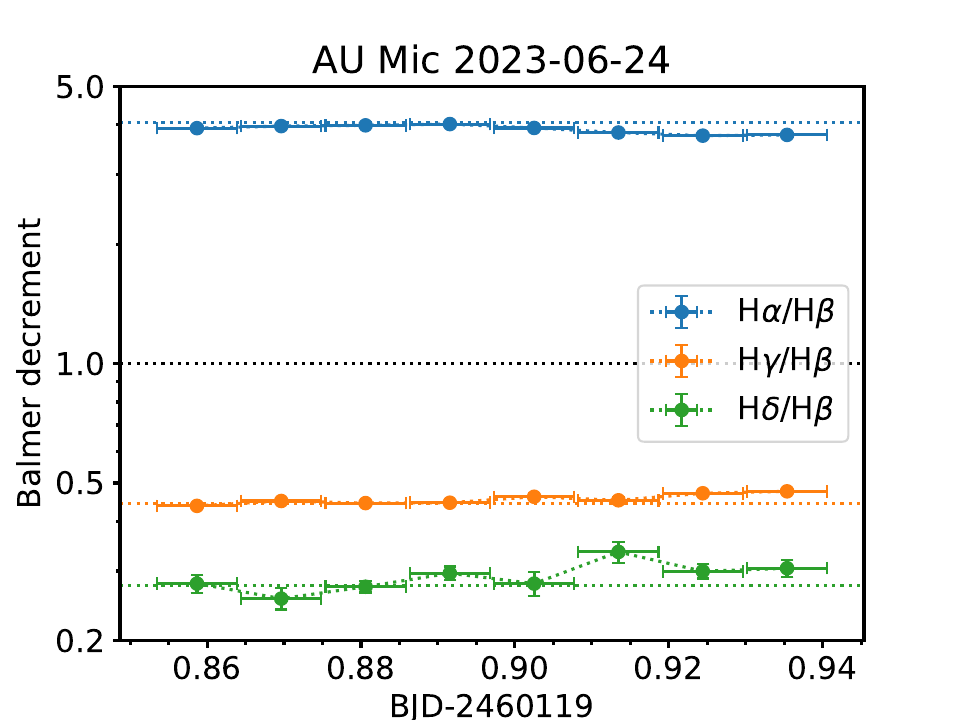}
    \caption{Left panel: Coordinated \textit{r'}-band photometry on 2023-06-24. Right panel: Balmer decrements relative to H$\beta$ along the flare.}
    \label{fig:phbd2023-06-24}
\end{figure*}

\begin{figure*}
	\includegraphics[width=\textwidth]{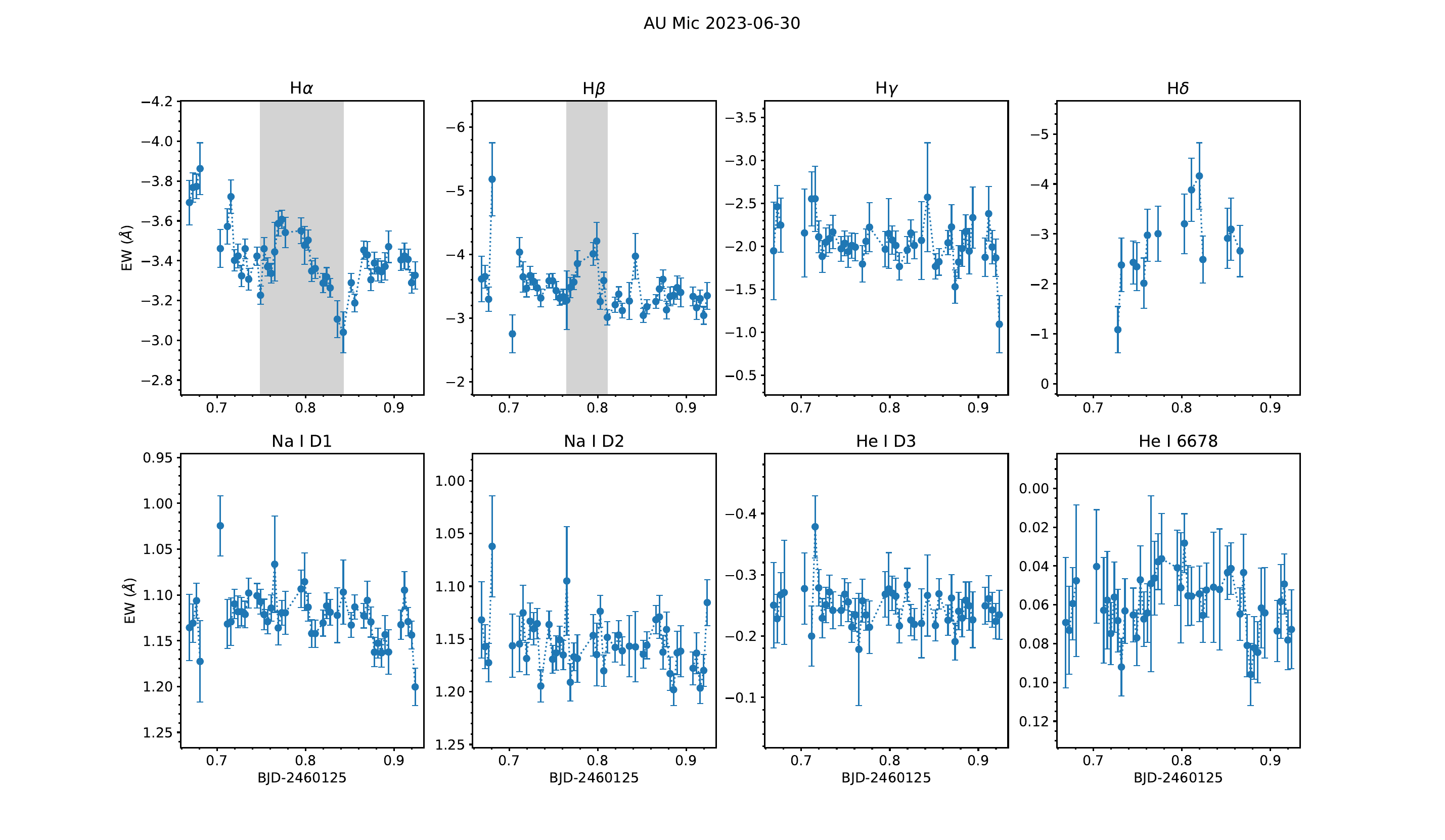}
    \caption{Flare \#18 on 2023-06-30 in all studied spectral lines. Significant detection in H$\alpha$ and H$\beta$.}
    \label{fig:2023-06-30}
\end{figure*}

\begin{figure*}
 \includegraphics[width=\columnwidth]{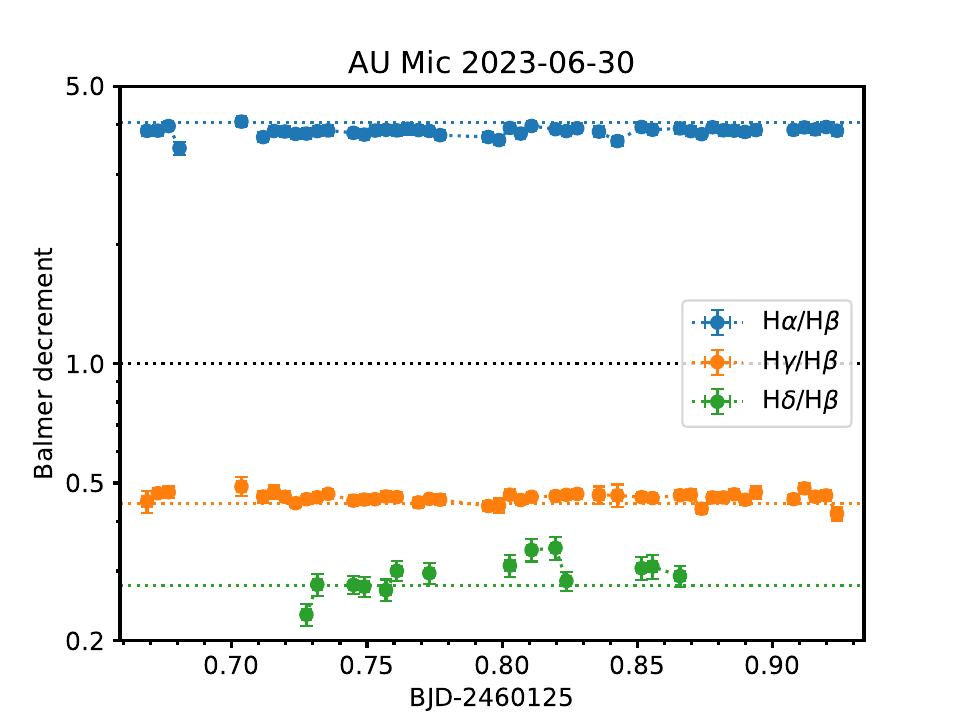}
    \caption{Balmer decrements relative to H$\beta$ along the flare.}
    \label{fig:bd2023-06-30}
\end{figure*}

\begin{figure*}
	\includegraphics[width=\textwidth]{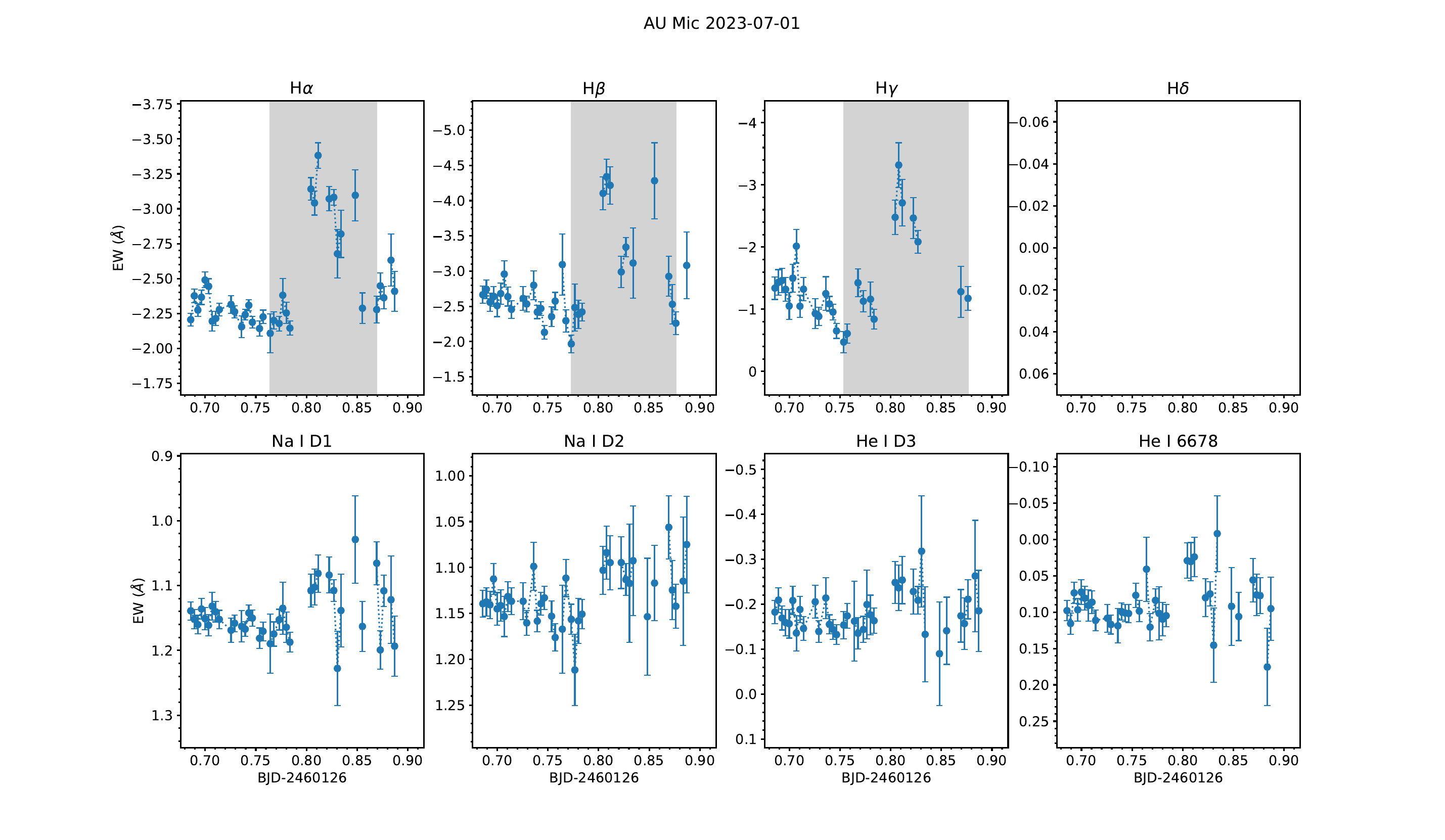}
    \caption{Flare \#19 on 2023-07-01 in all studied spectral lines. Significant detection in H$\alpha$, H$\beta$ and H$\gamma$.}
    \label{fig:2023-07-01}
\end{figure*}

\begin{figure*}
	\includegraphics[width=\columnwidth]{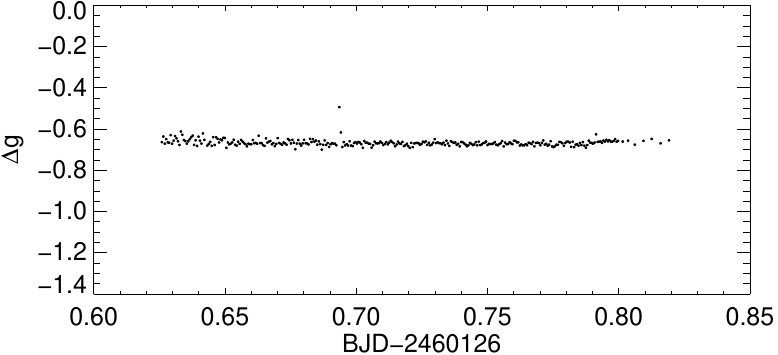}
 \includegraphics[width=0.7\columnwidth]{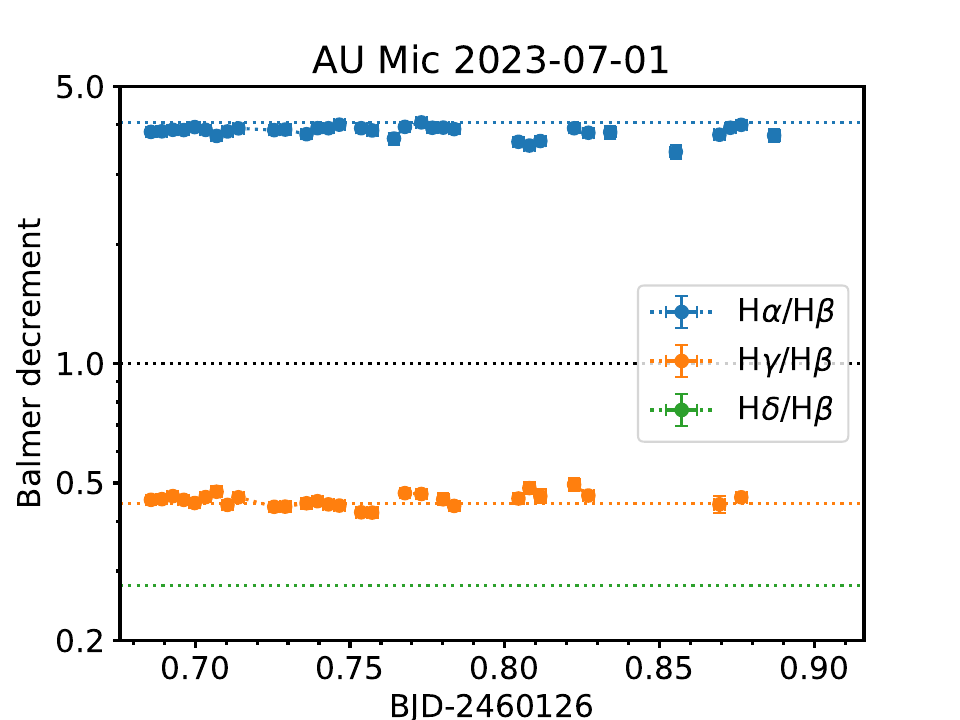}
    \caption{Left panel: Coordinated \textit{g'}-band photometry on 2023-07-01. The second half of the night was not usable. Right panel: Balmer decrements relative to H$\beta$ along the flare.}
    \label{fig:phbd2023-07-01}
\end{figure*}

\begin{figure*}
	\includegraphics[width=\textwidth]{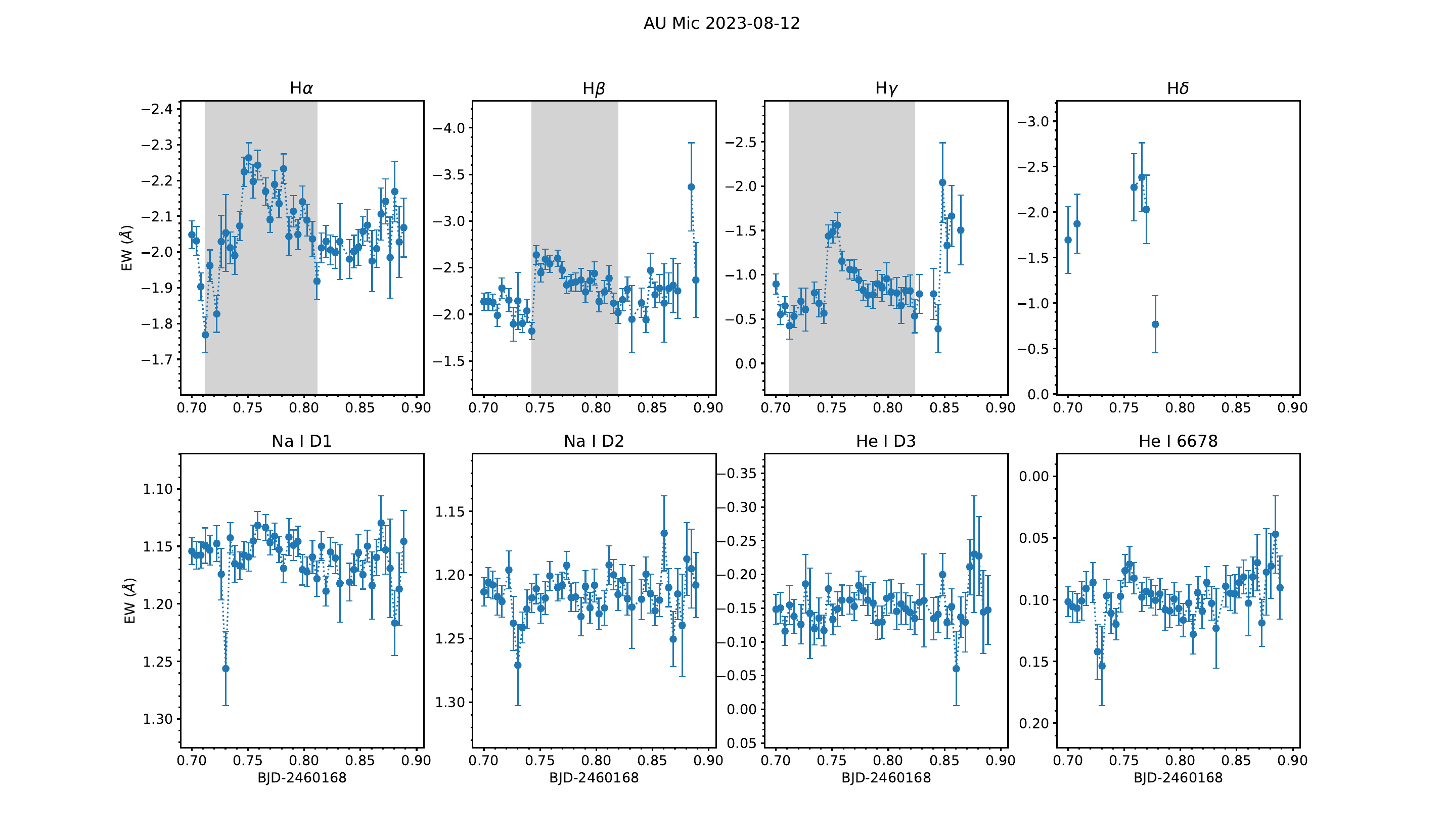}
    \caption{Flare \#20 on 2023-08-12 in all studied spectral lines. Significant detection in H$\alpha$, H$\beta$ and H$\gamma$.}
    \label{fig:2023-08-12}
\end{figure*}

\begin{figure*}
 \includegraphics[width=\columnwidth]{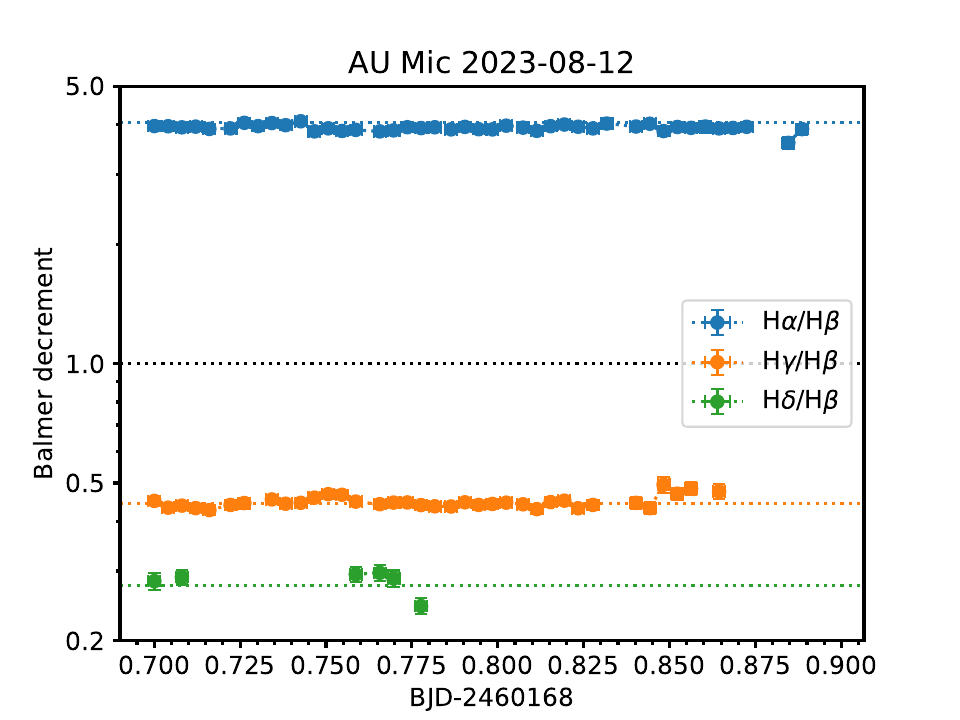}
    \caption{Balmer decrements relative to H$\beta$ along the flare.}
    \label{fig:bd2023-08-12}
\end{figure*}

\begin{figure*}
	\includegraphics[width=\textwidth]{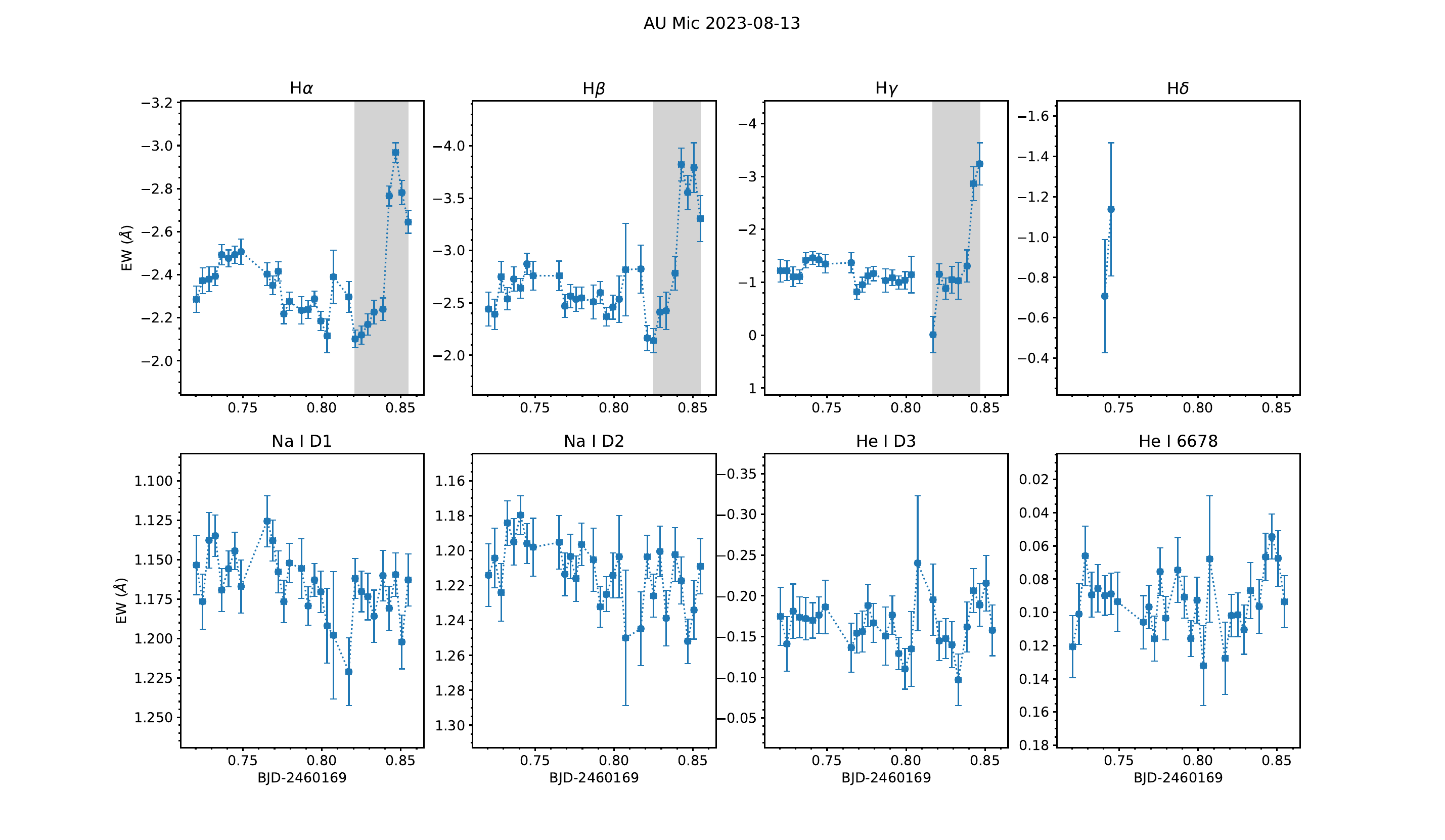}
    \caption{Flare \#21 on 2023-08-13 in all studied spectral lines. Significant detection in H$\alpha$, H$\beta$ and H$\gamma$.}
    \label{fig:2023-08-13}
\end{figure*}

\begin{figure*}
  \includegraphics[width=\columnwidth]{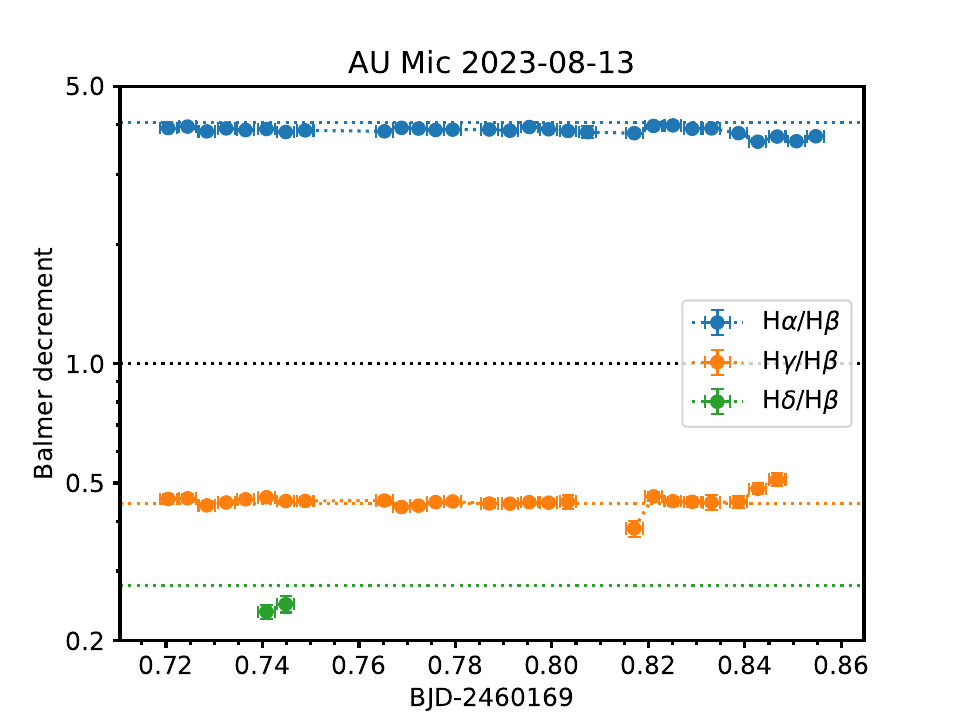}
    \caption{Balmer decrements relative to H$\beta$ along the flare.}
    \label{fig:bd2023-08-13}
\end{figure*}

\begin{figure*}
	\includegraphics[width=\textwidth]{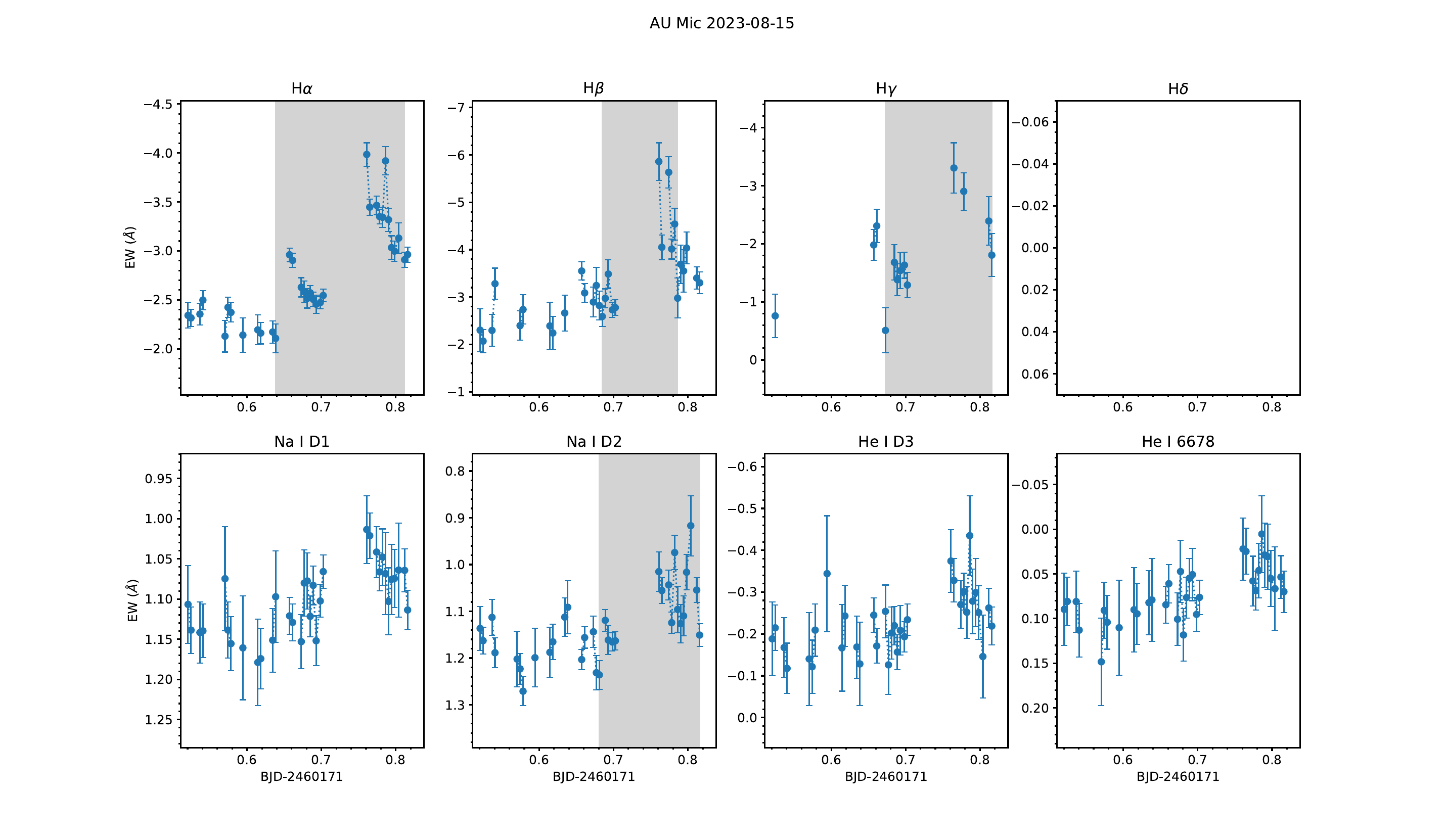}
    \caption{Flare \#22 on 2023-08-15 in all studied spectral lines. Significant detection in H$\alpha$, H$\beta$, H$\gamma$ and \ion{Na}{i}\,D2.}
    \label{fig:2023-08-15}
\end{figure*}

\begin{figure*}
	\includegraphics[width=\columnwidth]{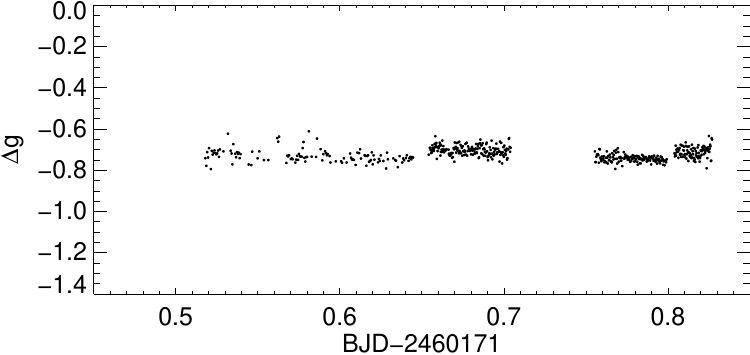}
 \includegraphics[width=0.7\columnwidth]{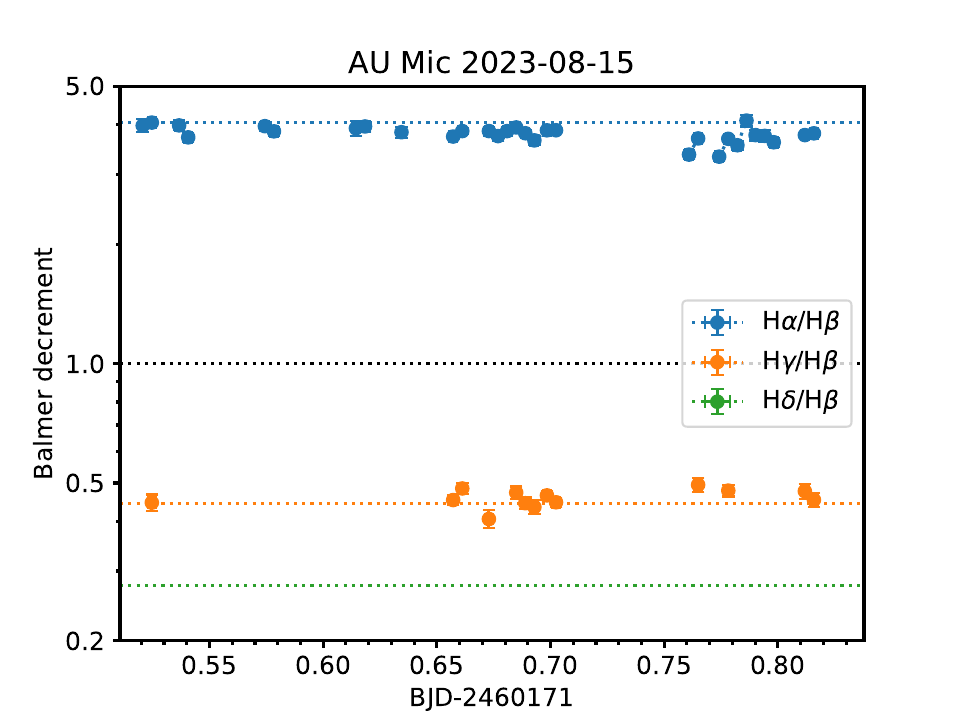}
    \caption{Left panel: Coordinated \textit{g'}-band photometry on 2023-08-15. Right panel: Balmer decrements relative to H$\beta$ along the flare.}
    \label{fig:phbd2023-08-15}
\end{figure*}

\begin{figure*}
	\includegraphics[width=\textwidth]{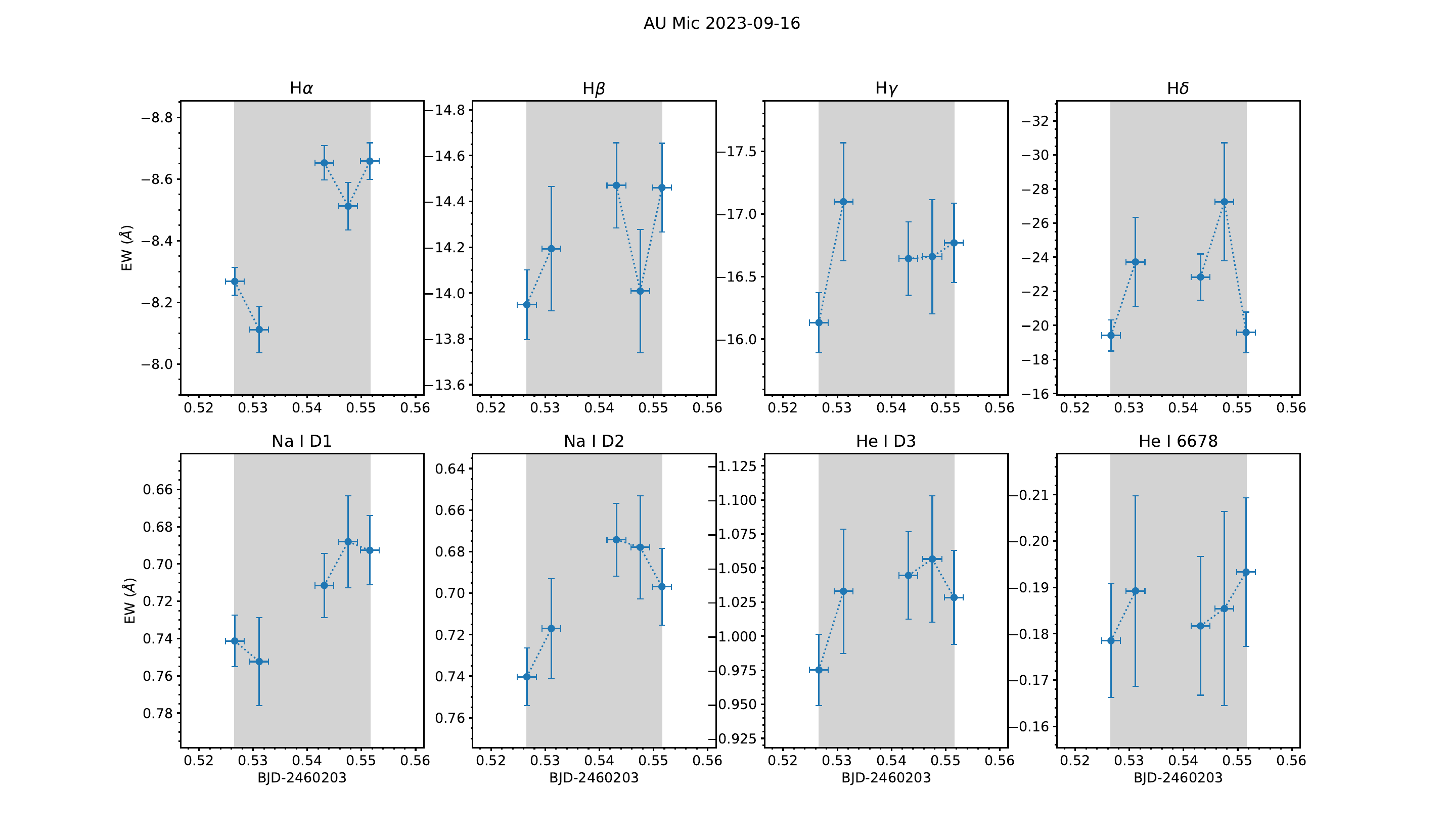}
    \caption{Flare \#23 on 2023-09-16 in all studied spectral lines. Significant detection in all spectral lines (in comparison to 2023-09-17, cf. Fig.\,\ref{fig:superflare}).}
    \label{fig:2023-09-16}
\end{figure*}

\begin{figure*}
 \includegraphics[width=\columnwidth]{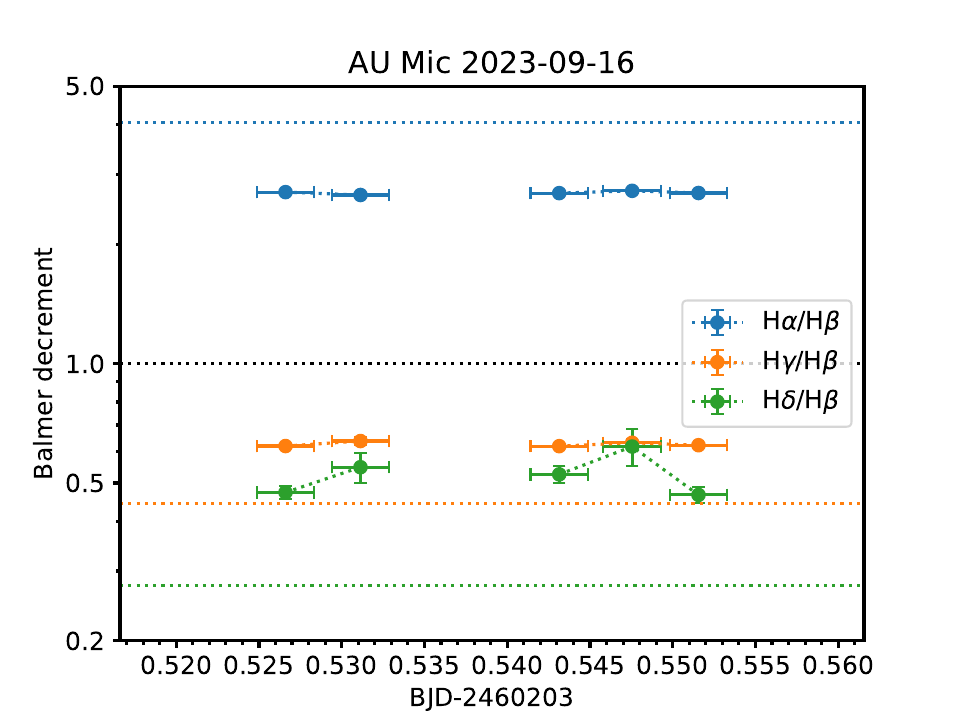}
    \caption{Balmer decrements relative to H$\beta$ along the flare.}
    \label{fig:bd2023-09-16}
\end{figure*}

\begin{figure*}
	\includegraphics[width=\textwidth]{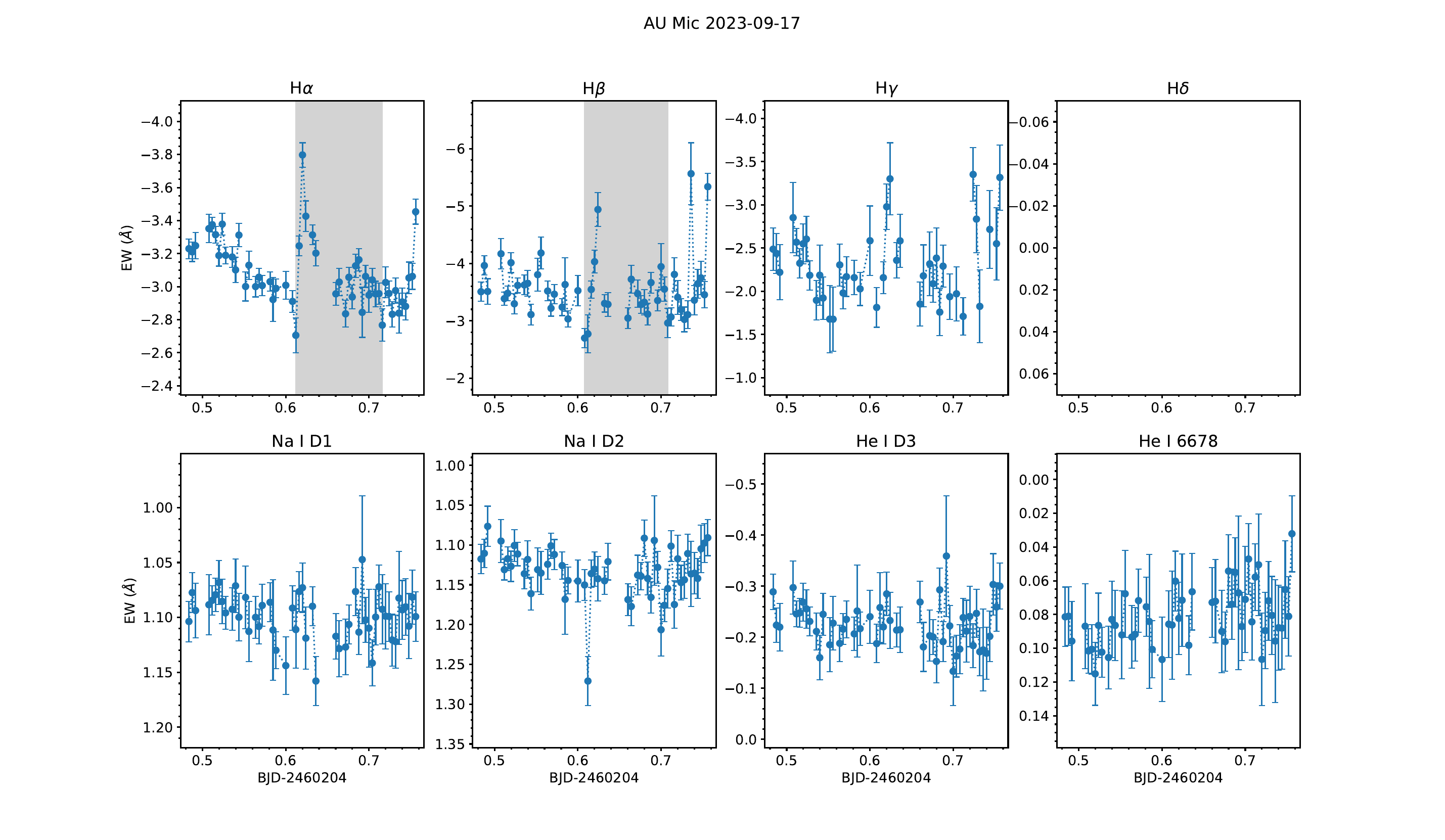}
    \caption{Flare \#24 on 2023-09-17 in all studied spectral lines. Significant detection in H$\alpha$ and H$\beta$.}
    \label{fig:2023-09-17}
\end{figure*}

\begin{figure*}
	\includegraphics[width=\columnwidth]{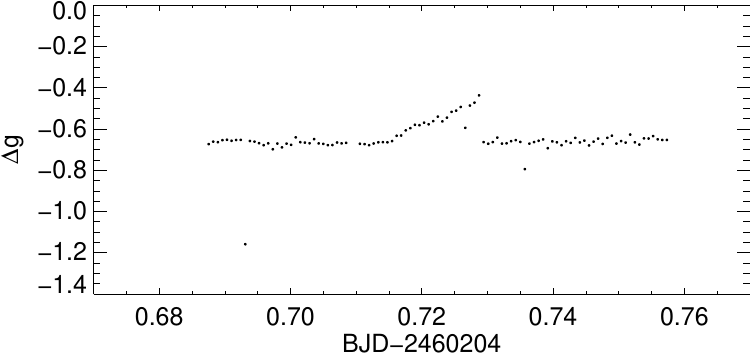}
    \includegraphics[width=0.7\columnwidth]{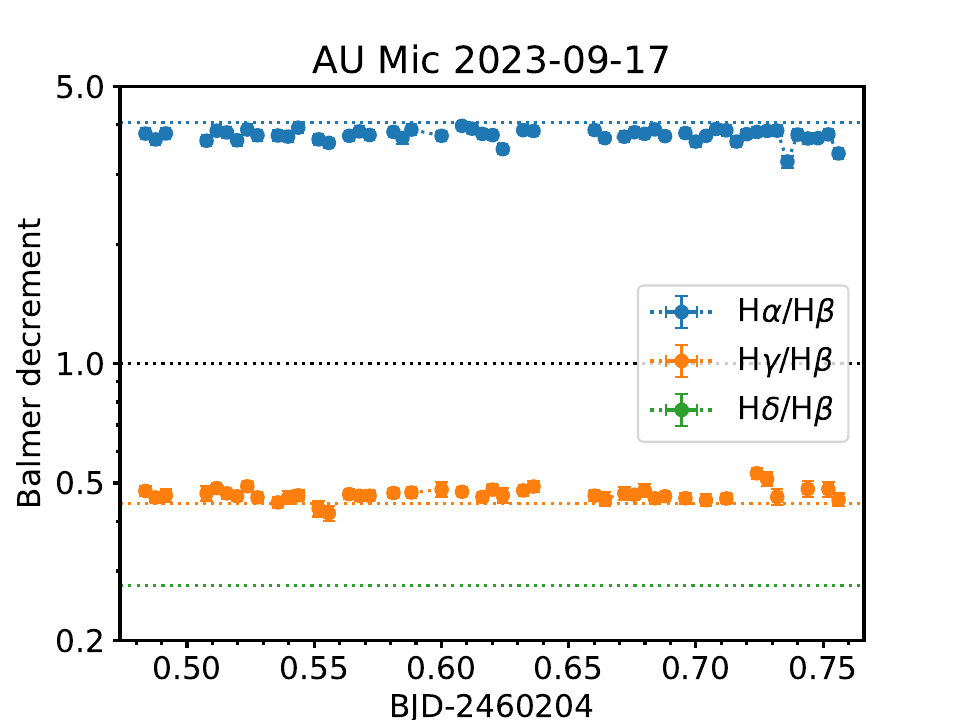}
    \caption{Left panel: Coordinated \textit{g'}-band photometry on 2023-09-17. The first half of the night was not usable. The ramp-like structure was caused by dome vignetting. Right panel: Balmer decrements relative to H$\beta$ along the flare.}
    \label{fig:phbd2023-09-17}
\end{figure*}

\begin{figure*}
	\includegraphics[width=\textwidth]{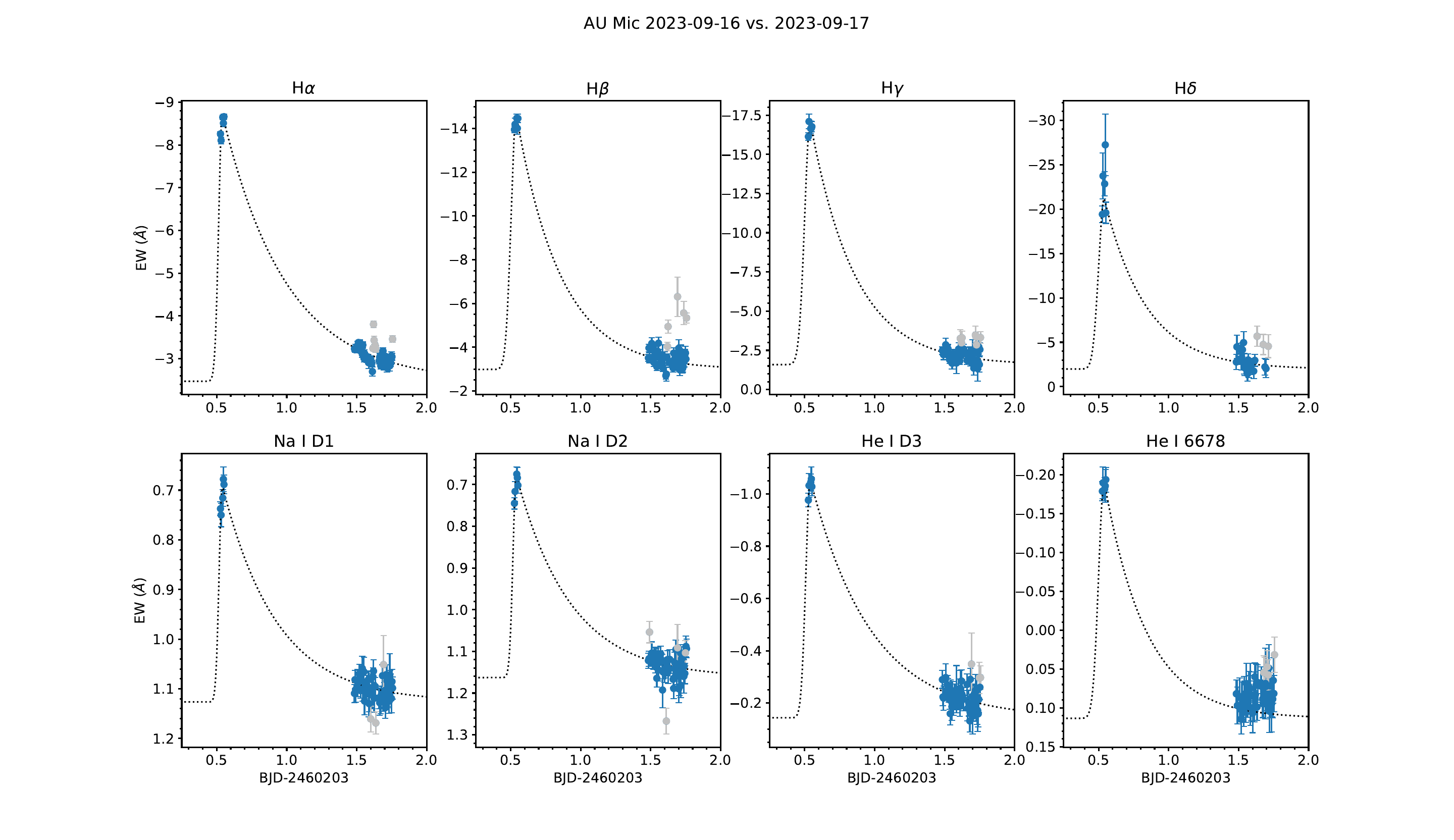}
    \caption{Nights of 2023-09-16 and 2023-09-17 plotted subsequently. The black dotted line displays the fits to the data using a flare model described in the text under the assumption that the data on 2023-09-17 still correspond to the decaying tail of the flare. The grey data points were omitted in the fits.}
    \label{fig:superflare}
\end{figure*}

\clearpage

\section{Table of flare parameters}
\label{app:param}

\begin{table*}
	\centering
	\caption{Flare parameters in all spectral lines. We only consider events with an H$\alpha$ amplitude (i.e. the maximum of the normalized net flux) larger than 5$\sigma$ and compute flare parameters for a given spectral line only if its amplitude exceeds 3$\sigma$. }
	\label{tab:flares}
	\begin{tabular}{cccccccccl}
		\hline
		flare & UT date & line & amplitude & rise & decay & duration & peak luminosity & energy & notes \\
        \# & & & (\%) & (min) & (min) & (min) & (10$^{27}$\,erg\,s$^{-1}$) & (10$^{31}$\,erg) & \\
		\hline
        1  & 2022-11-05 & H$\alpha$ & 1.65$\pm$0.30 & 11 & 174 & 185 & 1.74$\pm$0.35 & 0.86$\pm$0.09 & onset missed \\
        1  & 2022-11-05 & H$\beta$  & 6.47$\pm$1.45 & 11 & 185 & 195 & 1.70$\pm$0.41 & 0.96$\pm$0.12 & peak BD: 1.02/1/--/-- \\
        \hline
        2  & 2022-11-06 & H$\alpha$ & 3.96$\pm$0.53 & 22 & 152 & 174 & 4.34$\pm$0.71 & 2.52$\pm$0.16 & onset missed \\
        2  & 2022-11-06 & H$\beta$  & 16.44$\pm$3.52 & 33 & 119 & 152 & 4.44$\pm$1.05 & 1.93$\pm$0.22 & peak BD: 1.21/1/0.36/-- \\
        2  & 2022-11-06 & H$\gamma$ & 12.60$\pm$3.24 & 11 & 76  & 87  & 1.63$\pm$0.45 & 0.46$\pm$0.07 &  \\
        \hline
        3  & 2022-11-14 & H$\alpha$ & 2.47$\pm$0.38 & 22 & 32 & 54 & 2.70$\pm$0.49 & 0.38$\pm$0.06 & peak BD: 0.79/1/0.67/-- \\
        3  & 2022-11-14 & H$\beta$  & 12.30$\pm$ 2.02 & 32 & 43 & 76 & 3.41$\pm$0.65 & 0.86$\pm$0.09 & \\
        3  & 2022-11-14 & H$\gamma$ & 18.73$\pm$4.40 & 22 & 22 & 43 & 2.30$\pm$0.58 & 0.45$\pm$0.07 & \\
        \hline
        4  & 2022-11-14 & H$\alpha$ & 6.30$\pm$0.46 & 58 & 22 & 80 & 6.91$\pm$0.83 & 1.48$\pm$0.13 & decay missed \\
        4  & 2022-11-14 & H$\beta$  & 23.20$\pm$3.49 & 48 & 11 & 58 & 6.61$\pm$1.18 & 1.17$\pm$0.17 & peak BD: 1.05/1/--/-- \\
        4  & 2022-11-14 & \ion{Na}{i}\,D2 & 12.61$\pm$3.36 & 58 & 22 & 80 & 0.41$\pm$0.12 & 0.09$\pm$0.02 &  \\
        4  & 2022-11-14 & \ion{He}{i}\,D3 & 11.88$\pm$2.62 & 69 & 22 & 91 & 1.02$\pm$0.24 & 0.23$\pm$0.05 &  \\
        4  & 2022-11-14 & \ion{He}{i}\,6678 & 5.62$\pm$1.35 & 58 & 11 & 69 & 0.57$\pm$0.15 & 0.11$\pm$0.03 &  \\
        \hline
        5  & 2022-11-17 & H$\alpha$ & 2.99$\pm$0.35 & 22 & 119 & 141 & 3.27$\pm$0.49 & 1.42$\pm$0.09 & onset missed \\
        5  & 2022-11-17 & H$\beta$  & 10.32$\pm$1.65 & 11 & 119 & 130 & 2.83$\pm$0.53 & 1.06$\pm$0.10 & peak BD: 1.16/1/0.38/-- \\
        5  & 2022-11-17 & H$\gamma$ & 11.55$\pm$3.70 & 11 & 119 & 130 & 1.46$\pm$0.49 & 0.50$\pm$0.10 &  \\
        \hline
        6  & 2023-04-29 & H$\alpha$ & 2.58$\pm$0.36 & 111 & 20 & 131 & 2.77$\pm$0.47 & 1.09$\pm$0.09 & multi-peaked flare, decay missed \\
        6  & 2023-04-29 & H$\beta$  & 4.940$\pm$1.26 & 60 & 90 & 151 & 1.35$\pm$0.37 & 0.63$\pm$0.08 & peak BD: 2.69/1/0.57/0.69 \\
        6  & 2023-04-29 & H$\gamma$ & 10.92$\pm$2.28 & 20 & 90 & 111 & 1.34$\pm$0.31 & 0.39$\pm$0.05 &  \\
        6  & 2023-04-29 & H$\delta$ & 33.81$\pm$8.28 & 30 & 111 & 141 & 2.39$\pm$0.63 & 0.96$\pm$0.12 &  \\
        \hline
        7  & 2023-05-02 & H$\alpha$ & 1.38$\pm$0.28 & 30 & 60 & 90 & 1.51$\pm$0.34 & 0.48$\pm$0.06 & peak BD: 1.84/1/0.90/1.82 \\
        7  & 2023-05-02 & H$\beta$  & 2.96$\pm$1.12 & 20 & 60 & 80 & 0.82$\pm$0.32 & 0.25$\pm$0.05 & \\
        7  & 2023-05-02 & H$\gamma$ & 6.02$\pm$1.91 & 40 & 60 & 101 & 0.74$\pm$0.24 & 0.24$\pm$0.04 & \\
        7  & 2023-05-02 & H$\delta$ & 19.88$\pm$5.77 & 30 & 30 & 60 & 1.49$\pm$0.46 & 0.29$\pm$0.06 & \\
        \hline
        8  & 2023-05-02 & H$\alpha$ & 2.75$\pm$0.30 & 50 & 0 & 50 & 3.01$\pm$0.43 & 0.37$\pm$0.04 & only rise \\
        8  & 2023-05-02 & H$\beta$  & 6.45$\pm$1.17 & 20 & 0 & 20 & 1.78$\pm$0.36 & 0.13$\pm$0.02 & peak BD: 1.68/1/0.76/-- \\
        8  & 2023-05-02 & H$\gamma$ & 11.08$\pm$1.83 & 50 & 0 & 50 & 1.36$\pm$0.26 & 0.19$\pm$0.03 &  \\
        \hline
        9  & 2023-05-10 & H$\alpha$ & 5.43$\pm$0.24 & 16 & 128 & 144 & 6.04$\pm$0.63 & 2.21$\pm$0.11 & onset missed \\
        9  & 2023-05-10 & H$\beta$  & 12.65$\pm$0.92 & 16 & 110 & 126 & 3.59$\pm$0.43 & 1.09$\pm$0.08 & peak BD: 1.68/1/0.78/0.71 \\
        9  & 2023-05-10 & H$\gamma$ & 21.73$\pm$1.61 & 16 & 128 & 144 & 2.80$\pm$0.34 & 0.80$\pm$0.06 &  \\
        9  & 2023-05-10 & H$\delta$ & 31.74$\pm$5.39 & 16 & 128 & 144 & 2.56$\pm$0.50 & 0.83$\pm$0.10 &  \\
        9  & 2023-05-10 & \ion{Na}{i}\,D1 & 11.05$\pm$1.60 & 32 & 95 & 126 & 0.35$\pm$0.06 & 0.11$\pm$0.01 &  \\
        9  & 2023-05-10 & \ion{Na}{i}\,D2 & 9.10$\pm$1.60 & 16 & 128 & 144 & 0.27$\pm$0.05 & 0.11$\pm$0.01 &  \\
        9  & 2023-05-10 & \ion{He}{i}\,D3 & 6.21$\pm$1.18 & 16 & 110 & 126 & 0.53$\pm$0.11 & 0.21$\pm$0.03 &  \\
        9  & 2023-05-10 & \ion{He}{i}\,6678 & 2.86$\pm$0.69 & 0 & 144 & 144 & 0.29$\pm$0.07 & 0.08$\pm$0.02 &  \\
        \hline
        10 & 2023-05-26 & H$\alpha$ & 4.22$\pm$0.44  & 0 & 148 & 148 & 4.73$\pm$0.66 & 1.87$\pm$0.14 & only decay \\
        10 & 2023-05-26 & H$\beta$  & 15.69$\pm$1.76 & 0 & 114 & 114 & 4.60$\pm$0.67 & 1.05$\pm$0.11 & peak BD: 1.03/1/0.82/-- \\
        10 & 2023-05-26 & H$\gamma$ & 27.93$\pm$3.45 & 0 & 148 & 148 & 3.77$\pm$0.59 & 0.78$\pm$0.11 &  \\
        \hline
        11 & 2023-05-31 & H$\alpha$ & 3.45$\pm$0.25 & 268 & 32 & 300 & 3.78$\pm$0.45 & 3.11$\pm$0.13 & multi-peaked flare, decay missed \\
        11 & 2023-05-31 & H$\beta$  & 7.50$\pm$0.77 & 237 & 16 & 253 & 2.09$\pm$0.29 & 1.96$\pm$0.09 & peak BD: 1.81/1/0.91/1.3 \\
        11 & 2023-05-31 & H$\gamma$ & 15.48$\pm$1.74 & 268 & 32 & 300 & 1.91$\pm$0.28 & 1.67$\pm$0.09 &  \\
        11 & 2023-05-31 & H$\delta$ & 39.84$\pm$8.66 & 126 & 174 & 300 & 2.75$\pm$0.67 & 3.55$\pm$0.25 &  \\
        11 & 2023-05-31 & \ion{Na}{i}\,D1 & 7.11$\pm$1.64 & 268 & 32 & 300 & 0.22$\pm$0.05 & 0.24$\pm$0.02 &  \\
        11 & 2023-05-31 & \ion{Na}{i}\,D2 & 7.72$\pm$1.55 & 252 & 32 & 284 & 0.22$\pm$0.05 & 0.22$\pm$0.02 &  \\
        11 & 2023-05-31 & \ion{He}{i}\,D3 & 5.14$\pm$1.01 & 237 & 32 & 268 & 0.44$\pm$0.10 & 0.25$\pm$0.03 &  \\
        11 & 2023-05-31 & \ion{He}{i}\,6678 & 2.13$\pm$0.67 & 252 & 16 & 268 & 0.21$\pm$0.07 & 0.14$\pm$0.02 &  \\
        \hline
        12 & 2023-06-01 & H$\alpha$ & 6.69$\pm$0.20 & 34 & 158 & 192 & 7.39$\pm$0.73 & 3.15$\pm$0.13 & onset missed \\
        12 & 2023-06-01 & H$\beta$  & 18.91$\pm$0.91 & 34 & 189 & 223 & 5.22$\pm$0.55 & 2.60$\pm$0.12 & peak BD: 1.41/1/0.72/0.63 \\
        12 & 2023-06-01 & H$\gamma$ & 30.17$\pm$1.55 & 34 & 205 & 239 & 3.76$\pm$0.41 & 1.65$\pm$0.08 &  \\
        12 & 2023-06-01 & H$\delta$ & 41.69$\pm$6.14 & 34 & 189 & 223 & 3.31$\pm$0.59 & 1.45$\pm$0.15 &  \\
        12 & 2023-06-01 & \ion{Na}{i}\,D1 & 7.08$\pm$1.44 & 0 & 239 & 239 & 0.22$\pm$0.05 & 0.15$\pm$0.02 &  \\
        12 & 2023-06-01 & \ion{Na}{i}\,D2 & 9.19$\pm$1.35 & 32 & 110 & 142 & 0.27$\pm$0.05 & 0.13$\pm$0.01 &  \\
        12 & 2023-06-01 & \ion{He}{i}\,D3 & 8.35$\pm$0.94 & 34 & 158 & 192 & 0.71$\pm$0.10 & 0.39$\pm$0.03 &  \\
        12 & 2023-06-01 & \ion{He}{i}\,6678 & 3.81$\pm$0.58 & 34 & 158 & 192 & 0.38$\pm$0.07 & 0.20$\pm$0.02 &  \\
        \hline
	\end{tabular}
\end{table*}

\begin{table*}
	\centering
	\contcaption{}
	\begin{tabular}{cccccccccl}
		\hline
		flare & UT date & line & amplitude & rise & decay & duration & peak luminosity & energy & notes \\
        \# & & & (\%) & (min) & (min) & (min) & (10$^{27}$\,erg\,s$^{-1}$) & (10$^{31}$\,erg) & \\
		\hline
        13 & 2023-06-05 & H$\alpha$ & 5.02$\pm$0.30 & 16 & 273 & 289 & 5.50$\pm$0.62 & 3.89$\pm$0.21 & multi-peaked flare, onset missed \\
        13 & 2023-06-05 & H$\beta$  & 14.94$\pm$1.29 & 16 & 258 & 273 & 4.11$\pm$0.53 & 2.76$\pm$0.19 & peak BD: 1.34/1/0.84/0.79 \\
        13 & 2023-06-05 & H$\gamma$ & 28.08$\pm$3.13 & 16 & 210 & 226 & 3.45$\pm$0.51 & 1.74$\pm$0.16 &  \\
        13 & 2023-06-05 & H$\delta$ & 47.82$\pm$11.42 & 16 & 195 & 210 & 3.26$\pm$0.87 & 2.42$\pm$0.28 &  \\
        13 & 2023-06-05 & \ion{Na}{i}\,D1 & 7.34$\pm$2.11 & 16 & 258 & 273 & 0.24$\pm$0.07 & 0.19$\pm$0.04 &  \\
        13 & 2023-06-05 & \ion{Na}{i}\,D2 & 11.61$\pm$2.41 & 174 & 100 & 273 & 0.35$\pm$0.08 & 0.31$\pm$0.03 &  \\
        13 & 2023-06-05 & \ion{He}{i}\,D3 & 9.14$\pm$1.87 & 16 & 210 & 226 & 0.77$\pm$0.17 & 0.43$\pm$0.07 &  \\
        13 & 2023-06-05 & \ion{He}{i}\,6678 & 3.56$\pm$0.75 & 16 & 273 & 289 & 0.35$\pm$0.08 & 0.22$\pm$0.04 &  \\
        13 & 2023-06-05 & \textit{g'} & 27 & -- & -- & 11 & $\sim$752 & $\sim$11 & associated \textit{g'}-band flare  \\
		\hline
        14 & 2023-06-08 & H$\alpha$ & 2.16$\pm$0.29 & 63 & 95 & 158 & 2.33$\pm$0.38 & 1.46$\pm$0.11 & decay missed; preceding flare (decay \\
        14 & 2023-06-08 & H$\beta$  & 5.50$\pm$1.46 & 79 & 47 & 126 & 1.49$\pm$0.42 & 0.56$\pm$0.08 & only) not studied (bad data quality) \\
        14 & 2023-06-08 & H$\gamma$ & 10.81$\pm$3.08 & 158 & 0 & 158 & 1.29$\pm$0.39 & 0.55$\pm$0.07 & peak BD: 2.17/1/0.75/-- \\
        \hline
        15 & 2023-06-15 & H$\alpha$ & 4.99$\pm$0.34 & 79 & 126 & 205 & 5.48$\pm$0.64 & 2.27$\pm$0.15 & multi-peaked flare \\
        15 & 2023-06-15 & H$\beta$  & 14.49$\pm$1.19 & 79 & 110 & 189 & 4.05$\pm$0.51 & 1.83$\pm$0.13 & peak BD: 1.35/1/0.79/0.93 \\
        15 & 2023-06-15 & H$\gamma$ & 26.05$\pm$5.71 & 79 & 142 & 221 & 3.20$\pm$0.78 & 1.81$\pm$0.23 &  \\
        15 & 2023-06-15 & H$\delta$ & 48.83$\pm$8.27 & 32 & 110 & 142 & 3.78$\pm$0.75 & 1.12$\pm$0.17 &  \\
        15 & 2023-06-15 & \ion{Na}{i}\,D2 & 13.05$\pm$4.09 & 63 & 142 & 205 & 0.38$\pm$0.13 & 0.18$\pm$0.04 &  \\
        15 & 2023-06-15 & \ion{He}{i}\,D3 & 8.24$\pm$2.24 & 79 & 126 & 205 & 0.70$\pm$0.20 & 0.29$\pm$0.06 &  \\
        15 & 2023-06-15 & \ion{He}{i}\,6678 & 6.63$\pm$1.30 & 79 & 79 & 158 & 0.65$\pm$0.14 & 0.34$\pm$0.04 &  \\
        \hline
        16 & 2023-06-16 & H$\alpha$ & 3.82$\pm$0.23 & 205 & 47 & 252 & 4.14$\pm$0.46 & 1.80$\pm$0.08 & multi-peaked flare, but with large data gap \\
        16 & 2023-06-16 & H$\beta$  & 8.70$\pm$0.81 & 205 & 32 & 237 & 2.41$\pm$0.32 & 0.79$\pm$0.05 & and missing decay \\
        16 & 2023-06-16 & H$\gamma$ & 14.60$\pm$1.41 & 205 & 32 & 237 & 1.80$\pm$0.24 & 0.54$\pm$0.04 & peak BD: 1.71/1/0.74/0.56 \\
        16 & 2023-06-16 & H$\delta$ & 16.98$\pm$4.86 & 205 & 16 & 221 & 1.36$\pm$0.41 & 0.43$\pm$0.05 & \\
        16 & 2023-06-16 & \ion{He}{i}\,D3 & 8.16$\pm$1.33 & 221 & 47 & 268 & 0.69$\pm$0.13 & 0.10$\pm$0.02 &  \\
        16 & 2023-06-16 & \ion{He}{i}\,6678 & 2.47$\pm$0.82 & 205 & 63 & 268 & 0.25$\pm$0.08 & 0.10$\pm$0.02 &  \\
        \hline
        17 & 2023-06-24 & H$\alpha$ & 3.69$\pm$0.34 & 79 & 0 & 79 & 4.00$\pm$0.53 & 1.00$\pm$0.09 & only rise \\
        17 & 2023-06-24 & H$\beta$  & 9.88$\pm$1.14 & 47 & 16 & 63 & 2.68$\pm$0.40 & 0.65$\pm$0.07 & peak BD: 1.5/1/0.79/0.78 \\
        17 & 2023-06-24 & H$\gamma$ & 17.36$\pm$2.05 & 79 & 0 & 79 & 2.10$\pm$0.32 & 0.46$\pm$0.05 &  \\
        17 & 2023-06-24 & H$\delta$ & 39.03$\pm$11.98 & 63 & 32 & 95 & 2.72$\pm$0.89 & 0.76$\pm$0.15 &  \\
        17 & 2023-06-24 & \ion{Na}{i}\,D1 & 8.83$\pm$2.54 & 95 & 0 & 95 & 0.29$\pm$0.09 & 0.06$\pm$0.02 &  \\
        17 & 2023-06-24 & \ion{He}{i}\,D3 & 5.93$\pm$1.77 & 95 & 0 & 95 & 0.50$\pm$0.16 & 0.10$\pm$0.03 &  \\
        \hline
        18 & 2023-06-30 & H$\alpha$ & 2.98$\pm$0.61 & 35 & 100 & 135 & 3.35$\pm$0.75 & 1.69$\pm$0.16 & onset missed \\
        18 & 2023-06-30 & H$\beta$  & 9.20$\pm$2.47 & 49 & 17 & 66 & 2.65$\pm$0.75 & 0.62$\pm$0.08 & peak BD: 2.74/1/--/-- \\
        \hline
        19 & 2023-07-01 & H$\alpha$ & 7.03$\pm$0.97 & 68 & 83 & 151 & 7.53$\pm$1.25 & 3.11$\pm$0.29 & data gaps \\
        19 & 2023-07-01 & H$\beta$  & 19.83$\pm$2.44 & 50 & 99 & 149 & 5.25$\pm$0.81 & 2.74$\pm$0.28 & peak BD: 1.51/1/0.61/-- \\
        19 & 2023-07-01 & H$\gamma$ & 33.60$\pm$5.06 & 78 & 99 & 177 & 3.87$\pm$0.69 & 1.73$\pm$0.19 &  \\
        \hline
        20 & 2023-08-12 & H$\alpha$ & 2.78$\pm$0.37 & 56 & 87 & 143 & 2.92$\pm$0.48 & 1.61$\pm$0.08 & peak BD: 2.11/1/1.04/-- \\
        20 & 2023-08-12 & H$\beta$  & 6.90$\pm$1.20 & 6 & 105 & 110 & 1.81$\pm$0.36 & 0.80$\pm$0.05 & \\
        20 & 2023-08-12 & H$\gamma$ & 13.49$\pm$2.57 & 61 & 99 & 160 & 1.55$\pm$0.33 & 0.59$\pm$0.06 & \\
        \hline
        21 & 2023-08-13 & H$\alpha$ & 4.79$\pm$0.35 & 37 & 12 & 48 & 5.12$\pm$0.61 & 0.59$\pm$0.05 & decay missed \\
        21 & 2023-08-13 & H$\beta$  & 13.87$\pm$1.70 & 25 & 17 & 43 & 3.72$\pm$0.58 & 0.51$\pm$0.05 & peak BD: 1.63/1/1.40/-- \\
        21 & 2023-08-13 & H$\gamma$ & 40.34$\pm$7.79 & 43 & 0 & 43 & 4.40$\pm$0.96 & 0.48$\pm$0.07 &  \\
        \hline
        22 & 2023-08-15 & H$\alpha$ & 10.37$\pm$1.22 & 176 & 73 & 250 & 11.10$\pm$1.67 & 4.76$\pm$0.28 & multi-peaked flare, decay \\
        22 & 2023-08-15 & H$\beta$  & 25.98$\pm$4.73 & 109 & 36 & 146 & 7.24$\pm$1.49 & 1.10$\pm$0.14 & missed; several data gaps \\
        22 & 2023-08-15 & H$\gamma$ & 32.90$\pm$9.89 & 133 & 73 & 206 & 3.81$\pm$1.21 & 1.62$\pm$0.31 & peak BD: 1.53/1/--/-- \\
        22 & 2023-08-15 & \ion{Na}{i}\,D2 & 41.73$\pm$10.28 & 177 & 17 & 194 & 1.15$\pm$0.31 & 0.37$\pm$0.03 &  \\
        \hline
        23 & 2023-09-16 & H$\alpha$ & 28.97$\pm$0.59 & -- & -- & $>$36 & 32.35$\pm$3.12 & $>$6.99$\pm$1.18 & peak only, extreme event \\
        23 & 2023-09-16 & H$\beta$  & 77.86$\pm$2.11 & -- & -- & $>$36 & 23.30$\pm$2.30 & $>$5.03$\pm$0.86 & peak BD: 1.39/1/0.85/0.73 \\
        23 & 2023-09-16 & H$\gamma$ & 147.6$\pm$8.72 & -- & -- & $>$36 & 19.88$\pm$ 2.31 & $>$4.29$\pm$0.78 & see Section~\ref{sec:superflare} \\
        23 & 2023-09-16 & H$\delta$ & 180.4$\pm$48.2 & -- & -- & $>$36 & 17.07$\pm$5.60 & $>$3.69$\pm$1.31 &  \\
        23 & 2023-09-16 & \ion{Na}{i}\,D1 & 41.80$\pm$3.59 & -- & -- & $>$36 & 1.37$\pm$0.18 & $>$0.30$\pm$0.06 &  \\
        23 & 2023-09-16 & \ion{Na}{i}\,D2 & 51.63$\pm$3.90 & -- & -- & $>$36 & 1.59$\pm$0.19 & $>$0.34$\pm$0.06 &  \\
        23 & 2023-09-16 & \ion{He}{i}\,D3 & 36.29$\pm$2.20 & -- & -- & $>$36 & 3.13$\pm$0.35 & $>$0.67$\pm$0.12 &  \\
        23 & 2023-09-16 & \ion{He}{i}\,6678 & 13.74$\pm$1.55 & -- & -- & $>$36 & 1.38$\pm$0.20 & $>$0.30$\pm$0.06 &  \\
        \hline
        24 & 2023-09-17 & H$\alpha$ & 5.84$\pm$0.72 & 12 & 138 & 150 & 6.45$\pm$1.00 & 2.04$\pm$0.19 & preceding flare (decay only) from \#23? \\
        24 & 2023-09-17 & H$\beta$  & 17.64$\pm$2.79 & 23 & 121 & 144 & 4.96$\pm$0.91 & 1.43$\pm$0.15 & peak BD: 2.19/1/--/-- \\
        \hline
	\end{tabular}
\end{table*}

\clearpage


\bsp	
\label{lastpage}
\end{document}